\documentclass{JHEP3}
\usepackage{graphicx,epsf,epsfig}
\usepackage{amsmath}
\usepackage{amssymb}
\usepackage{amsfonts} 
\usepackage{mathrsfs}
\usepackage{bbm}

{
\def\eq#1{{eq.~(\ref{#1})}}

\def\vev#1{\left\langle #1\right\rangle}

\def\hbar{\hspace{0pt}\raisebox{1pt}{$-$} \hspace{-7pt} h}

\newcommand{\be}{\begin{equation}}
\newcommand{\ee}{\end{equation}}
\newcommand{\bd}{\begin{displaymath}}
\newcommand{\ed}{\end{displaymath}}
\newcommand{\bea}{\begin{eqnarray}}
\newcommand{\eea}{\end{eqnarray}}
\newcommand{\nn}{\nonumber}

\def\so10{$SO(10)$}

\title{Solving the SUSY Flavour and CP Problems with $SU(3)$ Family Symmetry}
\author{Stefan Antusch\\ 
Max-Planck-Institut f\"ur Physik (Werner-Heisenberg-Institut)\\
F\"ohringer Ring 6, D-80805 M\"unchen, Germany
\\ E-mail: \email{antusch@mppmu.mpg.de}} 
\author{Stephen
F. King\\ School of Physics and Astronomy, University of Southampton,
SO16 1BJ Southampton, United Kingdom\\ E-mail:
\email{sfk@hep.phys.soton.ac.uk}} \author{Michal Malinsk\'{y}\\ School
of Physics and Astronomy, University of Southampton, SO16 1BJ
Southampton, United Kingdom\\ E-mail:
\email{malinsky@phys.soton.ac.uk}}
\abstract{We show how the SUSY flavour and CP problems can be solved
  using gauged $SU(3)$ family symmetry previously introduced to describe 
quark and lepton masses and mixings, in particular neutrino
  tri-bimaximal mixing via constrained sequential dominance.
The Yukawa and soft trilinear and scalar mass squared matrices and
kinetic terms are expanded in powers of the flavons used to
spontaneously break the $SU(3)$ family symmetry, and the canonically
normalized versions of these matrices are constructed.
The soft mass matrices are then expressed in the Super-CKM basis,
and the leading order mass insertion parameters are calculated,
and are shown to satisfy the experimental constraints from 
flavour changing neutral current processes. Assuming that 
CP is spontaneously broken by the flavons, the next-to-leading
order effects responsible for CP violation are then estimated,
and the predictions for electric dipole moments are shown to be
an order of magnitude more suppressed than those predicted from
the constrained minimal supersymmetric standard model (CMSSM),
and may be further suppressed if the high energy trilinear soft 
parameter is assumed to be relatively small. We also predict
that, unlike in the CMSSM,
$\varepsilon_{K}'/\varepsilon_{K}$ may be dominated by the SUSY operator $O_8$.
We also discuss the additional constraints from unification,
which can lead to further predictions for flavour changing in our
scheme.}
\date{\today}

\begin{document}
\newpage
\section{Introduction}

The Flavour Problem in the Standard Model (SM) is one of the
deepest mysteries in physics. 
In the absence of neutrino mass and mixing, the flavour sector
of the SM involves ten parameters related to the quark sector
(six masses, three angles and one phase) plus the three charged
lepton masses. In the presence of neutrino mass and mixing
\cite{King:2003jb}, 
there could be a further nine parameters (three Majorana masses,
three angles and three phases), or more or less parameters depending
on the precise origin of neutrino masses and mixing.
The qualitative smallness of neutrino mixing has led many people
to conclude that neutrino mass must be associated with new physics
beyond the SM, although there is so far no consensus on the nature of that
new physics. A minimal Majorana approach is to consider higher
order non-renormalizable dimension five operators of the form
$\lambda_{ij} L_iL_j HH/M_{ij}$ \cite{Weinberg:1980bf}
where $L_i$ are lepton doublets,
$H$ are Higgs doublets, $\lambda_{ij}$ are Yukawa couplings and
$M_{ij}$ are some large mass scales. A minimal Dirac approach is to
conserve lepton number and 
add right-handed neutrinos $N_j$ and couple them to lepton
doublets as $h_{ij}L_iN_jH$ where $h_{ij}$ are (very small) Yukawa
couplings. There are of course many other approaches, some of which map onto
one or other of these minimal approaches, and some which do not.

It has recently been realised that $SU(3)$ family symmetry 
\cite{King:2001uz}, \cite{King:2003rf}
can lead to a solution to the Flavour Problem of the SM,
including answers to the five distinct questions:
Why are there three families of quarks and leptons?
Why are quark and charged lepton masses so peculiar?
Why are neutrino masses so small?
Why is lepton mixing so large compared to quark mixing?
What is the origin of CP violation?
The answer to the first question is provided by gauging the
$SU(3)$ symmetry, which puts the question of three families
on the same footing as that of the three quark colours.
The peculiar nature of the quark and charged lepton masses
is accounted for in terms of hierarchical textures 
based on powers of expansion parameters, each of which 
is not itself very hierarchical (typically of order 0.1).
The smallness of neutrino masses will be due to the
see-saw mechanism \cite{Minkowski:1977sc}, 
and the large lepton mixing will be
due to the sequential dominance (SD) mechanism \cite{King:1998jw},
\cite{King:1999mb}.
Indeed present neutrino oscillation data is consistent
with approximate tri-bimaximal lepton mixing \cite{Harrison:2002er},
and this can be readily achieved with constrained sequential
dominance (CSD) \cite{King:2005bj}, \cite{deMedeirosVarzielas:2005ax}, \cite{Antusch:2007vw}.
Finally the origin of CP violation can be due to 
the spontaneous breaking of the $SU(3)$ family symmetry
\cite{Ross:2002mr}, \cite{Ross:2004qn}.

In hierarchical models, $SU(3)$ family symmetry is broken
by so called flavon vacuum expectation values (VEVs) $\langle\phi \rangle$,
which may also have CP violating phases,
and the Yukawa couplings are generated in terms of powers
of some expansion parameters $\varepsilon = |\langle\phi \rangle|/M$ where $M$
is the so called messenger mass scale. If $\varepsilon <1$
then such small expansion parameters can serve to
describe the small Yukawa couplings in a hierarchical parametrization.
Such parametrizations have been proposed based on symmetric
Yukawa matrices with the $(1,1)$ elements being zero,
allowing the successful fermion mass relations to emerge. 
In fact, in the quark sector, since the up-type quark masses
are more hierarchical than the down-type quark masses,
two expansion parameters are required: $\varepsilon \approx 0.05$
and $\bar{\varepsilon} \approx 0.15$, with the quark Yukawa matrices
taking the approximate form \cite{Roberts:2001zy}:
\begin{eqnarray}
 Y^u \approx \begin{pmatrix}
0 & \varepsilon^3 & \varepsilon^3 \\
\varepsilon^3 & \varepsilon^2 & \varepsilon^2 \\
\varepsilon^3 & \varepsilon^2  & 1
\end{pmatrix}, \ \ \ \ 
Y^d \approx 
\begin{pmatrix}
0 & \bar\varepsilon^3 & \bar\varepsilon^3 \\
\bar\varepsilon^3 & \bar\varepsilon^2 & \bar\varepsilon^2 \\
\bar\varepsilon^3 & \bar\varepsilon^2  & 1
\end{pmatrix},
\end{eqnarray}
with independent undetermined order unity coefficients multiplying
each matrix element suppressed above. Recently a phenomenological fit
has been performed which determines the coefficients of the individual
matrix elements and the expansion parameters of such Yukawa textures 
precisely at high energies by comparing to the quark masses and
mixings run up to the high energy scale.
Such an analysis assumes low energy (TeV scale) supersymmetry (SUSY),
and includes possible low energy SUSY threshold effects \cite{Ross:2007az}.

With the addition of low energy SUSY
the Flavour Problem increases dramatically, due to the undetermined
superpartner masses, mixings and phases that must also be explained
\cite{Chung:2003fi}.
Indeed in SUSY extensions of the SM there are typically
about a hundred or so additional physical
parameters associated with the soft SUSY breaking Lagrangian,
depending on the precise nature of the SUSY SM and the origin
of neutrino masses and mixings in the SUSY context.
Moreover much of this parameter space is ruled out by the 
fact that high precision LEP measurements, Tevatron particle
searches, and high statistics experiments such as searches for
charged lepton flavour violation and electric dipole moments, as
well as Kaon and B 
flavour physics experiments are almost all consistent with 
the SM \cite{Chung:2003fi}. 
The only flavour changing neutral currents (FCNCs) that have
been observed are those consistent with the SM expectations.
The only significant crack in the SM edifice is the
hint of new physics coming from the anomalous magnetic moment of the
muon \cite{Miller:2007kk}. 
Many people argue that if SUSY was present at the TeV scale, there would
have been many signals of this by now, given the overwhelmingly large 
regions of soft SUSY parameter space in which SUSY by now would have
been discovered. Indeed fine-tuning arguments indicate that much of the
remaining parameter space of the minimal supersymmetric standard model
(MSSM) is unnatural or fine-tuned, mainly due to the failure
to discover the Higgs boson at LEP \cite{Kane:1998im}, 
and this is particularly 
accentuated if one requires the MSSM to give electroweak baryogenesis,
although both problems can be alleviated in the next-to-minimal
supersymmetric standard model (NMSSM) \cite{BasteroGil:2000bw}. 

In this paper we shall show how $SU(3)$
family symmetry can not only solve the Flavour Problem of the SM,
but also that of its SUSY extensions.
It has been also known for some time that, if the SM was extended
to include SUSY, then, in the $SU(3)$ family symmetry limit, the 
soft squark and slepton mass squared matrices would have a universal
form, proportional to unit matrices, enforced by the $SU(3)$ family
symmetry. However, in the $SU(3)$ family symmetry limit, the
Yukawa and soft trilinear matrices vanish, so in the real world 
the family symmetry must be spontaneously broken leading
simultaneously to flavour in the Yukawa sector, and violations
of universality in the soft SUSY breaking sector. In principle
therefore $SU(3)$ family symmetry can provide simultaneously
a solution to the Flavour Problem not only in the SM but also
in its SUSY extensions, since the violations of squark and slepton
soft mass universality are controlled by the same order parameters
$\varepsilon $ as are responsible for the origin of Yukawa couplings,
resulting in the prediction of (suppressed) FCNCs
\cite{Ross:2002mr}, \cite{Ross:2004qn}.
Furthermore, it has been postulated that CP is an exact symmetry
of the high energy (string) theory being spontaneously
broken (only) by phases in the flavon VEVs. In such a case 
$SU(3)$ family symmetry can explain the smallness of the phases
in soft SUSY breaking Lagrangian in general, and the 
suppression of electric dipole moments (EDMs) of the neutron
and electron in particular \cite{Ross:2002mr}, \cite{Ross:2004qn}.
These previous analyses were performed in the framework of a specific
scenario for SUSY breaking, namely SUGRA
\cite{Ross:2002mr}, \cite{Ross:2004qn}, and it is of interest
to know how much of the effects arise from the $SU(3)$
family symmetry and how much from the details of the SUGRA model.

In this paper we shall perform a detailed bottom-up operator analysis
of the soft SUSY breaking Lagrangian in terms of a 
spontaneously broken $SU(3)$ family symmetry, where 
the operator expansions are assumed to be $SU(3)$-symmetric. We shall make a careful
estimate of the mass insertion parameters describing flavour changing  
and CP violation, keeping track explicitly of all the coefficients,
including a careful treatment of canonical normalization effects.
Although our analysis is independent of the details of the 
mechanism of SUSY breaking (as long as the SUSY breaking scale is higher than the scale of the family symmetry breakdown so that the effective soft SUSY-breaking operators do obey the family symmetry)
the results do depend on the precise way that the family
symmetry is broken. For definiteness we shall assume that the
$SU(3)$ family symmetry is spontaneously broken in such a way
as to give rise to tri-bimaximal neutrino mixing, with small
corrections to tri-bimaximal lepton mixing coming from quark-like
charged lepton mixing angles predicted from theory, leading
to a prediction for $\theta_{13}$ and a neutrino mixing sum rule
\cite{King:2005bj}, \cite{Masina:2005hf}, \cite{Antusch:2005kw},
\cite{Antusch:2007rk}.

This approach requires three specific types of flavon fields,
each of which is an anti-triplet of the $SU(3)$ family symmetry,
and each of which has a particular type of vacuum alignment,
namely: $\vev{\phi_3} \sim (0,0,1)$, $\vev{\phi_{23}} \sim (0,1,1)$,
$\vev{\phi_{123}} \sim (1,1,1)$, up to phases. In practice, the desired
vacuum alignment must also ensure that $\vev{\phi_{23}^\dag}.\vev{ \phi_{123}}=0$,
in accordance with the CSD requirements
necessary to yield tri-bimaximal neutrino mixing
\cite{King:2005bj}, \cite{deMedeirosVarzielas:2005ax}.
Within such a framework, assuming such flavons,
we shall perform a bottom-up operator expansion
of the soft SUSY breaking Lagrangian, focussing on the predictions
for the soft squark and slepton mass squared matrices,
and the soft trilinear mass matrices. Since a similar operator
expansion also predicts the quark and lepton Yukawa matrices,
we are then able to construct the soft mass squared and trilinear
mass matrices in the basis in which the charged fermions are diagonal,
the so called super-CKM basis 
\cite{Chung:2003fi} (for the quarks) and an analogous
diagonal charged lepton basis, under certain common assumptions about the
messenger sectors. The mass-insertion $\delta$ parameters may then
be directly read-off in this basis, leading to the prediction
of (suppressed) FCNCs. We note that such predictions provide a
smoking gun signature of the $SU(3)$ family
symmetry in the SUSY spectrum, providing an indirect test of such
a symmetry which would be difficult to verify without SUSY.
One technicality worth mentioning is that
violations of $SU(3)$ family
symmetry also give rise to non-canonically normalized kinetic terms
\cite{King:2003xq}, \cite{King:2004tx} 
and we are always careful to include the effects of canonically normalising
these terms before interpreting the physical results.

Our analysis may be compared to the other related analyses in the
literature mentioned above \cite{Ross:2002mr}, \cite{Ross:2004qn}.
These previous analyses of the SUSY flavour and CP issues were
based on the gravity mediated or SUGRA type of SUSY breaking,
together with $SU(3)$ family symmetry.
In this paper we shall focus on the role played by the family symmetry if it is extended appropriately to the soft SUSY-breaking sector. As we shall see, $SU(3)$ in certain realizations can be powerful enough to significantly alleviate the SUSY flavour and CP problem of the low-scale supersymmetry.     
Although the details are model-dependent, many of the 
predictions will be relying only on the basic features of the $SU(3)$ symmetry.

In this paper we assume a specific scenario
involving the three flavons $\phi_3$, $\phi_{23}$, $\phi_{123}$
whose vacuum alignment gives rise to tri-bimaximal neutrino mixing,
since such a model has not been discussed previously, even in the SUGRA
context. Compared to previous models, the effect of the new flavon
$\phi_{123}$ is to greatly simplify the analysis, enabling a full
and detailed treatment in which we keep track explicitly of all
the dimensionless coefficients through the procedure of
canonical normalization and going to the SCKM basis.
In the previous models, the operator analysis was much more
complicated, rendering an explicit treatment intractable
\cite{Ross:2002mr}, \cite{Ross:2004qn}. Furthermore, we shall
show that in the tri-bimaximal models the EDMs have an additional
Cabibbo suppression factor compared to the previous models
\cite{Ross:2002mr}, \cite{Ross:2004qn}.

We may also compare the results of 
such models based on $SU(3)$ family symmetry to the conventional
minimal supergravity (mSUGRA) or constrained minimal supersymmetric
standard model (CMSSM) approaches. The latter assume
or postulate universal soft mass matrices at the high energy scale,
and any observed non-universality at low energy is due to
renormalization group (RG) running effects, including those due to 
the right-handed neutrino couplings entering the see-saw
mechanism. In the present paper we consider the corrections to
non-universality already at the high energy scale, due to the
$SU(3)$ violating effects of flavons. This implies that we predict
a higher amount of flavour changing than in mSUGRA or CMSSM, in
general, although still below current experimental limits.
However predicted EDMs in the effective $SU(3)$ family symmetry
approach are in fact smaller than those predicted in mSUGRA or CMSSM,
being suppressed by approximately one further power of the Cabibbo
angle. Similar results would also apply if 
$SU(3)$ family symmetry is replaced by a discrete subgroup 
\cite{deMedeirosVarzielas:2005qg}, 
which has the additional benefits of being more readily
obtained in string theory, yielding more readily the desired
flavon vacuum alignments, and not being subject to additional
sources of flavour violating $D$-terms. 
In any case we do not consider the
$D$-terms sources of flavour violation at all in this paper, 
since they have recently been considered elsewhere 
\cite{deMedeirosVarzielas:2006ma}.
Finally we note that other class of tri-bimaximal models based on
$SO(3)$ \cite{King:2005bj}, \cite{King:2006me} and
$\Delta(12)=A_{4}$ \cite{King:2006np} 
family symmetry have been proposed, and they 
will be considered in a future publication, along with a 
full analysis in the framework of an effective SUGRA approach.

The plan of the remainder of the paper is as follows: 
In section 2 we introduce $SU(3)$ family symmetry 
models with tri-bimaximal mixing, including our operator
analysis for the Yukawa matrices, the trilinear and scalar
soft matrices and the K\"ahler potential. We also discuss the
constraints of GUTs, and in particular $SO(10)$. We then go on to 
recast the results in the canonically normalized basis,
keeping explicit track of all the coefficients.
In section 3 we discuss the phenomenology arising from our
operator expansions. First we make some general remarks
about SUSY flavour changing, including the SCKM basis,
defining the mass insertion parameters, and giving a survey
of the experimental constraints on these parameters,
before discussing the SUSY Flavour and CP Problems.
We then perform a phenomenological analysis of the 
$SU(3)$ family symmetry models with tri-bimaximal mixing,
and give explicit forms for the soft mass matrices
in the SCKM basis, to leading order in our expansion 
parameters, with and without GUT constraints,
from which the mass insertion parameters can be readily
extracted and compared to the data, after including the effects
of RG running. CP violation is then discussed, which requires
a next-to-leading order operator analysis. The CP violation
relevant for third family phenomenology, EDMs and the Kaon system is 
subsequently discussed in detail. Finally section 4 concludes the paper.

 \paragraph{A note about notation:}\mbox{}\\
In what follows we shall work in three different bases - the (current)
basis in which the model is originally formulated, the canonical basis
in which the kinetic terms receive canonical form and finally the
physical super-CKM (SCKM) basis that is very convenient for SUSY
phenomenology.

The couplings entering the effective expansions in the defining basis
shall be denoted by a unique operator-specific {\it single-digit}
numerical subscript like e.g. $y_{1}$, $a_{2}$, $k_{3}$, $b_{4}$
etc. (appearing in the Yukawa, trilinear, K\"ahler and
soft-mass-squared sectors respectively). The corresponding matrix
structures shall be equipped with hats, i.e. $\hat Y$, $\hat A$, $\hat
K$ and $\hat m^{2}$ and so on.  Once we get to the canonical basis the
relevant matrices will lose the hats (i.e. we shall use $Y$, $A$,
$m^{2}$) and their entries shall be denoted by the same set of letters
as before but with {\it position-specific} (i.e. double) subscripts
(e.g. $y_{12}$, $a_{23}$, $b_{33}$ etc.).  Last, the SCKM basis
quantities will all receive tildes, i.e. $\tilde{Y}$ shall be diagonal
matrices with eigenvalues $\tilde{y}_{11}$, $\tilde{y}_{22}$ and
$\tilde{y}_{33}$; $\tilde{A}$ matrices will be given in terms of
$\tilde a_{ij}$ coefficients and so on.

Second, unless specified otherwise, we shall denote the SM matter $SU(2)_{L}$ doublet fields $Q$, $L$ by $f$ while $f^{c}$ will be used for the corresponding $SU(2)_{L}$ singlets $u^{c}$, $d^{c}$, $e^{c}$, $\nu^{c}$. On the other hand, $f$ as a superscript in the Yukawa and trilinear sector couplings (e.g. $y_{1}^{f}$, $a_{1}^{f}$ etc.) 
shall correspond to $u,d,e,\nu$ in the usual manner, i.e. for instance $y^{u}$ will show up in the up-type Yukawa sector couplings $\propto Q Y^{u} u^{c}H_{u}$ etc.  
We shall also drop all the charge conjugation matrices in the supersymmetric Yukawa couplings.
\section{$SU(3)$ and SUSY: Operator expansions and canonical normalization}

\subsection{$SU(3)$ predictions for Yukawa matrices\label{Yukawasection}}

In the MSSM, the Yukawa piece of the superpotential is given by \bea
W_{Y} & = & \varepsilon_{\alpha\beta} \left[ - \hat{H}_u^\alpha
\hat{Q}^{\beta i} Y^{u}_{ij} \hat{u}^{cj} - \hat{H}_d^\alpha
\hat{Q}^{\beta i} Y^{d}_{ij} \hat{d}^{cj} - \hat{H}_u^\alpha
\hat{L}^{\beta i} Y^{\nu}_{ij} \hat{\nu}^{cj} - \hat{H}_d^\alpha
\hat{L}^{\beta i} Y^{e}_{ij} \hat{e}^{cj} \right]. 
\eea 
If a family symmetry is employed, the Yukawa operators originate from
higher dimensional operators involving flavons  
which break the family symmetry, and one can write in general 
\be\label{Yukawaoperator}
W_Y=\sum_{\{f,f^{c}\}}\hat{H}
\hat{f}\left(\sum_{\Phi\Phi'}y^{ff^{c}}_{\Phi\Phi'}\frac{\hat\Phi\otimes\hat\Phi'}{M_{ff^{c},\Phi\Phi'}
^{2}}+\ldots\right)\hat{f}^{c}+\ldots
\ee 
where $\hat f$ and $\hat f^{c}$ stand for the flavour multiplets
of left chiral superfields in the model\footnote{The summation here is taken over the ``chirality'' pairs of $f$ and $f^{c}$ only, e.g. $\{f,f^{c}\}=\{Q,d^{c}\},\{L,e^{c}\}$ for $H=H_{d}$ etc. The ellipses stand for higher order flavon and/or SM singlet Higgs insertions.}, $\hat\Phi$, $\hat\Phi'$ etc. are generic symbols denoting flavon fields in a specific model and $\hat H$ is the Higgs superfield of the relevant hypercharge. The number of such multiplets
depends on the family symmetry imposed. 
In this paper we are mainly concerned with an
$SU(3)$ family symmetry, under which both left and right-handed fields
are triplets ${\bf 3}$,
so in this case $\hat f$ runs
only over the two triplets $\hat Q$, $\hat L$ and $\hat f^{c}$ denotes
only four structures - the triplets $\hat u^{c}$, $\hat d^{c}$, $\hat
e^{c}$ and $\hat \nu^{c}$
where $\hat Q=(\hat Q^{1}\ldots\hat Q^{3})$, 
$\hat L=(\hat L^{1}\ldots \hat L^{3})$ while 
$\hat f^{c}$ stands for $\hat u^c=(\hat u^{c1}\ldots \hat u^{c3})$,
$\hat d^c=(\hat d^{c1}\ldots \hat d^{c3})$ etc.
The tensor products of the $\hat \Phi$,
$\hat \Phi'$ etc. flavons are coupled appropriately to the matter sector
bilinears so that the whole structure is a family symmetry singlet.

Note also that there are in principle at least two distinct types of messengers entering the formula (\ref{Yukawaoperator}), in particular those transmitting the $SU(2)_{L}$ doublet nature of $f=Q, L$ to the Higgs VEV insertion point (for definiteness let's call them $\chi_{Q,L}$), and also the singlet ones propagating further the remaining $SU(3)_{c}\otimes U(1)_{Y}$  quantum numbers to $f^{c}=u^{c}$, $d^{c}$, $e^{c}$ and $\nu^{c}$ (to be called $\chi_{u,d,e,\nu}$). However, for the sake of simplicity, we shall use a generic symbol $M_{f}$ for both these classes and come back to this distinction only upon getting to physical implications\footnote{In this study we shall not, however, address the question of topology of the underlying messenger sector Feynman graphs giving rise to the operators under consideration. An interested reader can find more information for instance in \cite{King:2006me} and references therein. }.  

As discussed in the Introduction, 
it has been pointed out that the observed close-to
tri-bimaximal lepton mixing, along with the main features of the quark
and charged lepton masses and mixings, can be understood if there are
three flavons $\phi_3 , \phi_{23} , \phi_{123}$
which are antitriplets ${\bf \bar 3}$ with VEVs pointing in particular directions in SU(3) space, that appear in the following Yukawa operators 
(dropping superfield hats and using $f$ instead of the redundant $ff^{c}$ indices) \cite{deMedeirosVarzielas:2005ax}:
\bea
\label{TBYukawa} 
W_{Y}&=&
f^{i}f^{cj}
\frac{H}{M_{f}^{2}}\left[y_1^f(\phi_{123})_{i}(\phi_{23})_{j}+y_2^f(\phi_{23})_{i}(\phi_{123})_{j}+y_3^f(\phi_{3})_{i}(\phi_{3})_{j}+{y_{4}^f}(\phi_{23})_{i}(\phi_{23})_{j}\right]+\ldots
\nonumber
\eea
This approach requires a particular vacuum alignment in the $SU(3)$ space: 
$\vev{\phi_3} \sim (0,0,1)$, $\vev{\phi_{23}} \sim (0,1,1)$,
$\vev{\phi_{123}} \sim (1,1,1)$, up to phases. These structures should emerge from minimization of the relevant piece of the total scalar potential. However, a full-fledged discussion of how this can be achieved is beyond the scope of this work and  we shall defer an interested reader to the original paper \cite{deMedeirosVarzielas:2005ax} for further details.

Let us see now how the generic phase condition
$\vev{\phi_{23}^\dag}.\vev{\phi_{123}}=0$
(which is necessary to yield tri-bimaximal neutrino mixing in accordance with the constrained sequential dominance requirements \cite{King:2005bj}) can be mapped onto the phase structure of the vacuum.
Including the most general set of 
phases, the desired vacuum structure can be written as\footnote{Strictly speaking, we should in principle admit a phase difference between the two $SU(2)_{R}$ components of $\langle\phi_{3}\rangle$. However, it is easy to see that such an extra phase does not affect the results of the following analysis because it can enter only the 33 entries of either up- or down-type Yukawas (and the corresponding $A$-terms) and thus is effectively undone upon bringing the trilinear couplings to the SCKM basis. Moreover, the would-be physical effects in the soft masses are screened because  $\phi_{3}$ enters the soft masses and the K\"ahler potential in conjugated pairs only and only the subsequent SCKM rotation reveals such a phase; however, as we shall see in section \ref{sectionoffdiagonalities}, it is irrelevant for the physics because of the overall suppression of the off-diagonalities in the soft terms. Remarkably enough, even the CKM CP phase is essentially insensitive to such a phase difference because it can be almost entirely rotated away upon bringing the CKM mixing matrix into the standard form; for further comments see section \ref{sectionGUT}.}:
\begin{eqnarray}\label{VEVsinthegeneralbasis}
\!\!\!\langle\phi_{123}\rangle = \begin{pmatrix}
e^{i\omega_1} \\
e^{i(\omega_2 +\phi_{1})} \\
e^{i(\omega_3 +\phi_{2})} \\
\end{pmatrix}u_{1}\; ,
\quad
\langle\phi_{23}\rangle = \begin{pmatrix}
0 \\
e^{i\omega_2} \\
e^{i(\omega_3 + \phi_{3})} \\
\end{pmatrix}u_{2}\; ,
\quad
\langle\phi_{3}\rangle = \begin{pmatrix}
0 \\
0 \\
e^{i\omega_3} \\
\end{pmatrix}
\otimes 
\begin{pmatrix}
u_{3}^{u} & 0 \\
0 & u_{3}^{d} \\
\end{pmatrix}
\end{eqnarray}
where the phases $\omega_i$ can be removed by $SU(3)$
transformations, so may be regarded as unphysical.
Indeed the CSD condition that $\vev{\phi_{23}^\dag}.\vev{\phi_{123}}=0$
requires \cite{deMedeirosVarzielas:2005ax,King:2005bj}:
\be
\phi_2 -\phi_1 = \phi_3 - \pi \ (\mathrm{mod} \ 2\pi)
\label{CSDphases}
\ee
independently of $\omega_i$ which cancel.
A convenient choice of basis which we shall employ in this paper
is to simply set $\omega_i=0$, leading to 
\begin{eqnarray}\label{VEVsinthegoodbasis}
\langle\phi_{123}\rangle = \begin{pmatrix}
1 \\
e^{i\phi_{1}} \\
e^{i\phi_{2}} \\
\end{pmatrix}u_{1}\; ,
\quad
\langle\phi_{23}\rangle = \begin{pmatrix}
0 \\
1 \\
e^{i\phi_{3}} \\
\end{pmatrix}u_{2}\; ,
\quad
\langle\phi_{3}\rangle = \begin{pmatrix}
0 \\
0 \\
1 \\
\end{pmatrix}
\otimes 
\begin{pmatrix}
u_{3}^{u} & 0 \\
0 & u_{3}^{d} \\
\end{pmatrix}.
\end{eqnarray}
For the sake of the high-scale $D$-flatness\footnote{This can be seen from the following argument: splitting the set of the relevant hermitean generators $\{T^{(\phi)}\}$ (in a given representation $\phi$)  into the symmetric (and real) part $T_{S}^{(\phi)}$ and an antisymmetric (and thus imaginary) piece $T_{A}^{(\phi)}$ the $D$-flatness requires that the sum of all the $D$-terms like $\vev{\phi}^{\dagger}T_{S}^{(\phi)}\vev{\phi}=\vev{\phi}_{R}T_{S}^{(\phi)}\vev{\phi}_{R}+\vev{\phi}_{I}T_{S}^{(\phi)}\vev{\phi}_{I}$ and $\vev{\phi}^{\dagger}T_{A}^{(\phi)}\vev{\phi}=2i\vev{\phi}_{R}T_{A}^{(\phi)}\vev{\phi}_{I}$  (where $\vev{\phi}=\vev{\phi}_{R}+i\vev{\phi}_{I}$) over all relevant $\phi$'s are zero for all generators. Since the generators $\{T^{(\overline{\phi})}\}$ of the complex conjugated representation $\overline{\phi}$ obey $T_{S}^{(\phi)}=-T_{S}^{(\overline\phi)}$, $T_{A}^{(\phi)}=T_{A}^{(\overline\phi)}$ (from hermiticity) this cancellation can be achieved via extra contributions from a sector transforming as $\overline\phi$. Moreover, $\vev{\phi}_{R}=\vev{\overline\phi}_{R}$ and $\vev{\phi}_{I}=-\vev{\overline\phi}_{I}$ is needed up to a global phase.} one should add a set of extra
flavon triplets ${\bf 3}$ with VEVs:
\begin{eqnarray}
\langle\overline\phi_{123}\rangle = e^{i\psi_{1}}\begin{pmatrix}
1 \\
e^{-i\phi_{1}} \\
e^{-i\phi_{2}} \\
\end{pmatrix}u_{1}\; ,
\quad
\langle\overline\phi_{23}\rangle = e^{i\psi_{2}}\begin{pmatrix}
0 \\
1 \\
e^{-i\phi_{3}} \\
\end{pmatrix}u_{2}\; ,
\quad
\langle\overline\phi_{3}\rangle = e^{i\psi_{3}}\begin{pmatrix}
0 \\
0 \\
1 \\
\end{pmatrix}\otimes 
\begin{pmatrix}
u_{3}^{u} & 0 \\
0 & u_{3}^{d} \\
\end{pmatrix} .\nn
\end{eqnarray}
Due to their transformation properties the extra flavons
couple to the Yukawa and trilinear sectors at the higher order only
while they enter the soft masses and kinetic terms on the same
footing as the ``basic'' (i.e. antitriplet) ones.
\subsubsection{Charged sector Yukawa couplings \label{sectionrealsigma}} 
Writing flavon $SU(3)$ indices as $\phi_{i}$ (for antitriplets) and $\bar\phi^{i}$ (for triplets),
and inserting the flavon VEVs above yields Yukawa matrices\footnote{We use hats for the relevant matrices in the defining basis, i.e. prior to canonical normalization.},
\be
\hat Y^{f}_{ij} 
=  \frac{1}{M_{f}^{2}}\left[y_1^f\langle\phi_{123}\rangle_{i}\langle\phi_{23}\rangle_{j}+y_2^f\langle\phi_{23}\rangle_{i}\langle\phi_{123}\rangle_{j}+y_3^f\langle\phi_{3}\rangle_{i}\langle\phi_{3}\rangle_{j}+{y_{4}^f}\langle\phi_{23}\rangle_{i}\langle\phi_{23}\rangle_{j}\right]+\ldots\label{Yukawamatrix}
\ee
%
In matrix form, this  becomes
\bea\label{Eq:Y}
\hat Y^{f} &=& 
y_{3}^{f}(\varepsilon_3^{f})^{2} 
\left(
\begin{array}{ccc}
0 & 0 & 0 \\
0 & 0 & 0 \\
0 & 0 & 1
  \end{array}
\right) 
+ 
\varepsilon^{f}_{1}\varepsilon^{f}_{2} 
\left[
y_{1}^{f}
\left(
\begin{array}{ccc}
    0 & 1& e^{i\phi_{3}} \\ 
    0 & e^{i\phi_{1}}  & e^{i(\phi_{3}+\phi_{1})}\\ 
    0 & e^{i\phi_{2}} & e^{i(\phi_{3}+\phi_{2})}
\end{array}
\right)
+
  y_{2}^{f}
\left(
\begin{array}{ccc}
    0 & 0 & 0 \\ 
     1 & e^{i\phi_{1}}  & e^{i\phi_{2}}  \\ 
     e^{i\phi_{3}} & e^{i(\phi_{1}+\phi_{3})}  &e^{i(\phi_{2}+\phi_{3})}  
  \end{array}
\right)
\right] 
\nn \\
& + & 
y_{4}^f(\varepsilon^{f}_{2})^2 
\left(
\begin{array}{ccc}
    0 & 0 & 0 \\ 
    0 &  1 & e^{i\phi_{3}}  \\ 
    0 &  e^{i\phi_{3}}  &e^{2i\phi_{3}}  
  \end{array}
\right)+\ldots
\eea
where we have defined the real expansion parameters as:
\begin{eqnarray}
\varepsilon_3^{u,\nu}=\frac{u^u_{3}}{M_f}, \ 
\varepsilon_3^{d,e}=\frac{u^d_{3}}{M_f}, \ 
\varepsilon_2^f=\frac{u_{2}}{M_f}, \ 
\varepsilon_1^f=\frac{u_{1}}{M_f}. 
\label{epsi}
\end{eqnarray}

As mentioned above, the resulting Yukawa matrices for different charge sectors
depend on the messenger sector which is responsible for the
non-renormalizable operators via the
messenger scales $M_f$. We will (for definiteness)
assume the following hierarchy pattern\footnote{Here we come back to the generic distinction between the $SU(2)_{L}$-doublet messengers $\chi_{Q,L}$ and their singlet counterparts $\chi_{u,d,e,\nu}$. The current assumption admits great simplification of the resulting Yukawa hierarchy patterns along the lines of the recent studies \cite{Ross:2004qn}.}:
\begin{eqnarray}\label{Yukawamessengers}
M_{Q,L}\gg M_{u,\nu} \sim 3 M_{d,e} \; \quad \mathrm{with}\quad M_{u}=M_{\nu} \quad \mathrm{and} \quad M_{d}=M_{e}
\end{eqnarray}
along the lines of an underlying left-right symmetric framework. In such a case the Yukawa expansion (\ref{Eq:Y}) is governed by the $SU(2)_{L}$-singlet messengers and we can define
\begin{eqnarray}
\label{epsilons-Yukawa}
& & \frac{u_{2}}{M_{u,\nu}} =\varepsilon \; , \quad 
\frac{u_{2}}{M_{d,e}} = \bar\varepsilon \; , \quad
\frac{u_{1}}{M_{u,\nu}} = \varepsilon \bar\varepsilon \; , \quad 
\frac{u_{1}}{M_{d,e}} = \bar\varepsilon^2 \; , \quad 
\end{eqnarray}
where consistency with quark masses and mixing angles is obtained with $\varepsilon \approx 0.05, \bar\varepsilon\approx 0.12\sim 0.15$, see e.g. \cite{Roberts:2001zy}. 
The expansion parameters in Eq.~(\ref{epsi}) then become:   
\begin{eqnarray}\label{phenoepsilons}
\varepsilon^{u}_{1} \; = \; \varepsilon^{\nu}_{1} \;=\; \varepsilon\bar\varepsilon  \;,\quad
\varepsilon^{d}_{1} \; = \; \varepsilon^{e}_{1}\;=\; \bar\varepsilon^2  \;\quad \varepsilon^{u}_{2} \; = \; \varepsilon^{\nu}_{2} \; = \varepsilon \; ,\quad
\varepsilon^{d}_{2} \; = \; \varepsilon^{e}_{2} \;, = \; \bar\varepsilon \;
.
\end{eqnarray} 
We also assume $\varepsilon_{3}^{u}=$ $\varepsilon_{3}^{\nu}\equiv $ $\varepsilon_{3}$ and $\varepsilon_{3}^{e}=$ $\varepsilon_{3}^{d}\equiv $ $\overline\varepsilon_{3}$ where $\varepsilon_{3}\sim \overline\varepsilon_{3}\sim 0.5$. Though this is a relatively large expansion parameter and higher order insertions of $\phi_{3}$ and/or $\overline\phi_{3}$ are not strongly suppressed, they are harmless in the Yukawa sector because of the extra flavour symmetries preventing such insertions from entering the holomorphic superpotential to a high degree\footnote{Note that this argument can not be extended to the K\"ahler potential (or operators giving rise to the soft SUSY-breaking masses) which is not protected by holomorphy and can be in principle very sensitive to higher order $\phi_{3}$ and/or $\overline\phi_{3}$ insertions. However, as we shall argue in section \ref{sectsoftmasses}, such higher order effects can be always fully reabsorbed into definition of the Wilson coefficients of the operators under consideration.}. 

In powers of $\varepsilon$ and $\bar\varepsilon$, the leading contributions to the defining basis charged Yukawa matrices are given by:
\begin{eqnarray}\label{TBYukawa2}
 \hat Y^u = \begin{pmatrix}
0 & \varepsilon^2\bar \varepsilon\;  y_1^u& \varepsilon^2\bar \varepsilon\; y_1^u e^{i \phi_3}\\
\varepsilon^2\bar \varepsilon\;  y_2^u& \varepsilon^2 y_4^u & \varepsilon^2 y_4^u  e^{i \phi_3}\\
\varepsilon^2\bar \varepsilon\;  y_2^u e^{i \phi_3}& \varepsilon^2 y_4^u  e^{i \phi_3} & y_3^u \varepsilon_3^2
\end{pmatrix} +\ldots,
\eea
\bea
 \hat Y^d = 
\begin{pmatrix}
0 & \bar\varepsilon^3 y_1^d& \bar\varepsilon^3 y_1^d e^{i \phi_3}\\
\bar\varepsilon^3 y_2^d& \bar\varepsilon^2 y_4^d & \bar\varepsilon^2 y_4^d  e^{i \phi_3}\\
\bar\varepsilon^3 y_2^d e^{i \phi_3}& \bar\varepsilon^2 y_4^d  e^{i \phi_3} & y_3^d \bar\varepsilon_3^2
\end{pmatrix} +\ldots 
,\quad 
 \hat Y^e = \begin{pmatrix}
0 & \bar\varepsilon^3 y_1^e& \bar\varepsilon^3 y_1^e e^{i \phi_3}\\
\bar\varepsilon^3 y_2^e& \bar\varepsilon^2 y_4^e & \bar\varepsilon^2 y_4^e  e^{i \phi_3}\\
\bar\varepsilon^3 y_2^e e^{i \phi_3}& \bar\varepsilon^2 y_4^e  e^{i \phi_3} & y_3^e \bar\varepsilon_3^2
\end{pmatrix} +\ldots .\nn
\end{eqnarray}
\subsubsection{Neutrino Yukawa couplings}
Concerning the neutrino sector Yukawa, the fourth operator in expansion (\ref{Yukawamatrix}) disturbs in general the desired tri-bimaximal shape of the neutrino sector \cite{King:2005bj} unless $y_{4}^{\nu}\sim 0$ which, at the effective theory level, must be just assumed. With this at hand, one receives:
\be
\hat Y^\nu = \begin{pmatrix}
0 & \varepsilon^2\bar \varepsilon\;  y_1& \varepsilon^2\bar \varepsilon\;  y_1 e^{i \phi_3}\\
\varepsilon^2\bar \varepsilon\;  y_2& \varepsilon^2\bar \varepsilon\;  (y_1 e^{i \phi_1} + y_2 e^{i \phi_1}) & \varepsilon^2\bar \varepsilon\;  (y_1 e^{i (\phi_1 + \phi_3)} + y_2 e^{i \phi_2})\\
\varepsilon^2\bar \varepsilon\;  y_2 e^{i \phi_3}&\varepsilon^2\bar \varepsilon\;  (y_1 e^{i \phi_2} + y_2 e^{i (\phi_1+\phi_3)}) & y_3 \varepsilon_3^2
\end{pmatrix} +\ldots
\ee
This assumption, however, can be easily justified in an underlying unified model with $SO(10)$ or Pati-Salam gauge symmetry, where $y_{4}$ can be associated with a VEV with zero projection in the neutrino direction, see section \ref{sectionGUT} for further details.
\paragraph{Neutrino Majorana sector:}\mbox{}\\
In the model under consideration the Majorana masses originate from 
operators which involve the factors  
$f^{ci}f^{cj}(\phi_{23})_{i}(\phi_{23})_{j}$ and 
$f^{ci}f^{cj}(\phi_{123})_{i}(\phi_{123})_{j}$.
The neutrino Yukawa matrix and
Majorana mass matrix $M$ then have the leading order form \cite{deMedeirosVarzielas:2005ax}:
\begin{equation}
Y^{\nu}=
\left( \begin{array}{ccc}
0 & B & C_1\\
A & Be^{i \phi_1}+Ae^{i \phi_1} & C_2\\
Ae^{i \phi_3} & Be^{i \phi_2}+Ae^{i (\phi_1 + \phi_3 )} & C_3
\end{array}
\right)
, \ \ \ \
M=
\left( \begin{array}{ccc}
M_A & M_A e^{i \phi_1} & 0    \\
M_A e^{i \phi_1} & M_A e^{2i \phi_1}+M_B & 0    \\
0 & 0 & M_C
\end{array}
\right)\,,
\label{3by3ivo}
\end{equation}
where $A=y_2^{\nu}\varepsilon^3$, $B=y_1^{\nu}\varepsilon^3$, 
and the real positive Majorana masses satisfy $M_A<M_B<M_C$.
However it is not at all clear that the model corresponds to
SD since the right-handed neutrino mass matrix
is not diagonal. Moreover it is not clear that tri-bimaximal neutrino
mixing results since it does not satisfy
the usual CSD conditions. However 
the see-saw formula $Y^{\nu}M^{-1}Y^{\nu T}$
is left invariant by 
a transformation of the form \cite{King:2006hn}:
\be
Y^{{\nu}}  \rightarrow  Y^{{\nu}}
\,S^{-1},\ \ 
M  \rightarrow 
{S^T}^{-1} \,M\,S^{-1}, \ \ 
M^{-1}  \rightarrow 
{S}\,M^{-1}\,S^{T}
\label{S}
\ee
where $S$ is any non-singular (in general non-unitary) matrix,
and we see that under
\begin{equation}
S^{-1}=
\left( \begin{array}{ccc}
1 & -e^{i \phi_1} & 0    \\
0 & 1 & 0    \\
0 & 0 & 1
\end{array}
\right): \ \ \ 
Y^{\nu}\rightarrow
\left( \begin{array}{ccc}
0 & B & C_1\\
A & Be^{i \phi_1} & C_2\\
Ae^{i \phi_3} & Be^{i \phi_2} & C_3
\end{array}
\right)  \ \ \ \
M \rightarrow
\left( \begin{array}{ccc}
M_A & 0 & 0    \\
0 & M_B & 0    \\
0 & 0 & M_C
\end{array}
\right)\,,
\label{3by3ivotrans}
\end{equation}
where the transformed mass matrices
do satisfy the CSD conditions, providing the first and second
columns of the $Y^{\nu}$ satisfy $A^{\dagger} B=0$, which 
corresponds to the CSD phase condition in Eq.\ref{CSDphases}.
Clearly the transformed matrices
do not correspond to a change of basis, since the transformation is
non-unitary, but as shown in \cite{King:2006hn} the original see-saw matrices are in the
same invariance class as the transformed matrices and hence
lead to the same neutrino masses and mixing
angles. Hence we conclude that the original theory basis corresponds to CSD, 
and leads to tri-bimaximal neutrino mixing.

\subsubsection{Yukawa couplings in unified scenarios\label{sectionGUT}}
 The CSD pattern advocated in the previous section relies on $y^{\nu}_{4}=0$, which must be assumed at the effective theory level.
 However, this becomes automatic in unified models in which the 2-3 block is driven instead by an operator like: 
\be
O_{4}=f^{i}f^{cj}
\frac{y_{\Sigma}^f}{M_{f}^{2}M^{\Sigma}_{f}}(\phi_{23})_{i}(\phi_{23})_{j}\Sigma H+\ldots,
\ee 
where $\Sigma$ is a flavour-singlet Higgs field with zero Clebsch-Gordon coefficient in the neutrino direction. 

As mentioned above, this mechanism is easily realized in left-right symmetric frameworks \`a la Pati-Salam $SU(4)_{C}\otimes SU(2)_{L}\otimes SU(2)_{R}$ or $SO(10)$ GUT scenarios with $\Sigma=(15,1,3) $ under PS (or $45$ under $SO(10)$) symmetry. On top of the cancellation of the unwanted neutrino sector contribution, the Clebsches associated with $\Sigma$ lift nicely also the charged sector degeneracy (originally due to the coincidence of the various Yukawa couplings in the GUT-symmetry limit) like e.g. $y^{u}_{i}=y_{i}^{\nu}$ and $y^{d}_{i}=y_{i}^{e}$ in a class of left-right symmetric models, or 
\be\label{yGUT}
y_{i}\equiv y_i^u=y_i^d=y_i^e=y_i^{\nu}
\ee
 in Pati-Salam or $SO(10)$ with the minimal Higgs sector. This yields the generic Georgi-Jarskog \cite{Georgi:1979df} texture (accounting in particular for $m_{\mu}\sim 3 m_{s}$ at the GUT scale):
\bea
&& \hat Y^{u} \approx
\begin{pmatrix}
0 & \varepsilon^2\bar\varepsilon y_1& \varepsilon^2\bar\varepsilon y_1 e^{i \phi_3}\\
\varepsilon^2\bar\varepsilon y_2& \varepsilon^2 y_\Sigma C^{u} \sigma & \varepsilon^2 y_\Sigma C^{u} \sigma e^{i \phi_3}\\
\varepsilon^3 y_2 e^{i \phi_3}& \varepsilon^2 y_\Sigma C^{u} \sigma  e^{i \phi_3} & y_3 \varepsilon_3^2
\end{pmatrix}, \nn\\
& & \hat Y^{d,e} \approx
\begin{pmatrix}
0 & \bar\varepsilon^3 y_1& \bar\varepsilon^3 y_1 e^{i \phi_3}\\
\bar\varepsilon^3 y_2& \bar\varepsilon^2 y_\Sigma C^{d,e} \sigma & \bar\varepsilon^2 y_\Sigma C^{d,e} \sigma e^{i \phi_3}\\
\bar\varepsilon^3 y_2 e^{i \phi_3}& \bar\varepsilon^2 y_\Sigma C^{d,e} \sigma  e^{i \phi_3} & y_3 \bar\varepsilon_3^2
\end{pmatrix},\;\;
\label{Yukawaswithphi3phases}
\eea
with $\sigma\equiv \langle\Sigma\rangle/M^{\Sigma}$ (recall there is no distinction between $d,e$ in $SO(10)$ or Pati-Salam) and $C^{e}=3$ and $C^{d}=1$ in the charged-lepton and down-quark sectors respectively.

For the sake of simplicity, in what follows {\it we shall work in the framework of MSSM taken as an effective limit of any high-scale scenario, i.e. keep all the couplings in the different flavour sectors independent}, only assuming $y_{4}^{\nu}\sim 0$ in the effective theory (however, well motivated in various GUT scenarios).
Not only this makes our results more generic, but also admits imposing further GUT constraints at any point to derive model-specific conclusions.
We shall keep track of all the would-be Clebsch-Gordon coefficients associated with the Georgi-Jarlskog Higgs field $\Sigma$ in unified models by means of a simple identification
\be\label{ySigma4}
y_{4}^{f}=y_{\Sigma}C^{f}\sigma
\ee  
(with $C^{u,d,e,\nu}=-2, 1, 3, 0$) so that one can translate all the generic results given below in terms of the 'effective' $y_{4}^{f}$ couplings into the unified picture with $y_{\Sigma}$ instead.
\paragraph{Note on CKM CP violation:}\mbox{}\\
It has been pointed out \cite{PrivateCommunicationWithOscar} that if all the CP phases come from the flavon sector only, the current texture is unlikely to provide large enough CKM CP phase\footnote{Remarkably enough, the solution advocated (in a slightly different context) in \cite{Ross:2004qn} (i.e. an extra relative phase between the two $SU(2)_{R}$ components of the $\phi_{3}$ VEV) does not work in the current model because the net effect of such an extra phase on the diagonalization matrices can be reabsorbed upon getting the resulting CKM matrix into the standard form.} ($\delta_{CKM}$). The reason is that for a real $\sigma$ one can rotate away all the leading order $e^{i\phi_{3}}$ phase factors in (\ref{Yukawaswithphi3phases}) to end up with a leading-order phase only on the 33 entries of $\hat Y^{u,d}$, that gets reabsorbed upon bringing the resulting CKM matrix into the standard form \cite{Yao:2006px}, while the effects of the subleading phases are too suppressed to account for the measured value of $\delta_{CKM}$. 
In general, however, $\sigma$ is complex\footnote{This need not be straightforward due to the adjoint nature of $\Sigma$. However, if $\Sigma$ is just an effective description of a composite object one can generate an overall phase on its VEV from the misalignment of phases of the underlying degrees of freedom (recall that it is not entirely neutral and thus its components need not have their phases aligned from D-flatness). } and it can be shown its phase can account for the entire CKM CP-violating phase. 

In what follows, the phase of $\sigma$ will be mostly irrelevant and for the sake of simplicity we shall often work with $\sigma$ real. Nevertheless, we shall comment on the would-be effects of its non-zero phase whenever appropriate.

\subsection{$SU(3)$ predictions for soft-SUSY breaking parameters\label{softsector}}
Though successfully describing the basic features of the SM fermion masses and their mixing, the flavour models per se are difficult to test\footnote{Usually the situation is such that either there is a very limited set of parameters leading only to approximate fits of the SM fermion spectra and mixings or, on the other hand, a wider set of parameters admits perfect fits of the known measurables but does not lead to a clear-cut prediction that can be falsified.}. However, in supersymmetry, there are many additional constraints associated with the SUSY flavour and CP violating parameters coming from the soft-SUSY breaking part of the MSSM lagrangian, in particular from the trilinear scalar couplings and the soft-SUSY breaking scalar masses: 
\begin{eqnarray}
{\cal L}_\mathrm{soft} &=& \varepsilon_{\alpha\beta} \left[ 
- {H}_u^\alpha \tilde{Q}^{\beta i}  A^{u}_{ij} \tilde{u}^{cj}
- {H}_d^\alpha \tilde{Q}^{\beta i}  A^{d}_{ij} \tilde{d}^{cj}
- {H}_u^\alpha \tilde{L}^{\beta i}  A^{\nu}_{ij} \tilde{\nu}^{cj}
- {H}_d^\alpha \tilde{L}^{\beta i}  A^{e}_{ij} \tilde{e}^{cj} 
+ \text{H.c.}\right]\nonumber \\
& + & 
\tilde{Q}_{i\alpha}^{*} ( m_{Q}^2)^{i}_{j} \tilde{Q}^{\alpha j}
+ \tilde{u}^{c *}_i ( m_{u^{c}}^2)_{j}^{i} \tilde{u}^{cj}
+ \tilde{d}^{c *}_i ( m_{d^{c}}^2)_{j}^{i} \tilde{d}^{cj}
+ \tilde{L}_{i\alpha}^{*} ( m_{L}^2)^{i}_{j} \tilde{L}^{\alpha j}
+ \tilde{e}_i^{c *} ( m_{e^{c}}^2)_{j}^{i} \tilde{e}^{cj}\nonumber \\
& + & \tilde{\nu}_i^{c *} ( m_{\nu^{c}}^2)_{j}^{i} \tilde{\nu}^{cj}
 \; .
\end{eqnarray}
In flavour models,  these terms come from the generic operators of the form
\bea
{\cal L}^{A}_\mathrm{soft} & = &\int\mathrm{d}^{2}\theta  \sum_{X}\frac{\hat X}{M^{X}}  \hat{H}\sum_{\{f,f^{c}\}} \hat{f}\left(\sum_{\Phi,\Phi'}a_{\Phi\Phi'}^{ff^{c},X}\frac{\hat\Phi\otimes\hat\Phi'}{{M^{A,X}_{ff^{c},\Phi\Phi'}}^{2}}+\ldots\right)\hat{f}^{c}+\ldots\label{trilinearoperator}\\
{\cal L}^{m^{2}}_\mathrm{soft} & = &\int\mathrm{d}^{4}\theta\,  \frac{\hat X^{\dagger}\hat X}{{M^{X}}^{2}}\sum_{f}\hat f^{\dagger} \left(b^{f,X}_{0}\mathbbm{1}+\sum_{\Phi,\Phi'}b^{f,X}_{\Phi}\frac{\hat\Phi\otimes\hat\Phi^{\dagger}}{{M^{m,X}_{f,\Phi\Phi'}}^{2}}+\ldots\right)\hat f+
(\hat f \to \hat f^{c})+\ldots\;\;\;\;\;\;\;\;
\label{massoperator}
\eea
after the relevant fields develop their SUSY-breaking $F$-terms. In the formula above this sector is represented by a generic symbol $\hat{X}$ and typically corresponds to the SUSY-breakdown triggering hidden sector fields. The generic symbols $\Phi$, $\Phi'$ denote all the combinations of the $\phi_{A}$ (and $\overline{\phi}_{B}$) flavons in the model that are allowed by the extra symmetries. As before, the ellipses correspond to higher order terms. 

Note that in supergravity any superfield with nonzero VEV on its scalar component actually develops an $F$-term, c.f. \cite{Ross:2002mr}, and so do also the flavons. In such a case, there are non-zero contributions associated to the flavon sector $F$-terms emerging from even lower level operators like:
\bea
{\cal L}^{A}_\mathrm{soft} & \ni &\int\mathrm{d}^{2}\theta \hat{H}\sum_{\{f,f^{c}\}} \hat{f}\left(\sum_{\Phi,\Phi'}a_{\Phi\Phi'}^{ff^{c}}\frac{\hat\Phi\otimes\hat\Phi'}{{M^{A}_{ff^{c},\Phi\Phi'}}^{2}}+\ldots\right)\hat{f}^{c}+\ldots\nn\\
{\cal L}^{m^{2}}_\mathrm{soft} & \ni &\int\mathrm{d}^{4}\theta\,  \sum_{f}\hat f^{\dagger} \left(b^{f}_{0}\mathbbm{1}+\sum_{\Phi}b^{f}_{\Phi}\frac{\hat\Phi\otimes\hat\Phi^{\dagger}}{{M^{m}_{f,\Phi}}^{2}}+\ldots\right)\hat f+\;\;
(\hat f \to \hat f^{c})+\ldots\nn
\eea
A more detailed discussion of these matters is however beyond the scope of this paper and shall be covered in a separate publication \cite{inpreparation}.

The scales\footnote{$M^{X}$  are assumed to be larger than the scale of the family symmetry breaking so that the effective soft terms do follow the constraints imposed by the family symmetry.} $M^{X}$ in formulae (\ref{trilinearoperator}) and  (\ref{massoperator}) corresponds to the physics communicating the information about the SUSY breakdown (indicated by non-zero $F$-terms of the $\hat{X}$ superfields) into the visible sector; in gravity mediation one typically has $M^{X}\sim M_{Pl}$ for hidden sector superfields while in the case of a gauge mediation this scale could be significantly lower\footnote{The main constraint in both cases comes from the requirement that the soft SUSY-breaking scale given by $\vev{F_{X}}/M^{X}$ is in the desired (TeV) region.}. In any case, for the expansions  (\ref{trilinearoperator}) and  (\ref{massoperator}) to make sense, we must assume that the family symmetry breaking scale is below $M^{X}$.  

The flavour structure of the expansions (\ref{trilinearoperator}) and  (\ref{massoperator}) correspond to the case of an $SU(3)$ horizontal symmetry with all the fermion superfields 
$\hat Q$, $u^c$, $\hat d^c$, $\hat L$, $\hat e^c$ and $\hat \nu^c$ transforming as fundamental triplets. 
In the exact $SU(3)$ family symmetry limit the soft masses from formula (\ref{massoperator}) are universal:
\begin{eqnarray}
\hat m_{Q}^2 \propto \hat m_{u^{c}}^2 \propto \hat m_{d^{c}}^2 \propto \hat m_{L}^2 
\propto \hat m_{e^{c}}^2 \propto \hat m_{\nu^{c}}^2 \propto \mathbbm{1}
\end{eqnarray}
and the Yukawa couplings and trilinear terms vanish. Clearly, to be consistent with fermion masses and mixings, the $SU(3)$ family symmetry has to be broken, leading  to violations of universality, as we now discuss. 
\subsubsection{Trilinear soft couplings}
Let us thus consider the MSSM equipped with an $SU(3)$ family symmetry under which all the three families of matter (super)fields sharing the same gauge charges transform as fundamental triplets. Assuming that the extra flavon charges \cite{deMedeirosVarzielas:2005ax} admit just the ${\bf 3}_{f}\otimes {\bf 3}_{f^{c}}\otimes \bar {\bf 3}_{\phi}\otimes \bar{\bf 3}_{\phi}$ terms of the typical tri-bimaximal Yukawa structures given in formula (\ref{TBYukawa2}) but forbid the would-be singlets of the type ${\bf 3}_{f}\otimes {\bf 3}_{f^{c}}\otimes {\bf 3}_{\bar\phi}$ one arrives to the following expansion for the effective trilinear couplings:
\be
\label{Aexpansion}
\frac{\hat{A}^{f}_{ij}}{A_{0}} =  \frac{1}{{M^{A}_f}^2}\left(a^{f}_{1}{\langle\phi_{123}\rangle_{i} \langle\phi_{23}\rangle_{j}}+a^{f}_{2}{\langle\phi_{23}\rangle_{i} \langle\phi_{123}\rangle_{j}}+a^{f}_{3} {\langle\phi_{3}\rangle_{i} \langle\phi_{3}\rangle_{j}}+a_{4}^{f}{\langle\phi_{23}\rangle_{i} \langle\phi_{23}\rangle_{j}}\right)+\ldots
\ee
where, for the sake of simplicity, we have taken the masses of all the messenger fields relevant for this type of contractions to be the same (i.e. we put all the $M^{A,X}_{ff^{c},\Phi\Phi'}$ factors in formula  (\ref{trilinearoperator}) to a common value ${M^{A}_f}$ and similarly $a^{ff^{c},X}_{\Phi\Phi'}\equiv a^{f}_{i}$  assuming the fundamental dynamics behind all these operators is the same). This is quite natural because of the similar quantum structure of the operators behind these two types of contractions.
 
However, as pointed out above, up to ${\cal O}(1)$ coefficients $a^{f}_{i}$ that can be different from those ($y_{i}^{f}$) in the Yukawa sector (due to the effects of supersymmetry breaking)  the flavour structure of the trilinear terms (\ref{Aexpansion}) is similar to the Yukawa matrices $\hat Y$, c.f. (\ref{Yukawamatrix}). This suggests that also the masses of the messenger fields entering the denominators of these effective operators ($M_{f}^{A}$) can be identified with those entering the Yukawa sector\footnote{The SUSY-breaking effects are typically negligible for messenger scales well above the soft SUSY braking scale.} ($M_{f}$). Thus, in what follows we shall further assume\footnote{Without such assumptions on the trilinear sector messenger masses one obviously can not derive any physical predictions from the given model.} 
\be\label{softmessengers}
M_{f}^{A}=M_{f} \text{ for all }f.
\ee
The mass scale $A_{0}$ in the $A$-term expansion is taken at the order of a typical SUSY scale with $A_{0}=\langle F_{X}\rangle /M^{X}$.

Considering only the leading set of operators, the matrix structure of the trilinear soft terms in the effective $SU(3)$ flavour model under consideration (c.f. (\ref{VEVsinthegoodbasis}), (\ref{epsi}), (\ref{Aexpansion}) and (\ref{softmessengers})) is given by:
\bea\label{Eq:A}
\frac{\hat A^{f}}{A_{0}}&=& 
a_{3}^{f}(\varepsilon_3^{f})^{2} 
\left(
\begin{array}{ccc}
0 & 0 & 0 \\
0 & 0 & 0 \\
0 & 0 & 1
  \end{array}
\right) 
+ 
\varepsilon^{f}_{1}\varepsilon^{f}_{2} 
\left[
a_{1}^{f}
\left(
\begin{array}{ccc}
    0 & 1& e^{i\phi_{3}} \\ 
    0 & e^{i\phi_{1}}  & e^{i(\phi_{3}+\phi_{1})}\\ 
    0 & e^{i\phi_{2}} & e^{i(\phi_{3}+\phi_{2})}
\end{array}
\right)
+
  a_{2}^{f}
\left(
\begin{array}{ccc}
    0 & 0 & 0 \\ 
     1 & e^{i\phi_{1}}  & e^{i\phi_{2}}  \\ 
     e^{i\phi_{3}} & e^{i(\phi_{1}+\phi_{3})}  &e^{i(\phi_{2}+\phi_{3})}  
  \end{array}
\right)
\right] 
\nn \\
& + & 
a_{4}^f(\varepsilon^{f}_{2})^2
\left(
\begin{array}{ccc}
    0 & 0 & 0 \\ 
    0 &  1 & e^{i\phi_{3}}  \\ 
    0 &  e^{i\phi_{3}}  &e^{2i\phi_{3}}  
  \end{array}
\right)\,,
\eea
where the expansion parameters are those used in the Yukawa sector (\ref{epsi}). In the 'phenomenological' parametrization (\ref{phenoepsilons}) one can write:
\begin{eqnarray}
&& \hat A^u = A_{0}\begin{pmatrix}
0 & \varepsilon^2\bar \varepsilon\;  a_1^u& \varepsilon^2\bar \varepsilon\;  a_1^u e^{i \phi_3}\\
\varepsilon^2\bar \varepsilon\;  a_2^u& \varepsilon^2 a_4^u & \varepsilon^2 a_4^u  e^{i \phi_3}\\
\varepsilon^2\bar \varepsilon\;  a_2^u e^{i \phi_3}& \varepsilon^2 a_4^u  e^{i \phi_3} & a_3^u \varepsilon_3^2
\end{pmatrix} +\ldots 
, \\
&& \hat A^{d,e} = 
A_{0}\begin{pmatrix}
0 & \bar\varepsilon^3 a_1^{d,e}& \bar\varepsilon^3 a_1^{d,e} e^{i \phi_3}\\
\bar\varepsilon^3 a_2^{d,e}& \bar\varepsilon^2 a_4^{d,e} & \bar\varepsilon^2 a_4^{d,e}  e^{i \phi_3}\\
\bar\varepsilon^3 a_2^{d,e} e^{i \phi_3}& \bar\varepsilon^2 a_4^{d,e}  e^{i \phi_3} & a_3^{d,e} \bar\varepsilon_3^2
\end{pmatrix} +\ldots 
.\nn 
\end{eqnarray}
In a GUT limit (c.f. section \ref{Yukawasection}), it is again convenient to replace the last operator in (\ref{Aexpansion}) by the Georgi-Jarlskog structure:
\be
a_{\Sigma}\frac{\langle\phi_{23}\rangle \langle\phi_{23}\rangle\Sigma}{M^2M^{\Sigma}}\,,
\ee
and employ a matching condition along the lines of \eq{ySigma4}
\be\label{aSigma4}
a_{4}^{f}=a_{\Sigma}C^{f}\sigma
\ee  
so that the neutrino sector trilinear coupling obeys (in the GUT limit (\ref{yGUT}), i.e. $a_{i}\equiv$ $a^{u}_{i}=$ $a^{d}_{i}=$ $a^{e}_{i}=$ $a^{\nu}_{i}$):
\be \hat A^\nu = A_{0} \begin{pmatrix}
0 & \varepsilon^2\bar \varepsilon\;  {a}_1&\varepsilon^2\bar \varepsilon\;  {a}_1 e^{i \phi_3}\\
\varepsilon^2\bar \varepsilon\;  {a}_2& \varepsilon^2\bar \varepsilon\;  ({a}_1 e^{i \phi_1} + {a}_2 e^{i \phi_1}) & \varepsilon^2\bar \varepsilon\;  ({a}_1 e^{i (\phi_1 + \phi_3)} + {a}_2 e^{i \phi_2})\\
\varepsilon^2\bar \varepsilon\;  {a}_2^u e^{i \phi_3}& \varepsilon^2\bar \varepsilon\;  ({a}_1 e^{i \phi_2} + {a}_2 e^{i (\phi_1+\phi_3)}) & {a}_0 \varepsilon_{3}^2
\end{pmatrix} +\ldots
\ee
\subsubsection{Scalar soft mass-squared parameters \label{sectsoftmasses}}
Similarly, the soft masses allowed by the $SU(3)$ family symmetry can be written from (\ref{massoperator}) as (the generic subscript ${\!}_{A}$ runs over all the relevant flavon species, $f\equiv Q,L$ and $f^{c}\equiv$ $u^{c}, d^{c}, e^{c}, \nu^{c}$):
\begin{eqnarray}\label{softmassexpansion}
\hat m^{2}_{f,f^{c}} &=& 
m_{0}^{2}\left(b^{f,f^{c}}_{0}\mathbbm{1}+ \sum_{A}b^{f,f^{c}}_{A}\frac{\langle\phi_{A}\phi_{A}^{*}\rangle}{{M^{m}_{f,f^{c}}}^2}+\ldots\right)
\end{eqnarray}
where we have again made an assumption about the common origin of all the leading order operators, i.e. put all the ${M^{m,X}_{f,f^{c},\Phi}}$ coefficients in formula (\ref{massoperator}) to a common value $M_{f}^{m}$ and similarly $b_{\Phi}^{f(f^{c}),X}\equiv b_{i}^{f(f^{c})}$.
Since there is a lot more contractions allowed in this case (notice that due to hermiticity the most stringent constraint driving the Yukawa sector -- the saturation of the extra charges -- is trivial to satisfy) one should in general worry about some of the subleading terms as well, in particular those corresponding to extra $\phi_{3}\phi_{3}^{*}$ insertions. 

Inspecting thoroughly all the options one can conclude that the only relevant (and irreducible) set of leading order terms is\footnote{It can be shown (see Appendix \ref{Appendixhigherorder}) that the would-be effects of all the other potentially relevant operators can be hidden in redefining the expansion parameters $b_{i}$, $b_{i}'$ and $b_{i}''$ and thus becomes irrelevant in the order-of-magnitude analysis we aim at.}:
\begin{eqnarray}
(\hat m^{2}_{f,f^{c}})_{ij} &=& 
m_{0}^{2}\left( b_{0}^{f,f^{c}}\delta_{ij} + 
b^{f,f^{c}}_{1}\frac{\langle\phi_{123}\rangle_{j}\langle\phi_{123}^{*}\rangle_{i}}{{M^m_{f}}^2}+
b^{f,f^{c}}_{2}\frac{\langle\phi_{23}\rangle_{j}\langle\phi_{23}^{*}\rangle_{i}}{{M^m_{f}}^2}+ 
b^{f,f^{c}}_{3}\frac{\langle\phi_{3}\rangle_{j}\langle\phi_{3}^{*}\rangle_{i}}{{M^m_{f}}^2} 
\right.\nn \\
&+ &\label{softexpansionexplicit}
\left.  b'^{f,f^{c}}_{1}\frac{\langle\phi_{123}\rangle_{j}\langle\phi_{3}^{*}\rangle_{i}\langle\phi_{3}.\phi_{123}^{*}\rangle+\langle\phi_{3}\rangle_{j}\langle\phi_{123}^{*}\rangle_{i}\langle\phi_{123}.\phi_{3}^{*}\rangle}{{M^m_{f}}^4}\right. \\
&+ &\left.  b'^{f,f^{c}}_{2}\frac{\langle\phi_{23}\rangle_{j}\langle\phi_{3}^{*}\rangle_{i}\langle\phi_{3}.\phi_{23}^{*}\rangle+\langle\phi_{3}\rangle_{j}\langle\phi_{23}^{*}\rangle_{i}\langle\phi_{23}.\phi_{3}^{*}\rangle}{{M^m_{f}}^4}\right. \nn \\
&+ &\left. b''^{f,f^{c}}_{1}\frac{\varepsilon_{jkl} \varepsilon^{imn}\langle\phi_{123}\rangle_{k}\langle\phi_{3}\rangle_{l}\langle\phi_{123}^{*}\rangle^{m}\langle\phi_{3}^{*}\rangle^{n}}{{M^m_{f}}^4}\right.\nn \\
&+ &\left.  b''^{f,f^{c}}_{2}\frac{\varepsilon_{jkl} \varepsilon^{imn}\langle\phi_{23}\rangle_{k}\langle\phi_{3}\rangle_{l}\langle\phi_{23}^{*}\rangle^{m}\langle\phi_{3}^{*}\rangle^{n}}{{M^m_{f}}^4}+\ldots \right)+\mathrm{higher\;order\;terms} \nn\,,
\nonumber
\end{eqnarray}
where all the $b$-coefficients are real by hermiticity of $\tilde{f}_{i}^{*} (m^{2}_{f})_{ij} \tilde{f}_{j}$. 
The various messenger masses are $M^{m}_{Q}$, $M^{m}_{L}$ for the left-handed fields $f=Q,L$ and $M^{m}_{u}$, $M^{m}_{d}$, $M^{m}_{e}$, $M^{m}_{\nu}$ for $f^{c}=u^{c}, d^{c}, e^{c}$ and $\nu^{c}$ respectively. 

At this point, it is worth commenting on the would-be higher-order corrections due to multiple $\phi_{3}$ and/or $\bar\phi_{3}$ insertions, that (due to the relatively large associated expansion parameters $\varepsilon_{3}$ and $\bar\varepsilon_{3}$) could in principle alter the leading order structure of the expansion above. However, as described in detail in Appendix \ref{Appendixhigherorder}, the set of operators in (\ref{softexpansionexplicit}) is robust under further perturbations due to higher order $\phi_{3}$ and/or $\bar\phi_{3}$ insertions in the sense that any would-be higher order $\phi_{3}$ and/or $\bar\phi_{3}$ insertion can be accounted for by a mere redefinition of the $b$-coefficients in (\ref{softexpansionexplicit}).

In order to perform any quantitative analysis, yet further assumptions must be made about the messengers giving rise to the soft mass operators above. In particular, unless these are linked with the messengers in the Yukawa sector, one can suppress any off-diagonalities and non-universalities in the soft mass terms as much as desired by choosing the $M^m_{Q}$, $M^m_{L}$, $M^m_{u}$, $M^m_{d}$, $M^m_{e}$, $M^m_{\nu}$ much above the corresponding Yukawa (and trilinear) sector messenger masses $M_{u}$, $M_{d}$, $M_{e}$, $M_{\nu}$, c.f. (\ref{Yukawamessengers}) and (\ref{softmessengers}). 

However, a full-featured analysis of this question is far beyond the scope of this work. In what follows we shall resort to a minimal set of assumptions about the physics of this part of the messenger sector, much along the general lines of the simplest settings in which the Yukawa (and trilinear) sector messengers dominate the soft mass (and K\"ahler sector) operators:
\be\label{identifyingsoftmassmessengers}
M^{m}_{u}=M_{u}, \;\; M^{m}_{d}=M_{d}, \;\;  M_{e}^{m}=M_{e}, \;\;  M^{m}_{\nu}=M_{\nu}.
\ee
This is actually quite natural because of the similar quantum structure of the relevant operators that, at the level of an underlying theory, are often linked to each other, see e.g. \cite{Antusch:2007vw}.

As far as the 'doublet' soft-sector operators $\hat m^{2}_{Q}$ and $\hat m_{L}^{2}$ are concerned, the natural choice is less obvious\footnote{For a more detailed discussion of the relations between the Yukawa (and trilinear) sector messengers and the kinetic form (and the sector of soft masses) an interested reader is kindly deferred to \cite{Antusch:2007vw}.} and one can use the relative freedom in choosing $M^{m}_{Q}$ and $M^{m}_{L}$ to further suppress the left-handed contributions to SUSY FCNC. For the sake of simplicity, we shall adopt this strategy in what follows by taking  
\be
M^{m}_{Q}\sim M_{Q}\gg M_{u,d},\;\;\; M^{m}_{L}\sim M_{L}\gg M_{e,\nu} 
\ee
thus effectively decoupling the left-handed part of the messenger sector, c.f. (\ref{Yukawamessengers}). 

This allows us to use the same set of expansion parameters $\varepsilon_{1}^{f}$, $\varepsilon_{2}^{f}$ (or $\varepsilon$ and $\overline{\varepsilon}$ along the lines of the phenomenological fits)  for all the soft sector contractions, i.e\footnote{Note that these identifications admit future embedding into a class of left-right symmetric unified models like $SO(10)$ or Pati-Salam.}:  
\bea\label{epsilons-soft}\label{epsilons-soft2}
& & \frac{u_{1}}{M^m_{u,\nu}}\equiv\varepsilon_{1}^{u,\nu}  =  \varepsilon\bar\varepsilon\;, \frac{u_{1}}{M^m_{d,e}}\equiv\varepsilon_{1}^{d,e} =  \overline\varepsilon^{2}\;,
\qquad \frac{u_{1}}{M^m_{Q,L}}\equiv \varepsilon_{1}^{Q,L}\ll \varepsilon\bar\varepsilon \\
& & \frac{u_{2}}{M^m_{u,\nu}}\equiv\varepsilon_{2}^{u,\nu}  =  \varepsilon\;, \;\,\frac{u_{2}}{M^m_{d,e}}\equiv\varepsilon_{2}^{d,e} =  \overline\varepsilon\;,\;\,
\qquad \frac{u_{2}}{M^m_{Q,L}}\equiv \varepsilon_{2}^{Q,L}\ll \varepsilon,\bar\varepsilon \nn \\
& & \frac{u_{3}}{M^m_{u,\nu}}\equiv\varepsilon_{3}^{u,\nu}=\varepsilon_{3}\;,\, 
\frac{u_{3}}{M^m_{d,e}}\equiv\varepsilon_{3}^{d,e}= \overline\varepsilon_{3}\qquad\;\; \frac{u_{3}}{M^m_{Q,L}}\equiv \varepsilon_{3}^{Q,L}\ll \varepsilon_{3},\bar\varepsilon_{3} \nn 
\eea
With this identification at hand, the operator expansion (\ref{softexpansionexplicit}) can be recast in a matrix form as:
\begin{eqnarray}\label{Eq:M2}
\frac{\hat m^{2}_{f,f^{c}}}{m_{0}^{2}} &=& 
b_{0}^{f,f^{c}}
\left(\begin{array}{ccc}
1 & 0 & 0 \\
0 & 1 & 0 \\
0 &  0 & 1
\end{array}\right) 
+
b^{f,f^{c}}_{1} (\varepsilon^{f,f^{c}}_{1})^2 
 \left(
\begin{array}{ccc}
1 & e^{i\phi_{1}} & e^{i\phi_{2}} \\
e^{-i\phi_{1}} & 1 & e^{i(\phi_{2}-\phi_{1})} \\
e^{-i\phi_{2}} & e^{-i(\phi_{2}-\phi_{1})} & 1
  \end{array}
\right) 
\\
& + & 
b^{f,f^{c}}_{2} (\varepsilon^{f,f^{c}}_{2})^2
\left(
\begin{array}{ccc}
0 & 0 & 0 \\
0 & 1 & e^{i\phi_{3}} \\
0 & e^{-i\phi_{3}} & 1
  \end{array}
\right) 
+
b^{f,f^{c}}_{3} (\varepsilon_3^{f,f^{c}})^{2}  
\left(\begin{array}{ccc}
0 & 0 & 0 \\
0 & 0 & 0 \\
0 & 0 & 1
\end{array}\right)+\mathrm{higher\,\,order\,\,terms}\nn\,,
\end{eqnarray}
where it is understood $\varepsilon_i^{f,f^{c}}$ is to be replaced by $\varepsilon_{i}^{Q,L}$ for $f=Q,L$ and $\varepsilon_{i}^{u,d,e,\nu}$ for $f^{c}=u^{c}$, $d^{c}$, $e^{c}$ and $\nu^{c}$. 
Employing as before the $\varepsilon$, $\overline{\varepsilon}$ notation one is left with diagonal $\hat m^{2}_{Q}$ and  $\hat m^{2}_{L}$:
\be 
\hat  m^2_{Q} = m_0^2\, {b}_0^Q\, \mathbbm{1}
\; ,\quad  \hat m^2_{L} = m_0^2\, {b}_0^{\ell}\, \mathbbm{1}\, \qquad \mathrm{(to\;excellent\;approximation)}, 
\ee
while the $SU(2)_{L}$-singlet sector receives off-diagonal corrections as follows: 
\bea
&& 
\hat m^2_{u^c} \approx m_0^2 \left[{b}_0^{u^{c}} \mathbbm{1}
+
\left(\begin{array}{lll}
\varepsilon^2\overline\varepsilon^{2}\, {b}^{u^c}_{1}&  \varepsilon^2\overline\varepsilon^{2}\,  {b}^{u^c}_{1}  e^{i \phi_1}& \varepsilon^2\overline\varepsilon^{2}\, {b}^{u^c}_{1}  e^{i \phi_2}\\
.&  \varepsilon^2 {b}^{u^c}_{2} & \varepsilon^2 {b}^{u^c}_{2}e^{i \phi_3}\\
.& .&   \varepsilon_{3}^2{b}^{u^c}_3 
\end{array}\right)+\ldots\right] , \nn \\
&&
\hat m^2_{d^c} \approx m_0^2 \left[{b}_0^{d^{c}} \mathbbm{1}
+
\left(\begin{array}{lll}
\bar\varepsilon^4 {b}^{d^c}_{1}&  \bar\varepsilon^4 {b}^{d^c}_{1}  e^{i \phi_1}& \bar\varepsilon^4 {b}^{d^c}_{1}  e^{i \phi_2}\\
.&  \bar\varepsilon^2 {b}^{d^c}_{2} & \bar\varepsilon^2 {b}^{d^c}_{2}e^{i \phi_3}\\
.& .& \bar\varepsilon_{3}^2{b}^{d^c}_3 
\end{array}\right)+\ldots\right] , \nn \\
&& \hat m^2_{\nu^c} \approx m_0^2 \left[{b}_0^{\nu^{c}} \mathbbm{1}
+
\left(\begin{array}{lll}
 \varepsilon^2\overline\varepsilon^{2}\, {b}^{\nu^c}_{1}&  \varepsilon^2\overline\varepsilon^{2}\, {b}^{\nu^c}_{1}  e^{i \phi_1}&  \varepsilon^2\overline\varepsilon^{2}\, {b}^{\nu^c}_{1}  e^{i \phi_2}\\
.&  \varepsilon^2 {b}^{\nu^{c}}_{2} & \varepsilon^2 {b}^{\nu^{c}}_{2}e^{i \phi_3}\\
.& .&  \varepsilon_{3}^2{b}^{\nu^c}_3 
\end{array}\right)+\ldots\right] ,\nn \\
&& \hat m^2_{e^c} \approx m_0^2 \left[{b}_0^{e^{c}} \mathbbm{1}
+
\left(\begin{array}{lll}
 \bar\varepsilon^4 {b}^{e^c}_{1}&  \bar\varepsilon^4 {b}^{e^c}_{1}  e^{i \phi_1}& \bar\varepsilon^4 {b}^{e^c}_{1}  e^{i \phi_2}\\
.&  \varepsilon^2 {b}^{e^{c}}_{2} & \varepsilon^2 {b}^{e^{c}}_{2}e^{i \phi_3}\\
.& .&  \bar\varepsilon_{3}^2{b}^{e^c}_3 
\end{array}\right)+\ldots\right] 
 .
\end{eqnarray}
The dotted terms are readily obtained from hermiticity of all these soft mass-matrices.

In the various GUT limits there are extra correlations among the different $b$-coefficients (like $b_{i}^{f,f^{c}}\equiv b_{i}$ $\forall f,f^{c}$ in minimal Pati-Salam or $SO(10)$ case, c.f. formula (\ref{yGUT}) for the Yukawa couplings). The situation is, however, simpler than in the Yukawa sector as $\Sigma$ does not enter the irreducible part\footnote{The only effect of $\Sigma$  (being a flavour singlet) in $\hat m^{2}_{f,f^{c}}$ can arise from contractions  like $\Sigma^{\dagger}\Sigma$ entering as completely family-blind higher order corrections without any extra flavour violating effects.} of the soft mass-squared expansion (\ref{softexpansionexplicit}).
\subsection{The $SU(3)$ K\"ahler potential and effects of canonical normalization}
Whenever the K\"ahler potential of a given SUSY model is nontrivial there are extra effects coming from the canonical normalization procedure bringing the generic kinetic terms arising from the operator expansion like
\be
\label{kineticterm}
{\cal L}^{\hat f}_\mathrm{kin}  =  \int\mathrm{d}^{4}\theta\, \sum_{f}\hat f^{\dagger} \left(k_{0}^{f}\mathbbm{1}+\sum_{\Phi}k^{f}_{\Phi}\frac{\hat\Phi\otimes\hat\Phi^{\dagger}}{{M^{K}_{f,\Phi}}^{2}}+\ldots\right)\hat f+\mathrm{similarly \,\, for\,\,}f^{c}\nn
\ee
(or, equivalently in terms of the scalar $\tilde{f}$ ($\tilde{f}^{c}$) and fermionic ${f}$ (${f}^{c}$) degrees of freedom:  
 \begin{eqnarray}
{\cal L}^{\tilde{f}}_\mathrm{kin} &=& 
\partial_\mu \tilde{Q}_{i\alpha}^{*} (\hat{K}_{Q})_{ij} \partial^\mu\tilde{Q}_{\alpha j}+
\partial_\mu \tilde{u}^{c*}_{i} (\hat{K}_{u})_{ij} \partial^\mu\tilde{u}^{c}_{j}+
\partial_\mu \tilde{d}^{c*}_{i} (\hat{K}_{d})_{ij} \partial^\mu\tilde{d}^{c}_{j}+\ldots
\nonumber \\
{\cal L}^{f}_\mathrm{kin} &=& 
\overline{Q}_{i\alpha} (\hat{K}_{Q})_{ij}i\gamma^{\mu} \partial_\mu {Q}_{\alpha j}+
\overline{u^{c}}_{i} (\hat{K}_{u})_{ij}i\gamma^{\mu} \partial_\mu {u^{c}}_{j}+
\overline{d^{c}}_{i} (\hat{K}_{d})_{ij}i\gamma^{\mu} \partial_\mu {d^{c}}_{j}+\ldots
\end{eqnarray}
where $\hat{K}_{f} $ denotes the K\"ahler metric $\hat{K}_{f}\sim (\hat{K}_{f})_{\bar{a}b}(\phi,\phi^{*})f^{*}_{\bar{a}}f_{b}$ for a given field $f$) into the canonical form ${K}_{f} \sim \delta_{\bar{a}b}(f_{\mathrm{can}}^{*})_{\bar{a}}(f_{\mathrm{can}})_{b}$. As before, the ellipses stand for the higher order terms. 
Due to the common SUSY origin the kinetic terms of the scalars and the corresponding (Weyl) fermions in (\ref{kineticterm}) are the same. 

Moreover, since the symmetry properties of the K\"ahler metric are the same as those of the soft masses, the explicit form of the operator expansion  (\ref{kineticterm})  is completely analogous to (\ref{softmassexpansion}) 
and one can write the relevant expansion in the form:
\begin{eqnarray}\label{kahlergeneric}
(\hat{K}_{f,f^{c}})_{ij} &=& 
k_{0}^{f,f^{c}}\delta_{ij} + \sum_{A}k^{f,f^{c}}_{A}\frac{\langle\phi_{A}\phi_{A}^{*}\rangle_{ij}}{{M^{K}_{f,f^{c}}}^2}+\ldots 
\label{Kahlermetrics}
\end{eqnarray}
(where as before we assume common messenger scales $M^{K}_{f,f^{c},\Phi}\equiv M^{K}_{f,f^{c}}$ in the denominator) that differs from (\ref{softmassexpansion}) only by replacing the $b^{f,f^{c}}_{A}$ coefficients and messenger sector masses $M_{f,f^{c}}^{m}$ in  (\ref{softexpansionexplicit}) by $k_{A}^{f,f^{c}}$ and $M_{f,f^{c}}^{K}$ respectively. As before, we shall assume\footnote{For the same reasons that lead us to the formula (\ref{identifyingsoftmassmessengers}) above.}
\be
M_{f,f^{c}}^{K}=M_{f,f^{c}} 
\ee
so that one can again employ the same set of expansion parameters as in the other sectors, c.f. (\ref{epsilons-Yukawa}) and (\ref{epsilons-soft}).
Under this natural assumption one can recast $\hat{K}$ along the lines of the analogous soft-sector formula (\ref{Eq:M2}):
\begin{eqnarray}\label{Eq:K}
{\hat{K}_{f,f^{c}}} &=& 
k_{0}^{f,f^{c}}
\left(\begin{array}{ccc}
1 & 0 & 0 \\
0 & 1 & 0 \\
0 &  0 & 1
\end{array}\right) 
+
k^{f,f^{c}}_{1} (\varepsilon^{f,f^{c}}_{1})^2 
 \left(
\begin{array}{ccc}
1 & e^{i\phi_{1}} & e^{i\phi_{2}} \\
e^{-i\phi_{1}} & 1 & e^{i(\phi_{2}-\phi_{1})} \\
e^{-i\phi_{2}} & e^{-i(\phi_{2}-\phi_{1})} & 1
  \end{array}
\right) 
\\
& + & 
k^{f,f^{c}}_{2} (\varepsilon^{f,f^{c}}_{2})^2
\left(
\begin{array}{ccc}
0 & 0 & 0 \\
0 & 1 & e^{i\phi_{3}} \\
0 & e^{-i\phi_{3}} & 1
  \end{array}
\right) 
+
k^{f,f^{c}}_{3} (\varepsilon_3^{f,f^{c}})^{2}  
\left(\begin{array}{ccc}
0 & 0 & 0 \\
0 & 0 & 0 \\
0 & 0 & 1
\end{array}\right)+\mathrm{higher\,\,order\,\,terms}\nn\,,
\end{eqnarray}
where the same '$\varepsilon$' convention as in (\ref{Eq:M2}) has been adopted.

\subsubsection{Canonical normalization transformations}

Canonical normalization consists in redefining the defining basis field(s) ${f}$ and ${f}^{c}$ so that the original kinetic terms of the shape (for scalars for instance) ${\cal L}_\mathrm{kin}\sim\partial\tilde{f}^{\dagger}\hat{K}_{f}\partial\tilde{f}+\partial\tilde{f}^{c\dagger}\hat{K}^{f^{c}}\partial\tilde{f}^{c}$ receive the canonical form ${\cal L}^{\mathrm{can}}_{kin}\sim \partial  \tilde f^{\dagger}\partial \tilde f+\partial \tilde f^{c\dagger}\partial \tilde f^{c}$. This is achieved by transforming the defining superfields by $\hat{f}\to P_{f}\hat{f}\equiv \hat f_{\mathrm{can}}$ where $P_{f}$ is a matrix bringing the relevant K\"ahler metric $\hat K_{f}$ into the diagonal form (unless necessary from now on we shall suppress all the flavour indices): 
\be\label{Pmatrices}
P^{-1\dagger}\hat{K} P^{-1}=\mathbbm{1},\quad \mathrm{i.e.}\quad \hat{K}=P^{\dagger}P.
\ee 
Employing the $\varepsilon$, $\overline{\varepsilon}$ convention the leading order K\"ahler metric (\ref{Eq:K}) can be recast in a compact form:
\begin{eqnarray}
&& 
\hat  K_{Q} = k_0^Q \mathbbm{1}
 , \quad \hat K_{L} = k_0^{\ell} \mathbbm{1}
\qquad \mathrm{(to\;excellent\;approximation)}, 
\eea
with only the $SU(2)_{L}$-singlet sector featuring significant off-diagonalities:
\bea 
&&
\hat K_{u^c} \approx k_0^{u^{c}} \mathbbm{1}
+
\left(\begin{array}{lll}
 \varepsilon^2\overline\varepsilon^{2}\, k^{u^c}_{1}&  \varepsilon^2\overline\varepsilon^{2}\, k^{u^c}_{1}  e^{i \phi_1}&  \varepsilon^2\overline\varepsilon^{2}\, k^{u^c}_{1}  e^{i \phi_2}\\
.&  \varepsilon^2 k^{u^c}_{2} & \varepsilon^2 k^{u^c}_{2}e^{i \phi_3}\\
.& .&   \varepsilon_{3}^2k^{u^c}_3 
\end{array}\right)+\ldots , \nn \\
&&
\hat K_{d^c} \approx k_0^{d^{c}} \mathbbm{1}
+
\left(\begin{array}{lll}
\bar\varepsilon^4 k^{d^c}_{1}&  \bar\varepsilon^4 k^{d^c}_{1}  e^{i \phi_1}& \bar\varepsilon^4 k^{d^c}_{1}  e^{i \phi_2}\\
.&  \bar\varepsilon^2 k^{d^c}_{2} & \bar\varepsilon^2 k^{d^c}_{2}e^{i \phi_3}\\
.& .& \bar\varepsilon_{3}^2k^{d^c}_3 
\end{array}\right)+\ldots , \nn \\
&& \hat K_{\nu^c} \approx k_0^{\nu^{c}} \mathbbm{1}
+
\left(\begin{array}{lll}
 \varepsilon^2\overline\varepsilon^{2}\, k^{\nu^c}_{1}&   \varepsilon^2\overline\varepsilon^{2}\, k^{\nu^c}_{1}  e^{i \phi_1}&  \varepsilon^2\overline\varepsilon^{2}\, k^{\nu^c}_{1}  e^{i \phi_2}\\
.&  \varepsilon^2 k^{\nu^{c}}_{2} & \varepsilon^2 k^{\nu^{c}}_{2}e^{i \phi_3}\\
.& .&  \varepsilon_{3}^2k^{\nu^c}_3 
\end{array}\right)+\ldots ,\nn \\
&& \hat K_{e^c} \approx k_0^{e^{c}} \mathbbm{1}
+
\left(\begin{array}{lll}
 \bar\varepsilon^4 k^{e^c}_{1}&  \bar\varepsilon^4 k^{e^c}_{1}  e^{i \phi_1}& \bar\varepsilon^4 k^{e^c}_{1}  e^{i \phi_2}\\
.&  \varepsilon^2 k^{e^{c}}_{2} & \varepsilon^2 k^{e^{c}}_{2}e^{i \phi_3}\\
.& .&  \bar\varepsilon_{3}^2k^{e^c}_3 
\end{array}\right)+\ldots 
,
\end{eqnarray}
where as before the dotted terms can be reconstructed from hermiticity.
The matrices $P_{f}$ and $P_{f^c}$ are obtained\footnote{Recall that relation (\ref{Pmatrices}) fixes the $P$-matrices only up to a global unitary transformation $P\to UP$ that drops out in (\ref{Pmatrices}) and we adopt the convention in which $P$'s are hermitean at the leading order.} to leading order in $\varepsilon, \bar\varepsilon$ as
\be
P_{Q} = \begin{pmatrix}\label{Pf}
\sqrt{k_{0}^Q} &0 & 0\\
0& \sqrt{k_0^Q} &0 \\
0&0 & \sqrt{k_0^Q} 
\end{pmatrix}, \quad
P_{L} = \begin{pmatrix}
\sqrt{k_{0}^L} &0 & 0\\
0& \sqrt{k_0^L} &0 \\
0&0 & \sqrt{k_0^L} 
\end{pmatrix}\,,
\ee
while
\be\label{Pfc}
P_{u^c} = \begin{pmatrix}
\sqrt{k_0^{u^c}}  + \frac{ \varepsilon^2\overline\varepsilon^{2}\, k^{u^c}_{1}}{2\sqrt{k_0^{u^c}}}&  \frac{ \varepsilon^2\overline\varepsilon^{2}\, k^{u^c}_{1}  e^{i \phi_1}}{2\sqrt{k_0^{u^c}}}& \frac{ \varepsilon^2\overline\varepsilon^{2}\, k^{u^c}_{1}  e^{i \phi_2}}{\sqrt{k_0^{u^c}} + \sqrt{k_0^{u^c} + k^{u^c}_3 \varepsilon_3^2}}\\
\frac{ \varepsilon^2\overline\varepsilon^{2}\, k^{u^c}_{1}  e^{-i \phi_1}}{2\sqrt{k_0^{u^c}}}& \sqrt{k_0^{u^c}} + \frac{\varepsilon^2 k^{u^c}_{2}}{2\sqrt{k_0^{u^c}}} & \frac{\varepsilon^2 k^{u^c}_{2} e^{i \phi_3}}{\sqrt{k_0^{u^c}} + \sqrt{k_0^{u^c} + k^{u^c}_3 \varepsilon_3^2}}\\
\frac{ \varepsilon^2\overline\varepsilon^{2}\, k^{u^c}_{1}  e^{-i \phi_2}}{\sqrt{k_0^{u^c}} + \sqrt{k_0^{u^c} + k^{u^c}_3 \varepsilon_3^2}}& \frac{\varepsilon^2 k^{u^c}_{2}e^{-i \phi_3}}{\sqrt{k_0^{u^c}} + \sqrt{k_0^{u^c} + k^{u^c}_3 \varepsilon_3^2}}& \sqrt{k_0^{u^c} + k^{u^c}_3\varepsilon_3^2} 
\end{pmatrix}+\ldots, 
\ee
and the remaining $P_{f^c}$ matrices (for $f^{c}=d^{c}, e^{c}$ and $\nu^{c}$) are obtained from these by the following substitutions: 
\bea
P_{d^{c}} & : & k_{i}^{u^{c}}\to k_{i}^{d^{c}},\, \varepsilon, \varepsilon_{3}\to \overline\varepsilon, \overline\varepsilon_{3}\;,\quad
P_{e^{c}}  :  k_{i}^{u^{c}}\to k_{i}^{e^{c}},\, \varepsilon, \varepsilon_{3}\to \overline\varepsilon, \overline\varepsilon_{3}\;,\quad
P_{\nu^{c}}  :  k_{i}^{u^{c}}\to k_{i}^{\nu^{c}}.\nn  
\eea
Notice that due to the relatively large $\varepsilon_{3},\,\bar\varepsilon_{3}, $ the na\"\i ve factorization $P\sim \sqrt{k_{0}}(\mathbbm{1}+\Delta P)$ (with $|\Delta P|\ll \mathbbm{1}$, that could be  helpful in calculating $P^{-1}$ and $\Delta P$ from a similar expansion for the K\"ahler potential) is violated in the third family due to higher order $\varepsilon_{3}$, $\bar\varepsilon_{3}$,  effects.
\subsubsection{Canonical form of $Y$, $A$ and $m^{2}$\label{cansection}}
At the level of Yukawa, trilinear and soft mass matrices the transition to the canonically normalized quantities is achieved via  
\bea
Y^{f} =   (P^{-1}_{f})^{T}\hat Y^{f}P^{-1}_{f^{c}},\quad
A^{f} =   (P^{-1}_{f})^{T}\hat A^{f}P^{-1}_{f^{c}} \label{canonicalforms},\quad
m^{2}_{f,f^{c} }  =  (P^{-1}_{f,f^{c}})^{\dagger}\hat m^{2}_{f,f^{c}} P^{-1}_{f,f^{c}}
\eea
where the hats denote the defining basis matrices while the plain symbols stand for the canonically normalized ones. Expanding the K\"ahler metric 'square root' matrices $P_{f,f^{c}}$ in terms of the expansion parameters $\varepsilon, \overline{\varepsilon}$ as in (\ref{Pf}), (\ref{Pfc}), one can relatively easily calculate their inverse $P^{-1}_{f,f^{c}}$ that subsequently enter (\ref{canonicalforms}).
\paragraph{Canonical form of the Yukawa couplings:}\mbox{}\\
Utilizing the formulae (\ref{Pf}) and (\ref{Pfc}), the prescription  (\ref{canonicalforms}) yields the canonically normalized charged sector Yukawa matrices in the form:
\begin{eqnarray}
Y^u &=& 
\begin{pmatrix}
{\cal O}(\varepsilon^4\bar\varepsilon^{3})  & 
 y^{u}_{12} \varepsilon^2\bar\varepsilon  & 
 y^{u}_{13}  \varepsilon^2\bar\varepsilon  \\
 y^{u}_{21} \varepsilon^2\bar\varepsilon  & 
 y^{u}_{22} \varepsilon^2 & 
 y^{u}_{23} \varepsilon^2 \\
 y^{u}_{31}  \varepsilon^2\bar\varepsilon  & 
 y^{u}_{32} \varepsilon^2 & 
 y^{u}_{33} \varepsilon_{3}^2
\end{pmatrix}+\ldots, \nn \\
Y^d  &=& \label{Yukawacan}
\begin{pmatrix}
{\cal O}(\bar\varepsilon^7)  & 
 y^{d}_{12} \bar\varepsilon^3  & 
 y^{d}_{13} \bar\varepsilon^3 \\
 y^{d}_{21} \bar\varepsilon^3 & 
 y^{d}_{22} \bar\varepsilon^2 & 
 y^{d}_{23} \bar\varepsilon^2 \\
 y^{d}_{31} \bar\varepsilon^3 & 
 y^{d}_{32} \bar\varepsilon^2 & 
 y^{d}_{33} \bar\varepsilon_{3}^2
\end{pmatrix}+\ldots, \qquad
Y^e  = 
\begin{pmatrix}
{\cal O}(\bar\varepsilon^7)  & 
 y^{e}_{12} \bar\varepsilon^3  & 
 y^{e}_{13} \bar\varepsilon^3 \\
 y^{e}_{21} \bar\varepsilon^3 & 
 y^{e}_{22} \bar\varepsilon^2 & 
 y^{e}_{23} \bar\varepsilon^2 \\
 y^{e}_{31} \bar\varepsilon^3 & 
 y^{e}_{32} \bar\varepsilon^2 & 
 y^{e}_{33} \bar\varepsilon_{3}^2
\end{pmatrix}+\ldots, \nn
\end{eqnarray}
where for example (at leading order):
\begin{eqnarray}
y^u_{22} &=&  R^uy_4^u  \;,\quad 
y^u_{32} =  R^u \left(y_4^u e^{i\phi_3}  -  \frac{1}{2} 
{y}^u_3 \varepsilon_{3}^2\frac{{k}^{u^{c}}_{2}}{k_{0}^{u^{c}}}e^{- i\phi_3}\right)   \;,\label{yijcan}
\\
\quad
y^u_{12} & = &  R^u{y}_1^u  \;,\quad
y^u_{21} =  R^u{{y}_2^u}\;,\quad
y^u_{31} =  R^u{y}_2^u e^{i \phi_3}\;, \quad y^u_{13} =  R^u{y}_1^u\left(1-\frac{1}{2}\varepsilon_{3}^{2}\frac{k_{3}^{u^c}}{k_{0}^{u^c}}\right) e^{i \phi_3}\;,\nn \\
y^u_{23} &=&  R^u y_4^u\left(1  - \frac{1}{2}\varepsilon_{3}^2  \frac{{k}^{u^{c}}_{3}}{{k}_0^{u^{c}}}\right)e^{i \phi_3}\;, \quad y^u_{33} =  R^u{y}_3^u\left(1-\frac{1}{2}\varepsilon_{3}^{2}\frac{k_{3}^{u^c}}{k_{0}^{u^c}}\right)  \;,\nn
\end{eqnarray}
and $R^{u}\equiv \frac{1}{\sqrt{{k}_0^Q {k}_0^{u^{c}}}}$ is a universal rescaling coefficient. The $y_{ij}^{d}$ and $y_{ij}^{e}$ factors are obtained from (\ref{yijcan}) upon replacing $\{k_{i}^{u^{c}},\,y_{i}^{u},\,R^{u}\} \to \{k_{i}^{d^{c}},\,y_{i}^{d},\,R^{d}\}$ and  $\{k_{i}^{e^{c}},\,y_{i}^{e},\,R^{e}\}$ respectively.

One can see from the formulae above that the net effect of canonical normalization can be at the leading order described as a real rescaling of all the defining basis couplings followed by a unitary transformation that can substantially affects only the 23 sector entries. This can be seen from the fact that the 23 mixing angle in $P_{u^{c}}^{-1}$ is of the order of $\varepsilon^{2}$ so it can rotate the defining basis 33 Yukawa entry to the 32 position yielding a contribution comparable with the leading order $\sim \varepsilon^{2}$ term already present\footnote{This also justifies the presence of the $k_{2}^{u^{c}}$ coefficient in the relevant formula (\ref{yijcan}).} in $\hat Y_{23}$. However, the 12 and 13 rotations in $P_{u^{c}}^{-1}$ do bring only subleading effects to $\hat Y_{21}$ and $\hat Y_{31}$ and that is why they are absent in the leading order formulae (\ref{yijcan}). Finally, the 13, 23 and 33 entries are affected only by the rescaling due to the nontrivial 33 element of $P_{u^{c}}^{-1}$. 

\paragraph{Canonical form of the trilinear couplings:}\mbox{}\\
The effects of canonical normalisation in the trilinear sector are completely analogous to the corresponding Yukawas yielding the canonically normalized trilinear soft matrices: 
\begin{eqnarray}
A^u &=& 
\begin{pmatrix}
{\cal O}(\varepsilon^4\bar\varepsilon^{3})  & 
 a^{u}_{12} \varepsilon^2\bar\varepsilon  & 
 a^{u}_{13} \varepsilon^2\bar\varepsilon \\
 a^{u}_{21} \varepsilon^2\bar\varepsilon & 
 a^{u}_{22} \varepsilon^2 & 
 a^{u}_{23} \varepsilon^2 \\
 a^{u}_{31} \varepsilon^2\bar\varepsilon & 
 a^{u}_{32} \varepsilon^2 & 
 a^{u}_{33} \varepsilon_{3}^2
\end{pmatrix}+\ldots, \nn \\
A^d &=& 
\begin{pmatrix}\label{Acan}
{\cal O}(\bar\varepsilon^7)  & 
 a^{d}_{12} \bar\varepsilon^3  & 
 a^{d}_{13} \bar\varepsilon^3 \\
 a^{d}_{21} \bar\varepsilon^3 & 
 a^{d}_{22} \bar\varepsilon^2 & 
 a^{d}_{23} \bar\varepsilon^2 \\
 a^{d}_{31} \bar\varepsilon^3 & 
 a^{d}_{32} \bar\varepsilon^2 & 
 a^{d}_{33} \bar\varepsilon_{3}^2
\end{pmatrix}+\ldots, \qquad
A^e = 
\begin{pmatrix}
{\cal O}(\bar\varepsilon^7)  & 
 a^{e}_{12} \bar\varepsilon^3  & 
 a^{e}_{13} \bar\varepsilon^3 \\
 a^{e}_{21} \bar\varepsilon^3 & 
 a^{e}_{22} \bar\varepsilon^2 & 
 a^{e}_{23} \bar\varepsilon^2 \\
 a^{e}_{31} \bar\varepsilon^3 & 
 a^{e}_{32} \bar\varepsilon^2 & 
 a^{e}_{33} \bar\varepsilon_{3}^2
\end{pmatrix}+\ldots,\nn
\end{eqnarray}
where for example (at leading order):
\begin{eqnarray}
a^u_{22} &=&  R^ua_4^u  \;,\quad 
a^u_{32} =  R^u \left(a_4^u e^{i\phi_3}  -  \frac{1}{2} 
{a}^u_3 \varepsilon_{3}^2\frac{{k}^{u^{c}}_{2}}{k_{0}^{u^{c}}}e^{- i\phi_3}\right)   \;,\label{aijcan}
\\
\quad
a^u_{12} & = &  R^u{a}_1^u  \;,\quad
a^u_{21} =  R^u{{a}_2^u}\;,\quad
a^u_{31} =  R^u{a}_2^u e^{i \phi_3}\;, \quad a^u_{13} =  R^u{a}_1^u\left(1-\frac{1}{2}\varepsilon_{3}^{2}\frac{k_{3}^{u^c}}{k_{0}^{u^c}}\right) e^{i \phi_3}\;,\nn \\
a^u_{23} &=&  R^u a_4^u\left(1  - \frac{1}{2}\varepsilon_{3}^2  \frac{{k}^{u^{c}}_{3}}{{k}_0^{u^{c}}}\right)e^{i \phi_3}\;, \quad a^u_{33} =  R^u{a}_3^u\left(1-\frac{1}{2}\varepsilon_{3}^{2}\frac{k_{3}^{u^c}}{k_{0}^{u^c}}\right)  \;,\nn
\end{eqnarray}
while the relevant expressions for down-type quark and charged lepton sector quantities are obtained along the same lines as those in the Yukawa sector.

One can again appreciate the similarity of the canonically normalized Yukawa and trilinear couplings (\ref{yijcan}) and (\ref{aijcan}) coming from the common origin (\ref{Yukawaoperator}) and (\ref{trilinearoperator}). This will lead to great simplification upon getting to the SCKM basis, c.f. section \ref{Aandm2inSCKM}. 
\paragraph{Canonical form of the soft masses:}\mbox{}\\
Employing again prescription (\ref{canonicalforms}), the canonically normalized soft mass terms in the $SU(2)_{L}$-doublet sector read:
\be 
m^2_{Q} = m_0^2\, b^{Q}
\mathbbm{1}
,  \qquad
m^2_{L} = m_0^2 \, {b}^{\ell} 
\mathbbm{1},
\ee
while the $SU(2)_{L}$-singlet ones develop a non-diagonal structure
\bea
\label{mcan}
& & m^2_{u^{c}} \approx m_0^2 \left[{b}^{u^{c}} 
\mathbbm{1}
+
\left(\begin{array}{lll}
 \varepsilon^2\bar\varepsilon^{2} {b}^{u^{c}}_{11}&  \varepsilon^2\bar\varepsilon^{2} {b}^{u^{c}}_{12} &  \varepsilon^2\bar\varepsilon^{2} {b}^{u^{c}}_{13} \\
. &  \varepsilon^2 {b}^{u^{c}}_{22} & \varepsilon^2 {b}^{u^{c}}_{23}\\
. & . &  \varepsilon_{3}^2{b}^{u^{c}}_{33} 
\end{array}\right)+\ldots\right] , \nn 
\eea
\bea
& & m^2_{{d^c}} \approx m_0^2 \left[{b}^{d^c} 
\mathbbm{1}
+
\left(\begin{array}{lll}
\overline{\varepsilon}^4 {b}^{d^c}_{11}&  \overline{\varepsilon}^4 {b}^{d^c}_{12} & \overline{\varepsilon}^4 {b}^{d^c}_{13} \\
. &  \overline{\varepsilon}^2 {b}^{d^c}_{22} & \overline{\varepsilon}^2 {b}^{d^c}_{23}\\
. & . & \overline\varepsilon_{3}^2{b}^{d^c}_{33} 
\end{array}\right)+\ldots\right] , 
\\
&& 
m^2_{{\nu^c} } \approx m_0^2 \left[{b}^{\nu^c} 
\mathbbm{1}
+
\left(\begin{array}{lll}
 \varepsilon^2\bar\varepsilon^{2} {b}^{\nu^c}_{11}&   \varepsilon^2\bar\varepsilon^{2}{b}^{\nu^c}_{12} &  \varepsilon^2\bar\varepsilon^{2} {b}^{\nu^c}_{13} \\
. &  {\varepsilon}^2 {b}^{\nu^c}_{22} & {\varepsilon}^2 {b}^{\nu^c}_{23}\\
. & . &  \varepsilon_{3}^2{b}^{\nu^c}_{33} 
\end{array}\right)+\ldots\right], \nn\\
&& 
m^2_{{e^c} } \approx m_0^2 \left[{b}^{e^c} 
\mathbbm{1}
+
\left(\begin{array}{lll}
\overline{\varepsilon}^4 {b}^{e^c}_{11}&  \overline{\varepsilon}^4 {b}^{e^c}_{12} & \overline{\varepsilon}^4 {b}^{e^c}_{13} \\
. &  \overline{\varepsilon}^2 {b}^{e^c}_{22} & \overline{\varepsilon}^2 {b}^{e^c}_{23}\\
. & . &  \overline\varepsilon_{3}^2{b}^{e^c}_{33} 
\end{array}\right)+\ldots\right]. \nn
\end{eqnarray}
where the various coefficients above are given by
\begin{eqnarray}
b^{f}&=&\tfrac{{b}_0^f}{k_{0}^{f}}\;,\quad b^{f^{c}}=\tfrac{{b}_0^{f^{c}}}{k_{0}^{f^{c}}}\;,\quad b^{f^c}_{11} =  \tfrac{1}{{k}_0^{f^c}} \left({b}^{f^c}_{1} - \tfrac{{b}_0^{f^c}}{{k}_0^{f^c}} {k}^{f^c}_{1}\right) \;,\quad 
b^{f^c}_{22} =  \tfrac{1}{{k}_0^{f^c}} \left({b}^{f^c}_{2} - \tfrac{{b}_0^{f^c}}{{k}_0^{f^c}} {k}^{f^c}_{2}\right) \;, \nn \\
\label{canbcoefficients}
b^{f^c}_{12} & = &  \tfrac{1}{{k}_0^{f^c}} \left({b}^{f^c}_{1} - \tfrac{{b}_0^{f^c}}{{k}_0^{f^c}} {k}^{f^c}_{1}\right)  e^{i \phi_1}  \;,\quad
\\
b^{f^c}_{33} & = &   \tfrac{1}{{k}_0^{f^c}}\left({b}^{f^c}_3-\tfrac{{b}_{0}^{f^c}}{k_{0}^{f^c}}k_{3}^{f}\right)\left(1-\tfrac{{k}^{f^c}_3}{k_{0}^{f}}\varepsilon_{3}^{f2}\right) + {\cal O}(\varepsilon_{3}^{f4})
\nn \\
b^{f^c}_{13} &=&  \left\{\tfrac{1}{{k}_0^{f^c}} \left({b}^{f^c}_{1} - \tfrac{{b}_0^{f^c}}{{k}_0^{f^c}} {k}^{f^c}_{1}\right) -  \frac{1}{2{k}_0^{f2}} \left[ {b}^{f^c}_{1} {k}_3^{f^c} - {k}^{f^c}_{1} \left(2 \tfrac{{b}_0^{f^c}}{{k}_0^{f^c}}  {k}_3^{f^c} - {b}_3^{f^c} \right) \right]\varepsilon_{3}^{f2}\right\} e^{i \phi_2} + {\cal O}(\varepsilon_{3}^{f4})\nn \\
b^{f^c}_{23} &=&  \left\{\tfrac{1}{{k}_0^{f^c}} \left({b}^{f^c}_{2} - \tfrac{{b}_0^{f^c}}{{k}_0^{f^c}} {k}^{f^c}_{2}\right) -  \frac{1}{2{k}_0^{f2}} \left[ {b}^{f^c}_{2} {k}_3^{f^c} - {k}^{f^c}_{2} \left(2 \tfrac{{b}_0^{f^c}}{{k}_0^{f^c}}  {k}_3^{f^c} - {b}_3^{f^c} \right) \right]\varepsilon_{3}^{f2}\right\} e^{i \phi_3} + {\cal O}(\varepsilon_{3}^{f4}),\nn 
\end{eqnarray} 
and hermiticity of the soft mass matrices. The generic symbol $\varepsilon_{3}^f$ represents $\varepsilon_{3}$ or $\bar\varepsilon_{3}$, respectively.

There are two points worth commenting on here: the first concerns the case when the K\"ahler metric expansion coefficients in (\ref{kahlergeneric}) and those in the soft mass terms (\ref{softmassexpansion}) are mutually proportional to each other, i.e. $b_{i}=\alpha k_{i}\; \forall i$ where $\alpha$ is a universal constant hence  all the nonuniversal $b_{ij}$ coefficients in (\ref{canbcoefficients}) vanish\,! This could (hypothetically) happen if for instance both the K\"ahler metric and soft masses originate from a common factorizable\footnote{By factorizability we mean that the hidden sector fields couple to matter in a flavour-blind way like e.g. those in formula (\ref{generic}).} operator of the form (dropping all the flavour indices) \cite{King:2003xq}:
\be
{\cal L}\ni  \int\mathrm{d}^{4}\theta\, \left(1+\frac{\hat X^{\dagger}\hat X}{{M^X}^{2}}+\ldots\right)\hat f^{\dagger} \left(c_{0}\mathbbm{1}+\sum_{\Phi}c_{\Phi}\frac{\hat\Phi\otimes\hat\Phi^{\dagger}}{M_{f,\Phi}^{2}}+\ldots\right)\hat f  +\ldots\label{generic}\,,
\ee
If now $\hat X$ develops both the VEV on its scalar component $\tilde X$ as well as a non-zero $F$-term (like e.g. in supergravity) the flavour structure of the soft masses (driven by $F_{X}$) is identical to the flavour structure of the K\"ahler (driven by $\langle \tilde X^{\dagger}\tilde X\rangle$) and one obtains: 
\be
m_{0}^{2} b_{i}=\frac{|F_{X}|^{2}}{{M^X}^{2}}c_{i},\qquad k_{i}=\frac{|\langle X \rangle|^{2}}{{M^X}^{2}}c_{i},
\ee
and thus  $b_{i}=\alpha k_{i}\; \forall i$ with $\alpha=\frac{|F_{X}^{2}|}{m_{0}^{2}\vev{\tilde X}^{2}}$. This, however, should be expected since the SUSY-breaking sector in (\ref{generic}) factorized out of the flavour structure and thus the transmission of the SUSY-breaking to the visible sector is flavour blind, leading to the flavour-universal soft SUSY-breaking masses (after canonical normalization).  

Second, let us comment on the GUT limit of the soft-SUSY breaking
sector in the canonical basis. Employing for instance the Pati-Salam
condition (\ref{yGUT}) together with $b_{ij}^{f,f^{c}}\equiv b_{ij}$,
$b^{f,f^{c}}\equiv b$ and replacing all the 'bookkeeping' $y_{4}$ and
$a_{4}$ couplings via (\ref{ySigma4}) and (\ref{aSigma4}), one can see
that all the soft mass matrices coincide at leading order 
\be
m^{2}_{u^{c}}=m^{2}_{\nu^{c}}, \qquad m^{2}_{Q}=m^{2}_{L}, \qquad
m^{2}_{\tilde e^{c}}=m^{2}_{\tilde d^{c}} 
\ee 
(up to\footnote{This
difference comes from the potential sensitivity of the messenger
sector to the GUT-symmetry breaking -- indeed, messengers (being
vector-like) are not protected by chiral symmetry. One can find an
explicit example of this for instance in \cite{King:2006me}.}
extra $\varepsilon$ factors entering $m^{2}_{u^{c}}$,
$m^{2}_{\nu^{c}}$). This is, however, not entirely
the case for the Yukawa and trilinear couplings because of the
Clebsch-Gordon coefficient in the $y_{22}$, $y_{23}$ and $y_{32}$
Yukawas and similarly $a_{22}$, $a_{23}$ and $a_{32}$ in the trilinear
sector (at leading order):
\begin{eqnarray}
y_{22}^{f} &=&  R\,y_\Sigma \sigma C^f  \;,\\ 
y_{32}^{u,\nu} &= & R \left(y_\Sigma \sigma C^{u,\nu} e^{i\phi_3}  -  \frac{1}{2} 
{y}_3 \varepsilon_{3}^2\frac{{k}_{2}}{k_{0}}e^{- i\phi_3}\right)   \;,\label{yijcanGUT}
\quad
y^{u,\nu}_{23} =  R\,y_\Sigma \sigma C^{u,\nu} \left(1  - \frac{1}{2}\varepsilon_{3}^2  \frac{{k}_{3}}{{k}_0}\right)e^{i \phi_3}\;, \nn\\
y_{32}^{d,e} &= & R \left(y_\Sigma \sigma C^{d,e} e^{i\phi_3}  -  \frac{1}{2} 
{y}_3 \bar\varepsilon_{3}^2\frac{{k}_{2}}{k_{0}}e^{- i\phi_3}\right)   \;,
\quad
y^{d,e}_{23} =  R\,y_\Sigma \sigma C^{d,e} \left(1  - \frac{1}{2}\bar\varepsilon_{3}^2  \frac{{k}_{3}}{{k}_0}\right)e^{i \phi_3}\;, \nn
\end{eqnarray}
and similarly for the A-terms.
\section{Solving SUSY flavour and CP problems with $SU(3)$ family symmetry}
With all the relevant ingredients at hand we can now approach a detailed study of phenomenology of the tri-bimaximal $SU(3)$ flavour model under consideration. 
\subsection{General remarks}
The first step is to rotate all the canonically normalized quantities we have obtained in the last section into the SCKM basis \cite{Hall:1985dx} that makes the quark and (charged) lepton Yukawa matrices diagonal. The major benefit of this operation is that all the parameters in the soft sector are then (at least in principle) physical. 

On the technical side, this is achieved by redefining the matter fields by means of unitary transformations $f_{L}\to U_{L}^{f}f_{L}$, $f_{R}\to U_{R}^{f}f_{R}$ that act on the Yukawa matrices as:
\begin{eqnarray}\label{SCKMbasis}
Y_u &\rightarrow& U_\mathrm{L}^{u \dagger} Y_u  U_\mathrm{R}^{u}\equiv \tilde{Y}_{u} \; , \quad
Y_d \;\rightarrow\; U_\mathrm{L}^{d \dagger} Y_d  U_\mathrm{R}^{d} \equiv \tilde{Y}_{d}\; ,\quad
Y_e \;\rightarrow\; U_\mathrm{L}^{e \dagger} Y_e  U_\mathrm{R}^{e}\equiv \tilde{Y}_{e} \; ,
\end{eqnarray} 
where the tilded matrices correspond to the relevant quantities in the SCKM basis. Note that (by definition) $\tilde Y_{u,d,e}$ are all diagonal with (conventionally) positive eigenvalues.

Once electroweak symmetry is broken, the $SU(2)_{L}$ doublets $Q_{L}, L_{L}$ decompose into $u_{L}, d_{L}$ and $\nu_{L}, e_{L}$ components and so do the corresponding superpartners $\tilde{Q}$ and $\tilde{L}$. Thus, each doublet soft mass term $m^{2}_{Q,L}$ corresponds to two physical scalar masses, i.e. one is left with four SCKM soft mass matrices corresponding to the up- and down- type rotations imposed onto $m^{2}_{Q,L}$:  
\bea
(\tilde{m}^{2}_{u})_{LL}\equiv U^{u\dagger}_{L}m^{2}_{Q}U^{u}_{L},\quad (\tilde{m}^{2}_{d})_{LL}\equiv U^{d\dagger}_{L}m^{2}_{Q}U^{d}_{L}, \\
(\tilde{m}^{2}_{e})_{LL}\equiv U^{e\dagger}_{L}m^{2}_{L}U^{e}_{L},\quad (\tilde{m}^{2}_{\nu})_{LL}\equiv U^{\nu\dagger}_{L}m^{2}_{L}U^{\nu}_{L}\nn, 
\eea
and similarly for the right-handed mass insertions:
\bea
(\tilde{m}^{2}_{u})_{RR}\equiv U^{u\dagger}_{R}m^{2}_{u^{c}}U^{u}_{R},
\quad (\tilde{m}^{2}_{d})_{RR}\equiv U^{d\dagger}_{R}m^{2}_{d^{c}}U^{d}_{R}, \\
(\tilde{m}^{2}_{e})_{RR}\equiv U^{e\dagger}_{R}m^{2}_{e^{c}}U^{e}_{R},
\quad (\tilde{m}^{2}_{\nu})_{RR}\equiv U^{\nu\dagger}_{R}m^{2}_{\nu^{c}}U^{\nu}_{R}\nn.
\eea
Notice that the rotations imposed onto $m^{2}_{L}$ and $m^{2}_{\nu^{c}}$ in order to get the relevant neutrino soft mass matrices in the SCKM basis are {\it not} the large (tri-bimaximal in the left-handed sector) mixings $\propto U_{MNS}$ diagonalizing the physical light neutrino mass matrix (coming from seesaw) but the (relatively small) mixing matrices $U_{L}^{\nu}$,  $U_{R}^{\nu}$ diagonalizing the neutrino Yukawa coupling itself (that in unified framework are close to all the other mixings around).

On the similar grounds one must rotate the canonically normalized trilinear couplings to get  
\bea
\tilde{A}_{u}\equiv U^{u\dagger}_{L}A_{u}U^{u}_{R},
\quad \tilde{A}_{d}\equiv U^{d\dagger}_{L}A_{d}U^{d}_{R}, \\
\tilde{A}_{e}\equiv U^{e\dagger}_{L}A_{e}U^{e}_{R},
\quad \tilde{A}_{\nu}\equiv U^{\nu\dagger}_{L}A_{\nu}U^{\nu}_{R}\nn,
\eea
plus hermitean conjugated formulae for the corresponding $RL$ quantities.
 
SUSY CP and flavour violation is then induced by a misalignment of the {\it full} sfermion and fermion mass matrices that apart from the soft factors defined above contain extra Yukawa and $D$-term (in the LL and RR sectors) and $\mu$-term (in the LR and RL entries) contributions:
\be
m^{2}_{\tilde{f}}\equiv \begin{pmatrix}
{m^2_{\tilde{f}}}_{LL} & {m^2_{\tilde{f}}}_{LR} \\
{m^2_{\tilde{f}}}_{RL} & {m^2_{\tilde{f}}}_{RR} 
\end{pmatrix}
=\begin{pmatrix}
(\tilde{m}^{2}_{f})_{LL}+\tilde Y_{f}\tilde Y_{f}^{\dagger}v_{u,d}^{2}+\tilde D^{f}_{LL} & \tilde A_{f}v_{u,d}-\mu Y_{f}v_{d,u}\\
\tilde A^{\dagger}_{f}v_{u,d}-\mu \tilde Y_{f}^{\dagger}v_{d,u} & (\tilde{m}^{2}_{f})_{RR}+\tilde Y_{f}^{\dagger}\tilde Y_{f}v_{u,d}^{2}+\tilde D^{f}_{RR} \\
\end{pmatrix}
\ee
where the various tilded terms correspond to the SCKM quantities defined above and $\tilde D_{LL,RR}$ denote the effects of the $D$-terms, that are however strongly model dependent and shall be neglected in this study.
It is very convenient to define a set of dimensionless ``mass insertion'' parameters as follows \cite{Gabbiani:1996hi}:
\begin{eqnarray}\label{deltas}
(\delta^f_{LL})_{ij} = \frac{\left({m^2_{\tilde{f}}}_{LL}\right)_{ij}}{\langle m_{\tilde f}\rangle^2_{LL}} \; , \quad
(\delta^f_{RR})_{ij} = \frac{\left({m^2_{\tilde{f}}}_{RR}\right)_{ij}}{\langle m_{\tilde f}\rangle^2_{RR}} \; , \quad
(\delta^f_{LR})_{ij} = \frac{\left({m^2_{\tilde{f}}}_{LR}\right)_{ij}}{\langle m_{\tilde f}\rangle^2_{LR}} \; , \quad
\mathrm {etc.}
\end{eqnarray}
with an average squared squark (or slepton) mass $\langle m_{\tilde{f}}\rangle^2_{AB}\equiv \sqrt{(m_{\tilde{f}AA}^{2})_{ii}(m_{\tilde{f}BB}^{2})_{jj}}$. 
In the mass insertion approximation, these quantities can be used
directly to estimate rates for flavour and CP violating processes at the loop-level and bounds on the $(\delta^{f}_{XY})_{ij}$'s (which typically depend on $\tan\beta\equiv v_{u}/v_{d}$ and $\langle\tilde m_{f}\rangle^2$) have been derived in the literature; for further details see Tables \ref{Tabledeltad}-\ref{Tabledeltal} and references therein.
%

\subsubsection{Experimental constraints\label{experiment}}
Since $\tilde{m}^2_{{\nu}} \sim  \tilde{m}^2_{{e}}$ under the assumptions we made, in what follows we shall consider constraints on 
\begin{eqnarray}
(\delta^\ell_{LL})_{ij} \equiv (\delta^e_{LL})_{ij} \sim (\delta^\nu_{LL})_{ij} \; ,
\end{eqnarray} 
as is commonly done in the literature.
A sample compilation of the various experimental constraints on $\delta$'s available in the literature is given in Tables \ref{Tabledeltad}, \ref{Tabledeltau} and \ref{Tabledeltal}. The extra assumptions made throughout compiling these tables are commented on in the relevant captions.
\begin{table}
\begin{tabular}{|c||r|r|r|cc|}
\hline
\multicolumn{1}{|c||}{${\delta^{d}}$} & \multicolumn{1}{c|}{$LL$} & \multicolumn{1}{c|}{$LR/RL$} & \multicolumn{1}{c|}{$RR$} & \multicolumn{2}{c|}{source}\\
\hline
\hline
$|\delta_{12}|$& 
$\mbox{}_{LL} : 1.4\times 10^{-2}$ &  
$\mbox{}_{LR}  : 9.0\times 10^{-5}$ &  
$\mbox{}_{RR} : 9.0\times 10^{-3}$ &  
\cite{Ciuchini:2007ha} & $\Delta m_{K},\varepsilon, ...$
\\ 
$|\mathrm{Re}\delta^{2}_{12}|^\frac{1}{2}$ & 
$\mbox{}_{LL^2} : 4.0\times 10^{-2}$ &  
$\mbox{}_{LR^2}  : 4.4\times 10^{-3}$ &  
$\mbox{}_{LLRR} : 2.8\times 10^{-3}$ &  
\cite{Gabbiani:1996hi,Masiero:1999ub} & $\Delta m_{K}$
\\
$|\mathrm{Im}\delta^{2}_{{12}}|^\frac{1}{2}$ & 
$\mbox{}_{LL^2} : 3.2\times 10^{-3}$ &  
$\mbox{}_{LR^2}  : 3.5\times 10^{-4}$ &  
$\mbox{}_{LLRR} : 2.2\times 10^{-4}$ & 
\cite{Gabbiani:1996hi,Masiero:1999ub} &  $\varepsilon$
\\
$|\mathrm{Im}\delta_{{12}}|$ & 
$\mbox{}_{LL} : 4.8\times 10^{-1}$ &  
$\mbox{}_{LR}  : 2.0\times 10^{-5}$ &  
\multicolumn{1}{c|}{$-$}  & 
\cite{Gabbiani:1996hi,Masiero:1999ub} &  $\varepsilon'/\varepsilon$
\\
\hline
$|\delta_{13}|$& 
$\mbox{}_{LL} : 9.0\times 10^{-2}$ &  
$\mbox{}_{LR}  : 1.7\times 10^{-2}$ &  
$\mbox{}_{RR} : 7.0\times 10^{-2}$ &  
\cite{Ciuchini:2007ha} & $\Delta m_{B_{d}},2\beta$
\\
$|\mathrm{Re}\delta^{2}_{13}|^\frac{1}{2}$ & 
$\mbox{}_{LL^2} : 9.8\times 10^{-2}$ &  
$\mbox{}_{LR^2}  : 3.3\times 10^{-2}$ &  
$\mbox{}_{LLRR} : 1.8\times 10^{-2}$ &  
\cite{Gabbiani:1996hi} & $\Delta m_{B_{d}}$
\\
$|\mathrm{Re}\delta_{13}|$ & 
$\mbox{}_{LL} : 1.4\times 10^{-1}$ &  
$\mbox{}_{LR}  : 5.2\times 10^{-2}$ &  
$\mbox{}_{RR}  : 2.1\times 10^{-2}$ &   
\cite{Becirevic:2001jj} &  $\Delta m_{B_{d}}$
\\
$|\mathrm{Im}\delta_{13}|$ & 
$\mbox{}_{LL} : 3.0\times 10^{-1}$ &  
$\mbox{}_{LR}  : 2.3\times 10^{-2}$ &  
$\mbox{}_{RR}  : 9.0\times 10^{-3}$ &    
\cite{Becirevic:2001jj} &  $B_{d}-\overline{B}_{d}$
\\
\hline
$|\delta_{23}|$& 
$\mbox{}_{LL} : 1.6\times 10^{-1}$ &  
$\mbox{}_{LR}  : 4.5\times 10^{-3}$ &  
$\mbox{}_{RR} : 2.2\times 10^{-1}$ &  
\cite{Ciuchini:2007ha} & $\Delta m_{B_{s}}$
\\
$\mathrm{Re}\delta_{23}$& 
$\mbox{}_{LL} : 5.0\times 10^{-1}$ &  
$\mbox{}_{LR}  : 2.5\times 10^{-2}$ &  
$\mbox{}_{RR} : 5.0\times 10^{-1}$ &  
\cite{Lunghi:2001af} & $b\to s\gamma$
\\
& 
$\mbox{}_{LL} : 3.0\times 10^{-1}$ &  
$\mbox{}_{LR}  : 2.0\times 10^{-2}$ &  
$\mbox{}_{RR} : 2.0\times 10^{-1}$ &  
\cite{Silvestrini:2005zb} & $b\to s l^{+}l^{-}$
\\
$|\mathrm{Im}\delta_{23}|$& 
\multicolumn{1}{c|}{$-$}  &  
$\mbox{}_{LR}  : 1.5\times 10^{-2}$ &  
\multicolumn{1}{c|}{$-$}  &   
\cite{Lunghi:2001af} & $b\to s\gamma$\\
& 
$\mbox{}_{LL} : 3.0\times 10^{-1}$ &  
$\mbox{}_{LR}  : 1.8\times 10^{-2}$ &  
$\mbox{}_{RR} : {\cal O}(1)\;\;\;\;\;\;\;\;\; $ &  
\cite{Silvestrini:2005zb} & $b\to s l^{+}l^{-}$
\\ 
\hline
$|\mathrm{Re}\delta_{11}|$& 
\multicolumn{1}{c|}{$-$}  & 
$\mbox{}_{LR}  : 1.6\times 10^{-3}$ &  
\multicolumn{1}{c|}{$-$}  & 
\cite{Gabbiani:1996hi} & $\Delta m_{d}$
\\ 
$|\mathrm{Im}\delta_{11}|$& 
\multicolumn{1}{c|}{$-$}  & 
$\mbox{}_{LR}  : 3.0\times 10^{-6}$ &  
\multicolumn{1}{c|}{$-$}  & 
\cite{Gabbiani:1996hi}& $d_{n}$
\\ 
$ $& 
\multicolumn{1}{c|}{$-$}  & 
$\mbox{}_{LR}  : 1.1\times 10^{-6}$ &  
\multicolumn{1}{c|}{$-$}  & 
\cite{Abel:2001vy} & $d_{n}$
\\ 
$ $& 
\multicolumn{1}{c|}{$-$}  & 
$\mbox{}_{LR}  : 6.7\times 10^{-8}$ &  
\multicolumn{1}{c|}{$-$}  & 
\cite{Abel:2001vy} & $d_{Hg}$
\\
\hline
$|\mathrm{Re}\delta_{22}|$& 
\multicolumn{1}{c|}{$-$}  & 
$\mbox{}_{LR}  : 2.4\times 10^{-2}$ &  
\multicolumn{1}{c|}{$-$}  & 
\cite{Gabbiani:1996hi} & $\Delta m_{s}$
\\ 
$|\mathrm{Im}\delta_{22}|$& 
\multicolumn{1}{c|}{$-$}  & 
$\mbox{}_{LR}  : 6.6\times 10^{-6}$ &  
\multicolumn{1}{c|}{$-$}  & 
\cite{Abel:2001vy} & $d_{n}$
\\ 
$ $& 
\multicolumn{1}{c|}{$-$}  & 
$\mbox{}_{LR}  : 5.6\times 10^{-6}$ &  
\multicolumn{1}{c|}{$-$}  & 
\cite{Abel:2001vy} & $d_{Hg}$
\\
\hline
$|\mathrm{Re}\delta_{33}|$& 
\multicolumn{1}{c|}{$-$}  & 
$\mbox{}_{LR}  : 7.3\times 10^{-1}$ &  
\multicolumn{1}{c|}{$-$}  & 
\cite{Gabbiani:1996hi} & $\Delta m_{b}$
\\
\hline
\end{tabular}
\caption{\label{Tabledeltad}Various experimental constraints on the mass insertion parameters associated with the down-type squark soft parameters $\tilde{m}_d$ ($LL$ and $RR$ case) and $A^{d}$ ($LR/RL$ case). The given numbers correspond to an average down squark mass $\langle\tilde{m}_{d}\rangle=500\,\mathrm{GeV}$ in one mass insertion approximation. In all $m_{\tilde{g}}/\tilde{m}=1$ is assumed. The actual numbers correspond to $\tan\beta$ in the range $5\lesssim\tan\beta\lesssim15$, but can vary with the details of the underlying model. Note also that unlike the $RR/LL$ limits, the $LR/RL$ bounds are essentially $\tan\beta$ insensitive, see e.g. \cite{Gabbiani:1996hi}.}
\end{table}

\begin{table}
\begin{tabular}{|c||r|r|r|cc|}
\hline
\multicolumn{1}{|c||}{$\delta^{u}$} & \multicolumn{1}{c|}{$LL$} & \multicolumn{1}{c|}{$LR/RL$} & \multicolumn{1}{c|}{$RR$} & \multicolumn{2}{c|}{source}\\
\hline
\hline
$|\mathrm{Re}\delta_{12}|^\frac{1}{2}$ & 
$\mbox{}_{LL^2} : 1.0\times 10^{-1}$ &  
$\mbox{}_{LR^2}  : 3.1\times 10^{-2}$ &  
$\mbox{}_{LLRR} : 1.7\times 10^{-2}$ &  
\cite{Gabbiani:1996hi} & $\Delta m_{D}$
\\ 
\hline
$|\mathrm{Im}\delta_{11}|$& 
\multicolumn{1}{c|}{$-$}  & 
$\mbox{}_{LR}  : 5.9\times 10^{-6}$ &  
\multicolumn{1}{c|}{$-$}  & 
\cite{Gabbiani:1996hi}& $d_{n}$
\\ 
$ $& 
\multicolumn{1}{c|}{$-$}  & 
$\mbox{}_{LR}  : 1.5\times 10^{-6}$ &  
\multicolumn{1}{c|}{$-$}  & 
\cite{Abel:2001vy} & $d_{n}$
\\ 
$ $& 
\multicolumn{1}{c|}{$-$}  & 
$\mbox{}_{LR}  : 6.7\times 10^{-8}$ &  
\multicolumn{1}{c|}{$-$}  & 
\cite{Abel:2001vy} & $d_{Hg}$
\\
\hline
\end{tabular}
\caption{\label{Tabledeltau}Various experimental constraints on the mass insertion parameters associated with the down-type squark soft parameters $\tilde{m}_u$ ($LL$ and $RR$ case) and $A^{u}$ ($LR/RL$ case). The given numbers correspond to an average down squark mass $\langle\tilde{m}_{u}\rangle=500\,\mathrm{GeV}$ in one insertion approximation. In all cases we assume $m_{\tilde{g}}/\tilde{m}=1$. As before, the given bounds correspond to $\tan\beta$ in the range $5\lesssim\tan\beta\lesssim15$. Note that the third generation bounds are entirely absent due to the elusiveness of the top sector.}
\end{table}
\begin{table}
\begin{tabular}{|c||r|r|r|cc|}
\hline
\multicolumn{1}{|c||}{$\delta^{\ell}$} & \multicolumn{1}{c|}{$LL$} & \multicolumn{1}{c|}{$LR/RL$} & \multicolumn{1}{c|}{$RR$} & \multicolumn{2}{c|}{source}\\
\hline
\hline
$|\delta_{12}|$& 
$\mbox{}_{LL} : 6.0\times 10^{-4}$ &  
$\mbox{}_{LR}  : 1.0\times 10^{-5}$ &  
\multicolumn{1}{c|}{$-$}  & 
\cite{Ciuchini:2007ha} & $\mu\to e\gamma$
\\
$ $& 
$\mbox{}_{LL} : 2.0\times 10^{-3}$ &  
$\mbox{}_{LR}  : 3.5\times 10^{-5}$ &  
$\mbox{}_{RR} : 9.0\times 10^{-2}$ &  
\cite{Ciuchini:2007ha} & $\mu\to  e e e $
\\
$ $& 
$\mbox{}_{LL} : 2.0\times 10^{-4}$ &  
$\mbox{}_{LR}  : 3.5\times 10^{-5}$ &  
\multicolumn{1}{c|}{$-$}  & 
\cite{Ciuchini:2007ha} & $\mu\to e\,\,\mathrm{in}\,\, {}^{22}\mathrm{Ti}$
\\
\hline
$|\delta_{13}|$& 
$\mbox{}_{LL} : 1.5\times 10^{-1}$ &  
$\mbox{}_{LR}  : 4.0\times 10^{-2}$ &  
\multicolumn{1}{c|}{$-$}  & 
\cite{Ciuchini:2007ha} & $\tau\to e\gamma$
\\
$ $& 
\multicolumn{1}{c|}{$-$}  & 
$\mbox{}_{LR}  : 5.0\times 10^{-1}$ &  
\multicolumn{1}{c|}{$-$}  & 
\cite{Ciuchini:2007ha} & $\tau\to  e e e $
\\
\hline
$|\delta_{23}|$& 
$\mbox{}_{LL} : 1.2\times 10^{-1}$ &  
$\mbox{}_{LR}  : 3.0\times 10^{-2}$ &  
\multicolumn{1}{c|}{$-$}  & 
\cite{Ciuchini:2007ha} & $\tau\to \mu\gamma$
\\
$ $& 
\multicolumn{1}{c|}{$-$}  & 
$\mbox{}_{LR}  : 5.0\times 10^{-1}$ &  
\multicolumn{1}{c|}{$-$}  & 
\cite{Ciuchini:2007ha} & $\tau\to  \mu e e $
\\
\hline
$|\mathrm{Re}\delta_{11}|$& 
\multicolumn{1}{c|}{$-$}  & 
$\mbox{}_{LR}  : 8.0\times 10^{-3}$ &  
\multicolumn{1}{c|}{$-$}  & 
\cite{Gabbiani:1996hi} & $\Delta m_{e}$
\\ 
$|\mathrm{Im}\delta_{11}|$& 
\multicolumn{1}{c|}{$-$}  & 
$\mbox{}_{LR}  : 3.7\times 10^{-7}$ &  
\multicolumn{1}{c|}{$-$}  & 
\cite{Gabbiani:1996hi}& $d_{e}$
\\ 
$ $& 
\multicolumn{1}{c|}{$-$}  & 
$\mbox{}_{LR}  : 1.6\times 10^{-7}$ &  
\multicolumn{1}{c|}{$-$}  & 
\cite{Abel:2001vy} & $d_{e}$
\\
\hline
\end{tabular}
\caption{\label{Tabledeltal}Various experimental constraints on the mass insertion parameters associated with the charged lepton soft parameters $\tilde{m}_l$ ($LL$ and $RR$ case) and $A^{\ell}$ ($LR/RL$ case). The average slepton mass used in deriving these bounds is somewhat smaller than the one in the squark sector, typically $\langle\tilde{m}_{l}\rangle=200\,\mathrm{GeV}$. As usual, $m_{\tilde{g}}/\tilde{m}=1$  and $5\lesssim\tan\beta\lesssim15$ is assumed.
While  the LL bounds generically scale as $1/\tan\beta$ due to the chirality flip in the relevant amplitudes, the LR bounds are essentially $\tan\beta$-insensitive. For more details see e.g. \cite{Ciuchini:2007ha} and references therein.}
\end{table}

In the remaining parts of this section we shall estimate these ``mass insertions'' in theory under consideration and compare them to the bounds. Let us remark that in most cases these bounds are obtained in one insertion approximation and thus ignore the possibility of intricate cancellations. Nevertheless, it is useful to compare theory predictions with these bounds to decide whether there is a generic conflict with constraints coming from SUSY flavour and CP violation.     
 
\subsubsection{SUSY flavour problem}
The data in the Tables \ref{Tabledeltad}, \ref{Tabledeltau} and \ref{Tabledeltal} show that there are strong experimental
constraints on the off-diagonal squark and slepton masses
in the basis where the charged Yukawa matrices are diagonal.
These constraints are particularly strong for the first and
second families of squarks and sleptons. For example 
from the experimental limit on the branching ratio for
$\mu \rightarrow e \gamma $ of about $10^{-11}$,
one deduces that $(\delta^l_{LL})_{12}$, i.e. the ratio of the 12 element of the 
slepton doublet mass squared matrix to the average diagonal
element mass squared element
must be less than about $6\times 10^{-4}$. 
Notice also that in general the constraints involving the third family are
much weaker which is clearly the consequence of the particular difficulty of the heavy flavour physics experiments. 

In general, when one writes down the soft SUSY breaking Lagrangian
there is no {\it a priori} reason why the off-diagonal elements
should be any smaller than the diagonal elements, yet phenomenology
is telling us that they must be. This is the SUSY flavour problem.
In the CMSSM one postulates that
the soft mass matrices must be universal, i.e. proportional
to the unit matrix, a property that is preserved in all Yukawa
bases, although not preserved by radiative corrections
due to flavour violating Yukawa couplings. Thus the CMSSM
predicts small violations of universality at low energies
due to RGE running effects, in the case of leptons due to the
effects of the see-saw mechanism. However the CMSSM is not
a theory but an {\it ansatz}, although its assumptions may be
realized in specific frameworks such as minimal supergravity
(mSUGRA) under certain assumptions about the hidden sector couplings
that break SUSY.

In the present section we shall see that $SU(3)$ family symmetry 
provides an alternative resolution to the SUSY flavour problem\footnote{For previous attempts in this direction see e.g. \cite{Ramage:2003pf} and references therein.},
in which the non-universalities in soft masses are linked to the
Yukawa couplings, leading to small SUSY flavour violation
involving the first and second families, and larger
violations of universality involving the third family,
consistently with the constraints in the Tables. 

\subsubsection{SUSY CP problem\label{CPsection}}
Another facet of the SUSY flavour issue is the so called SUSY CP problem stemming from the fact that in general there could be large extra CP phases coming from the soft SUSY breaking sector of the MSSM. However, the Standard Model accounts for the observed CP violating effects to such a level of accuracy that one must impose stringent bounds on such extra contributions to avoid conflict with experiment. This is, however, often at odds with naturalness. Let us focus on two particularly interesting manifestations of the issue emerging in the CP violating electric dipole moments of electron and neutron and CP violation in rare decays, that are perhaps the most promising channels to see the ``Beyond Standard Model'' CP violating effects. 
\paragraph{Electric dipole moments:}\mbox{}\\
The measured values of the CP violating EDMs of electron, neutron and mercury ($d_{e}<6.3\times 10^{-26} e\, \mathrm{cm}$, $d_{n}<4.3\times 10^{-27} e\, \mathrm{cm}$ and $d_{Hg}<2.1\times 10^{-28} e\, \mathrm{cm}$, (all of them at 90\% C.L.) c.f. for instance \cite{Abel:2001vy}, \cite{Gabbiani:1996hi} and references therein) impose strong constraints on the phases of the (first generation) elements of the LR (and RL) pieces of the soft-SUSY breaking mass matrices driven in the MSSM by the $A$-terms and the $\mu$ parameter, in particular
\be\label{deltadefinition}
\left(\delta^{u,d(l)}_{LR}\right)_{11}\equiv \frac{1}{\langle \tilde m_{u,d(l)}\rangle^{2}_{LR}}\left[\tilde{A}^{u,d(l)}_{11}v_{u,d}-\mu \tilde Y^{u,d(l)}_{11} v_{d,u}\right]
\ee
in the form $\mathrm{Im}(\delta^{u,d}_{LR})_{11}\lesssim10^{-7}$ (for neutralino masses comparable to the masses of relevant scalars).

The standard approach to this issue is imposing the CMSSM universal boundary condition (at the SUSY breaking scale) connecting the trilinear couplings $A^{f}$ to the corresponding Yukawa sector via $A^{f}=A_{0}Y^{f}$ which in the super-CKM basis leads to
\be
\tilde A^{f}=A_{0}\tilde{Y}^{f}=A_{0}
   \left(\begin{matrix} 
      |\tilde{y}_{11}^{f}| &  &\\
       & |\tilde y_{22}^{f}| & \\
      & & |\tilde y_{33}^{f}|\\
   \end{matrix}\right).
\ee
Thus, the experimental bounds $\mathrm{Im}(\delta_{11}^{u,d})_{LR}\lesssim 10^{-6}$,  $\mathrm{Im}(\delta_{11}^{\ell})_{LR}\lesssim 10^{-7}$ (for the low scale values $\langle\tilde m_{q}\rangle\sim 500\,\mathrm{GeV}$, $\langle\tilde m_{l}\rangle\sim 200\,\mathrm{GeV}$ corresponding roughly to $m_{0}\sim 100$ GeV at the GUT scale, c.f. section \ref{runningsect})  can be satisfied provided 
\bea
|\mathrm{Im}(\delta_{11}^{u})_{LR}| & \sim & 0.05\times  \frac{\mathrm{Im}A_{0}v}{m_{0}^{2}}\frac{\tan\beta}{\sqrt{1+\tan^{2}\beta}}|\tilde y_{11}^{u}|  \lesssim  10^{-6}\nn\\
|\mathrm{Im}(\delta_{11}^{d})_{LR}| & \sim &  0.05 \times \frac{\mathrm{Im}A_{0}v}{m_{0}^{2}}\frac{1}{\sqrt{1+\tan^{2}\beta}}|\tilde y_{11}^{d}| \lesssim 10^{-6}\label{mSUGRAEDMs}\\ 
|\mathrm{Im}(\delta_{11}^{l})_{LR}| & \sim &  0.2 \times \frac{\mathrm{Im}A_{0}v}{m_{0}^{2}}\frac{1}{\sqrt{1+\tan^{2}\beta}}|\tilde y_{11}^{l}| \lesssim 10^{-7}\nn 
\eea
where the numerical factors 0.05 and 0.2 account for the RG evolution discussed in more detail in section \ref{runningsect}. For $|\tilde{y}_{11}^{d}|\sim 10^{-3}$, $|\tilde y^{u}_{11}|\sim 10^{-4}$, $|\tilde y_{11}^{e}|\sim 10^{-4}$, $\tan\beta\sim {\cal O}(10)$ this requires approximately $\mathrm{Im}(A_{0}/m_{0}^{2})\lesssim 10^{-2}$
and $\mathrm{Arg}(\mu) < 10^{-1}$, which is not very natural. 

\paragraph{CP violation in rare decays:}\mbox{}\\
Another manifestation of the presence of the SUSY CP phases is the possibly large contributions to the direct and indirect CP violation in rare decays. Since the most stringent constraints come from the neutral kaon system, let us focus here on the $\varepsilon_{K}$ and $\varepsilon_{K}'$ parameters.

While the indirect CP violation $\varepsilon_{K}$ parameter is well under control in the Standard Model,  the situation of the direct CP violating CP parameter $\varepsilon'_{K}$ is still not entirely clear. This is mainly due to an intricate interplay between the two leading contributions coming from the so called $O_{6}$ and $O_{8}$ operators \cite{Gabbiani:1996hi} corresponding to  the structures
\be
O_{6}=\overline{d^{\alpha}_{L}}\gamma^{\mu}s_{L}^{\beta}\sum_{q=u,d,s}\overline{q_{R}}^{\beta}\gamma_{\mu}q_{R}^{\alpha}\qquad O_{8}=\frac{g}{8\pi^{2}}m_{s}\overline{d^{\alpha}_{L}}\sigma^{\mu\nu}t^{A}_{\alpha\beta}G^{A}_{\mu\nu}s^{\beta}_{R} 
\ee
The estimates given in the literature, c.f. for instance \cite{Ciuchini:1997kd}, \cite{Bertolini:1997nf}, \cite{Buras:1999rb}, \cite{Bertolini:2000dy} and references therein, are subject to large uncertainties in the relevant hadronic matrix elements and the general tendency is to somewhat underestimate the measured value of $\varepsilon'_{K}/\varepsilon_{K}$. As has been pointed out by Masiero and Murayama \cite{Masiero:1999ub}, a significant SUSY contribution to this measurable - still compatible with the experimental limits - can emerge for instance in frameworks beyond the CMSSM. This is because the SUSY parts of the Wilson coefficients associated with these operators are sensitive to different pieces of the soft SUSY breaking sector:
\be
C_{6}\propto \frac{\alpha_{s}^{2}}{\langle m_{\tilde{q}}^{2}\rangle}(\delta_{LL}^{d})_{12}P_{6}(x), \qquad C_{8}\propto  \frac{\alpha_{s}\pi}{\langle m_{\tilde{q}}^{2}\rangle}\left[(\delta_{LL}^{d})_{12}P^{LL}_{8}(x)+\frac{m_{\tilde{g}}}{m_{s}}(\delta_{LR}^{d})_{12}P^{LR}_{8}(x)\right]
\ee
Here $P_{6}(x)$ and $P_{8}(x)$ are polynomial  factors of ${\cal O}(1)$ depending on $x=m_{\tilde{g}}^{2}/\langle m_{\tilde{q}}^{2}\rangle$. Notice in particular the enhancement of the $\delta_{LR}^{d}$ contribution to $C_{8}$ due to the large ratio of the gluino to strange quark masses $m_{\tilde{g}}/m_{s}$. This makes $C_{8}$ quite sensitive to the 12 off-diagonal term in the relevant trilinear coupling as a potential source of dominance of $O_{8}$ over $O_{6}$ violating their destructive interference observed in the Standard Model approach.  

\subsection{$SU(3)$ family symmetry predictions}

Remarkably enough, the flavour model considered in this work fits nicely the set of criteria proposed in \cite{Masiero:1999ub} - namely there is a flavour symmetry protecting the flavour and CP violation - recall that in the current model CP is a symmetry of the lagrangian that gets spontaneously broken by the flavon (and a GUT-scale Higgs) VEVs. Second, the Yukawa textures are hierarchical and the CKM mixing is dominated by the down-quark sector contributions (leading to a significant $A_{12}^{d}$ in the SCKM basis). Moreover, the expected tight connection between $\varepsilon'_{K}/\varepsilon_{K}$ and neutron EDM anticipated in \cite{Masiero:1999ub} is realized as both these phenomena turn out to be dependent on a single flavon phase factor $\phi_{1}$, c.f. formulae  (\ref{VEVsinthegoodbasis}), (\ref{numbers}) and (\ref{epsilonprimenumbers}).

\subsubsection{The SCKM rotations - leading order 3 $\times$ 3 analysis}
To assess the SUSY flavour and CP violation in the current model one should consider the soft mass matrices in the SCKM basis , where the mass insertion $\delta$'s are defined (\ref{deltas}). 

The mixing angles parametrizing the relevant super-CKM rotation matrices $U_L^u$ and $U_R^u$ in (\ref{SCKMbasis}) can be (for hierarchical Yukawa matrices) readily read off from (\ref{Yukawacan}):
\be
\theta^{f,L}_{12} \sim  \frac{Y^{f}_{12}}{Y^{f}_{22}},\quad \theta^{f,L}_{13} \sim  \frac{Y^{f}_{13}}{Y^{f}_{33}}, \quad \theta^{f,L}_{23} \sim  \frac{Y^{f}_{23}}{Y^{f}_{33}},\quad 
\theta^{f,R}_{12} \sim  \frac{Y^{f*}_{21}}{Y^{f*}_{22}},\quad \theta^{f,R}_{13} \sim  \frac{Y^{f*}_{31}}{Y^{f*}_{33}}, \quad \theta^{f,R}_{23} \sim  \frac{Y^{f*}_{32}}{Y^{f*}_{33}}, 
\ee
These rotations (by definition) bring the Yukawa matrices to diagonal form
\be\label{Yukawaeigenvalues}
\tilde{Y}^{u}=R^{u} \begin{pmatrix}
-\frac{y_{1}^{u}y_{2}^{u}}{y_{4}^{u}}\varepsilon^{2}\bar\varepsilon^{2} & 0 & 0\\
0 & y_{4}^{u}\varepsilon^{2} & 0\\
0 & 0 & y_{3}^{u}\varepsilon_{3}^{2} \\
\end{pmatrix},\qquad 
\tilde{Y}^{d}=R^{d} \begin{pmatrix}
-\frac{y_{1}^{d}y_{2}^{d}}{y_{4}^{d}}\overline\varepsilon^{4} & 0 & 0 \\
0 & y_{4}^{d}\overline\varepsilon^{2} & 0\\
0 & 0 & y_{3}^{d}\overline\varepsilon_{3}^{2} \\
\end{pmatrix}, 
\ee
\be
\tilde{Y}^{e}=R^{e} \begin{pmatrix}
-\frac{y_{1}^{e}y_{2}^{e}}{y_{4}^{e}}\overline\varepsilon^{4} & 0 & 0\\
0 & y_{4}^{e}\overline\varepsilon^{2} & 0\\
0 & 0 & y_{3}^{e}\overline\varepsilon_{3}^{2} \\
\end{pmatrix}
\ee
with the minus sign at the 11 entry corresponding to the seesaw origin of the first generation masses.
Note that if there is an underlying GUT symmetry (\ref{yGUT}) one can replace $y_{4}$ from (\ref{ySigma4}) to get (below the GUT scale)
\be
\tilde{Y}^{u}=R\! \begin{pmatrix}
-\frac{y_{1}y_{2}}{y_\Sigma\sigma C^{u}}\varepsilon^{2}\bar\varepsilon^{2} & 0 & 0\\
0 & y_\Sigma\sigma C^{u}\varepsilon^{2} & 0\\
0 & 0 & y_{3}\varepsilon_{3}^{2}
\end{pmatrix} \;\;\; \mathrm{and} \;\;\;
\tilde{Y}^{d,e}=R\! \begin{pmatrix}
-\frac{y_{1}y_{2}}{y_\Sigma\sigma C^{d,e}}\overline\varepsilon^{4} & 0 & 0\\
0 & y_\Sigma\sigma C^{d,e}\overline\varepsilon^{2} & 0 \\
0 & 0 & y_{3}\overline\varepsilon_{3}^{2}
\end{pmatrix},
\ee
where the role of the Clebsches $C^{d}=1$ and $C^{e}=3$ in disentangling the down-quark and charged lepton spectra is obvious.

The SCKM rotations $U_{L,R}^f$ are then given (at leading order) by
\bea
U_L^u &\approx& \begin{pmatrix}
1 & t^{u,L}_{12} \bar \varepsilon& t^{u,L}_{13}  \varepsilon^2\bar\varepsilon \\
-t^{u,L}_{12}\bar\varepsilon & 1 & t^{u,L}_{23} \varepsilon^2  \\
{\cal O}(\varepsilon^3\bar\varepsilon)&-t^{u,L}_{23} \varepsilon^2  e^{-i \phi_{3}}
& e^{-i \phi_{3}}   
\end{pmatrix} \label{SCKMrotations},\\
U_R^u &\approx& \begin{pmatrix}
1 & t^{u,R}_{12} \bar \varepsilon& t^{u,R}_{13}  \varepsilon^2\bar\varepsilon\ e^{-2i\phi_{3}} \\
-t^{u,R}_{12}\bar\varepsilon & 1 & t^{u,R}_{23} \varepsilon^2\ e^{-2i\phi_{3}}  \\
-t_{31}^{u,R}  \varepsilon^2\bar\varepsilon\ e^{-i\phi_{3}}  &t^{u,R}_{32} \varepsilon^2  e^{-i \phi_{3}}
& e^{-i \phi_{3}}   
\end{pmatrix}\nn ,
\eea
where the ${\cal O}(1)$ coefficients $t_{ij}^{u,L}$ and  $t_{ij}^{u,R}$ obey:
\begin{eqnarray}
	t^{u,L}_{12} \; & = & \; \frac{ {y}^u_1}{ y_4^u} \;,\quad 
	t^{u,L}_{13} \;  =  \; \frac{y_{1}^{u}} { y_3^u \varepsilon_{3}^2} \;,\quad
	t^{u,L}_{23} \; =   \; \frac{y_{4}^{u}} { y_3^u \varepsilon_{3}^2}\;,\label{LHrotations}\\
	t^{u,R}_{12} \; & = & \; \frac{ {y}^u_2}{ y_4^u} \;,\quad 
	t^{u,R}_{13} \;  =  \; \frac{y_{2}^{u}} { y_3^u \varepsilon_{3}^2}\left(1+\frac{1}{2}\varepsilon_{3}^2\frac{k_{3}^{u^{c}}}{k_{0}^{u^{c}}}\right) \;,\quad
	t^{u,R}_{31} \;  =  \; \frac{1}{2}\frac{y_{2}^{u}k_{2}^{u^{c}}} { y_4^u k_{0}^{u^{c}}}\left(1+\frac{1}{2}\varepsilon_{3}^2\frac{k_{3}^{u^{c}}}{k_{0}^{u^{c}}}\right) \;,\nn\\
	t^{u,R}_{23} \; &=&   \; \frac{y_{4}^{u}} { y_3^u \varepsilon_{3}^2}+\frac{1}{2}\left(\frac{k_{3}^{u^{c}}}{k_{0}^{u^{c}}}\frac{y_{4}^{u}}{y_{3}^{u}}-\frac{k_{2}^{u^{c}}}{k_{0}^{u^{c}}}e^{2i\phi_{3}}\right)\;, 	
	t^{u,R}_{32} \; = \; -\frac{y_{4}^{u}} { y_3^u \varepsilon_{3}^2}e^{2i\phi_{3}}-\frac{1}{2}\left(\frac{k_{3}^{u^{c}}}{k_{0}^{u^{c}}}\frac{y_{4}^{u}}{y_{3}^{u}}e^{2i\phi_{3}}-\frac{k_{2}^{u^{c}}}{k_{0}^{u^{c}}}\right)\;,\nn
\end{eqnarray}
while the corresponding down-quark (and charged lepton) sector quantities are readily obtained from 
$U^{u}_{L,R}$ upon replacing $\varepsilon\to\bar\varepsilon$, $\varepsilon_{3}\to\bar\varepsilon_{3}$, $k^{u^{c}}_{i}\to k^{d^{c}}_{i} (k^{u^{c}}_{i})$, $y_{i}^{u}\to y_{i}^{d} (y_{i}^{e})$ respectively.  
We have enforced a phase convention such that the unphysical phases are removed from the CKM matrix $V_{CKM}=U_{L}^{u\dagger}U_{L}^{d}$ (and $V_{CKM}$ happens to be real at the leading order). 

Notice that the diagonal nature of the $P_{Q,L}$ transformation leads to essentially no K\"ahler sector dependence of the left-handed rotation angles (\ref{LHrotations}) at the leading order. 
Moreover, one can expect that the similarity of entries of $U^{f}_{L,R}$ matrices with the corresponding trilinear sector off-diagonalities shall cancel most of the canonical normalization effects in the A-terms, in particular if the A-terms and Yukawa couplings come from a common source, c.f. similar discussion of the soft masses and K\"ahler metric around formula (\ref{generic}).  

\subsubsection{The SCKM form of $A$ and $m^{2}$\label{Aandm2inSCKM}}
'Sandwiching' the canonically normalized A-terms (\ref{Acan}) and soft masses (\ref{mcan}) between the relevant SCKM rotations (\ref{SCKMrotations}) along the lines of formulae (\ref{SCKMbasis}) one readily obtains the SCKM form of these quantities.   
\paragraph{The SCKM form of the trilinear couplings:}\mbox{}\\
The trilinear coupling matrices, in the SCKM basis, are given by $\tilde A_{f}  \equiv  U^{f\dagger}_{L}A_{f}U^{f}_{R}$ which yields:
\begin{eqnarray}
\tilde A_{u} &  = & A_{0} \begin{pmatrix}
 \tilde{a}^{u}_{11} \  \varepsilon^2\bar \varepsilon^{2}  & 
 \tilde{a}^{u}_{12}\  \varepsilon^2\bar \varepsilon  & 
 \tilde{a}^{u}_{13} \  \varepsilon^2\bar \varepsilon \\
 \tilde{a}^{u}_{21} \  \varepsilon^2\bar \varepsilon& 
 \tilde{a}^{u}_{22} \varepsilon^2 & 
 \tilde{a}^{u}_{23} \varepsilon^2 \\
 \tilde{a}^{u}_{31} \  \varepsilon^2\bar \varepsilon& 
 \tilde{a}^{u}_{32} \varepsilon^2 & 
 \tilde{a}^{u}_{33} \varepsilon_{3}^2 
\end{pmatrix} +\ldots
, \end{eqnarray}
\begin{eqnarray}
\tilde A_{d}&= &  
A_{0} \begin{pmatrix}
 \tilde{a}^{d}_{11} \bar\varepsilon^4  & 
 \tilde{a}^{d}_{12} \bar\varepsilon^3 & 
 \tilde{a}^{d}_{13} \bar\varepsilon^3 \\
 \tilde{a}^{d}_{21} \bar\varepsilon^3 & 
 \tilde{a}^{d}_{22} \bar\varepsilon^2 & 
 \tilde{a}^{d}_{23} \bar\varepsilon^2 \\
 \tilde{a}^{d}_{31} \bar\varepsilon^3 & 
 \tilde{a}^{d}_{32} \bar\varepsilon^2 & 
 \tilde{a}^{d}_{33} \bar\varepsilon_{3}^2 
\end{pmatrix}+\ldots ,
\quad \tilde A_{e} =
A_{0} \begin{pmatrix}
 \tilde{a}^{e}_{11} \bar\varepsilon^4  & 
 \tilde{a}^{e}_{12} \bar\varepsilon^3 & 
 \tilde{a}^{e}_{13} \bar\varepsilon^3 \\
 \tilde{a}^{e}_{21} \bar\varepsilon^3 & 
 \tilde{a}^{e}_{22} \bar\varepsilon^2 & 
 \tilde{a}^{e}_{23} \bar\varepsilon^2 \\
 \tilde{a}^{e}_{31} \bar\varepsilon^3 & 
 \tilde{a}^{e}_{32} \bar\varepsilon^2 & 
 \tilde{a}^{e}_{33} \bar\varepsilon_{3}^2 
\end{pmatrix}+\ldots \nn,
\end{eqnarray}
where
\begin{eqnarray}
 \tilde{a}^{u}_{11} &=& R^u \frac{1}{ {y}_4^{u} } 
\left(  \frac{{a}_4^{u}}{{y}_4^{u}}  {y}_1^u  {y}_2^u - {a}_2^u {y}_1^u -  {a}_1^u {y}_2^u   \right),\quad
 \tilde{a}^{u}_{22} =  R^u{a}_4^{u}  \;,\quad
  \tilde{a}^{u}_{33} =  R^ua_3^u,  \nn \\
 \tilde{a}^{u}_{12} &=&  R^u\left({a}_1^u - \frac{ {a}_4^{u} }{ {y}_4^{u} } {y}_1^u\right)\;,\quad
 \tilde{a}^{u}_{21} =  R^u\left({a}_2^u - \frac{ {a}_4^{u} }{ {y}_4^{u} } {y}_2^u\right)\;,\nn \\
 \label{aijSCKM}
 \tilde{a}^{u}_{13} &=&  R^u\left({a}_1^u-\frac{a_4^{u}}{y_{4}^{u}}y_{1}^{u}\right)\left(1
 - \frac{1}{2}\varepsilon_{3}^2 \tfrac{{k}^{u^{c}}_{3}}{{k}_0^{u^{c}}}+\ldots\right)\;,
 \tilde{a}^{u}_{31} =  R^u\left({a}_2^u-\frac{a_4^{u}}{y_{4}^{u}}y_{2}^{u}\right)e^{2i\phi_{3}}, \\
 %
 \tilde{a}^{u}_{23} &=& R^u \left({a}_4^{u} -  \frac{ a_3^u}{ {y}_3^u}  {y}_4^{u} \right)\left(1
 - \frac{1}{2}\varepsilon_{3}^2 \tfrac{{k}^{u^{c}}_{3}}{{k}_0^{u^{c}}}+\ldots\right) \nn,\;\tilde{a}^{u}_{32} = R^u \left({a}_4^{u} -  \frac{ a_3^u}{ {y}_3^u}  {y}_4^{u} \right)e^{2i\phi_{3}} \;, \nn
\end{eqnarray}
and analogously for the coefficients $ \tilde{a}^{d}_{ij},\tilde{a}^{e}_{ij}$. 

It is worth pointing out that in the limit of the defining basis trilinear couplings being all proportional to the corresponding Yukawas by a common factor (i.e. proportional in the matrix sense) the off-diagonalities in the SCKM form of the A-terms (\ref{aijSCKM}) all drop while the diagonal elements converge to the relevant Yukawa sector eigenvalues (\ref{Yukawaeigenvalues}), up to a global factor. The canonical normalization effect boils down to real rescaling $R$ common to both Yukawa and trilinear sectors. Second, the unitary parts of the canonical normalization transformations $P_{f^{c}}$ drop out as they should (notice that there is no factor proportional to $k_{1,2}^{f^{c}}$ above), because their net effects correspond to just a common change of basis in the Yukawa and A-sector. This, in turn, provides a nontrivial consistency check of all our results.  
\paragraph{The SCKM form of the soft masses:}\mbox{}\\
Due to the unitary nature of the SCKM transformation, the left-handed soft mass terms remain essentially diagonal even in the SCKM basis (to an excellent approximation):
\begin{eqnarray}
(m_{\tilde{u}}^{2})_{LL}&\equiv&
(U_{L}^{u})^{\dagger}m^2_{Q}U_{L}^{u} =  m_0^2 b^{Q}
\mathbbm{1}
, \nn \\
(m_{\tilde{d}}^{2})_{LL}&\equiv &
(U_{L}^{d})^{\dagger}m^2_{Q}U_{L}^{d} =   m_0^2 b^{Q}
\mathbbm{1}
 ,  \label{softmassSCKM2}
\\
(m_{\tilde{l}}^{2})_{LL}&\equiv &
(U_{L}^{\ell})^{\dagger}m^2_{L}U_{L}^{\ell} =  m_0^2 b^{\ell}
\mathbbm{1}
, \nn  
\eea
while the right-handed ones receive nontrivial contributions from the SCKM rotations, c.f. (\ref{SCKMrotations}). Keeping only the leading terms (in powers of $\varepsilon$ and $\overline\varepsilon$ where  $\varepsilon<\overline\varepsilon$) one can write:
\bea
(m_{\tilde{u}}^{2})_{RR}=
(U_{R}^{u})^{\dagger}m^2_{u^{c}}U_{R}^{u} & \sim &   m_0^2 \left[\,{{b}}^{u^c} 
\mathbbm{1}
+
\left(\begin{array}{lll}
\varepsilon^2\bar\varepsilon^{2} \,{\tilde{b}}^{u^c}_{11}&  \varepsilon^2\bar\varepsilon \,{\tilde{b}}^{u^c}_{12} &\varepsilon^2\bar\varepsilon \,{\tilde{b}}^{u^c}_{13} \\
. & \varepsilon^2 \,{\tilde{b}}^{u^c}_{22} & \varepsilon^2 \,{\tilde{b}}^{u^c}_{23}\\
. & . & \varepsilon_{3}^2\,{\tilde{b}}^{u^c}_{33} 
\end{array}\right)+\ldots\right] ,\nn \\
(m_{\tilde{d}}^{2})_{RR}=
(U_{R}^{d})^{\dagger}m^2_{d^{c}}U_{R}^{d} & \sim &   m_0^2 \left[\,{{b}}^{d^c} 
\mathbbm{1}
+
\left(\begin{array}{lll}
\overline{\varepsilon}^4 \,{\tilde{b}}^{d^c}_{11}&  \overline{\varepsilon}^3 \,{\tilde{b}}^{d^c}_{12} & \overline{\varepsilon}^3 \,{\tilde{b}}^{d^c}_{13} \\
.& \overline{\varepsilon}^2 \,{\tilde{b}}^{d^c}_{22} & \overline{\varepsilon}^2 \,{\tilde{b}}^{d^c}_{23}\\
. & . & \overline{\varepsilon}_{3}^2\,{\tilde{b}}^{d^c}_{33} 
\end{array}\right)+\ldots\right] ,  \label{softmassSCKM1} \\
(m_{\tilde{l}}^{2})_{RR}=
(U_{R}^{\ell})^{\dagger}m^2_{e^{c}}U_{R}^{\ell} & \sim &  m_0^2 \left[b^{e^{c}}
\mathbbm{1}
+
\left(\begin{array}{lll}
\bar\varepsilon^4 \,{\tilde{b}}^{e^c}_{11}&  \bar\varepsilon^4 \,{\tilde{b}}^{e^c}_{12} & \bar\varepsilon^4 \,{\tilde{b}}^{e^c}_{13} \\
. & \bar\varepsilon^2 \,{\tilde{b}}^{e^c}_{22} & \bar\varepsilon^2 \,{\tilde{b}}^{e^c}_{23}\\
. & . &\bar \varepsilon_{3}^2\,{\tilde{b}}^{e^c}_{33} 
\end{array}\right)+\ldots\right] , \nn 
\end{eqnarray}
where again the universal pieces remain intact because of the unitarity of the SCKM rotations (and the relevant $b^{f,f^{c}}$ coefficients are identical to those in the canonically normalized case, c.f. (\ref{canbcoefficients})) while the non-universal coefficients are given by:
\bea
\tilde b_{11}^{u^{c}} &= & \frac{1}{k_{0}^{u^{c}}}\left(b_{1}^{u^{c}}- \frac{b_{0}^{u^{c}}}{k_{0}^{u^{c}}}k_{1}^{u^{c}}\right)+ \frac{1}{k_{0}^{u^{c}}}\left(b_{2}^{u^{c}}- \frac{b_{0}^{u^{c}}}{k_{0}^{u^{c}}}k_{2}^{u^{c}}\right)\left(\frac{y_{2}^{u}}{y_{4}^{u}}\right)^{2},\quad
\tilde b_{22}^{u^{c}} = \frac{1}{k_{0}^{u^{c}}}\left(b_{2}^{u^{c}}- \frac{b_{0}^{u^{c}}}{k_{0}^{u^{c}}}k_{2}^{u^{c}}\right),\nn\\
\tilde b_{33}^{u^{c}} &= & \frac{1}{k_{0}^{u^{c}}}\left(b_{3}^{u^{c}}- \frac{b_{0}^{u^{c}}}{k_{0}^{u^{c}}}k_{3}^{u^{c}}\right)\left(1-\frac{k_{3}^{u^{c}}}{k_{0}^{u^{c}}}\varepsilon_{3}^{2}\right)\nn,\qquad
\tilde b_{12}^{u^{c}} = - \frac{1}{k_{0}^{u^{c}}}\left(b_{2}^{u^{c}}- \frac{b_{0}^{u^{c}}}{k_{0}^{u^{c}}}k_{2}^{u^{c}}\right)\frac{y_{2}^{u}}{y_{4}^{u}},\label{softmassSCKM3}\\
\tilde b_{13}^{u^{c}} &= &- \frac{1}{k_{0}^{u^{c}}}\left(b_{2}^{u^{c}}- \frac{b_{0}^{u^{c}}}{k_{0}^{u^{c}}}k_{2}^{u^{c}}\right)\frac{y_{2}^{u}}{y_{4}^{u}}\left(1-\frac{1}{2}\frac{k_{3}^{u^{c}}}{k_{0}^{u^{c}}}\varepsilon_{3}^{2}\right),\label{SCKMbs}\\
\tilde b_{23}^{u^{c}} &= &\frac{1}{k_{0}^{u^{c}}}\left[\left(b_{2}^{u^{c}}- \frac{b_{0}^{u^{c}}}{k_{0}^{u^{c}}}k_{2}^{u^{c}}\right)-\left(b_{3}^{u^{c}}- \frac{b_{0}^{u^{c}}}{k_{0}^{u^{c}}}k_{3}^{u^{c}}\right)\frac{y_{4}^{u}}{y_{3}^{u}}e^{-2i\phi_{3}}\right]\left(1-\frac{1}{2}\frac{k_{3}^{u^{c}}}{k_{0}^{u^{c}}}\right).\nn
\eea
The analogous formulae for the relevant down-type quark (and charged lepton) sector structures are again obtained from (\ref{softmassSCKM2})-(\ref{softmassSCKM3}) upon replacing $\varepsilon\to\bar\varepsilon$, $\varepsilon_{3}\to\bar\varepsilon_{3}$, $k^{u^{c}}_{i}\to k^{d^{c}}_{i} (k^{u^{c}}_{i})$, $y_{i}^{u}\to y_{i}^{d} (y_{i}^{e})$ respectively.
Notice that, as discussed below (\ref{canbcoefficients}), all these coefficients vanish if the K\"ahler metric is proportional to the defining basis soft masses, i.e. $b_{ij}^{f^{c}}=0$. 

We have already seen in section \ref{cansection} that if there is an underlying GUT symmetry, all the (untilded) $b^{f,f^{c}}$, $b_{ij}^{f,f^{c}}$ coefficients tend to align at the leading order yielding:
\be
b^{Q}=b^{u^{c}}=b^{d^{c}}=b^{\ell}=b^{\nu^{c}}=b^{e^{c}}, \quad
b^{Q}_{ij}=b^{u^{c}}_{ij}=b^{d^{c}}_{ij}=b^{\ell}_{ij}=b^{\nu^{c}}_{ij}=b^{e^{c}}_{ij}.
\ee
However, since the SCKM rotations do also feel the GUT symmetry breaking (via the Georgi-Jarlskog Higgs field $\Sigma$), this is no longer the case for the $\tilde b^{f,f^{c}}_{ij}$ coefficients above -- one should take into account the Clebsches associated with the $y_{4}$ coupling in formulae  (\ref{softmassSCKM3}), c.f. also (\ref{ySigma4}) and (\ref{aSigma4}). 

\subsubsection{Leading order predictions for $\delta$ parameters\label{predictions}}
With all this at hand, the leading order $\delta$'s (at the high scale) can now be read off from the SCKM form of the soft mass matrices and trilinear couplings given above. 

Concerning the experimental limits on the left-handed sector off-diagonal terms $(\delta_{LL}^{f})_{ij}$, the freedom to push up the $SU(2)_{L}$-doublet messenger masses $M_{Q,L}$ almost freely (c.f. formula (\ref{Yukawamessengers})) admits for suppressing all the off-diagonal $(\delta_{LL})_{ij}$ parameters as much as desired in order to satisfy all the relevant bounds. In principle, we can even assume $(\delta_{LL}^{f})_{ij}=0$ at the family symmetry breaking scale. In such a case the most stringent constraints in the current framework are those associated with $(\delta_{RR}^{f})_{ij}$ and $(\delta_{LR,RL}^{f})_{ij}$ parameters.

Using the magnitudes of the universal soft mass terms to approximate the average squark and slepton masses entering the relevant prescription (\ref{deltas}) one can estimate (in terms of $\varepsilon$ and $\bar\varepsilon$) \begin{eqnarray}
&&(\delta^u_{RR})_{12} \approx \frac{\tilde b^{u^{c}}_{12}}{ b^{u^c}}\varepsilon^2\bar\varepsilon \; , \quad
(\delta^u_{RR})_{13} \approx \frac{\tilde b^{u^{c}}_{13}}{ b^{u^c}}\varepsilon^2\bar\varepsilon \; , \quad
(\delta^u_{RR})_{23} \approx \frac{\tilde b^{u^{c}}_{23}}{ b^{u^c}}\varepsilon^2 e^{i\Psi_{u}} \; \nn,\\
&&(\delta^d_{RR})_{12} \approx \frac{\tilde b^{d^{c}}_{12}}{ b^{d^c}}\bar\varepsilon^3 \; , \quad\;\;
(\delta^d_{RR})_{13} \approx \frac{\tilde b^{d^{c}}_{13}}{ b^{d^c}}\bar\varepsilon^3 \; , \quad
(\delta^d_{RR})_{23} \approx \frac{\tilde b^{d^{c}}_{23}}{ b^{d^c}}\bar\varepsilon^2 e^{i\Psi_{d}}\; ,\\
&&(\delta^e_{RR})_{12} \approx \frac{\tilde b^{e^{c}}_{12}}{ b^{e^c}}\bar\varepsilon^3 \; , \quad\;\;
(\delta^e_{RR})_{13} \approx \frac{\tilde b^{e^{c}}_{13}}{ b^{e^c}}\bar\varepsilon^3 \; , \quad
(\delta^e_{RR})_{23} \approx \frac{\tilde b^{e^{c}}_{23}}{ b^{e^c}}\bar\varepsilon^2e^{i\Psi_{e}} \; \nn,
\end{eqnarray}
where $\bar\varepsilon^2 \approx 2 \times 10^{-2}$, $\varepsilon^2 \approx 2 \times 10^{-3}$, $\bar\varepsilon^3 \approx 3 \times 10^{-3}$ and $\varepsilon^{2}\bar\varepsilon \approx 4 \times 10^{-4}$. Notice that there are ${\cal O}(1)$ phases emerging already at the leading order in the 23 sector (c.f. formulae (\ref{SCKMbs})) that we have denoted by generic symbols $\Psi_{f}$.

Concerning $\delta_{LR}$ and $\delta_{RL}$, these come from the trilinear couplings and Eq. \ref{deltas}. The flavour-off-diagonal ones obey:
\begin{eqnarray}
&&(\delta^u_{LR})_{12} \approx \frac{A_{0}v}{m_{0}^{2}} \frac{\tilde a^{u}_{12}}{ \sqrt{b^Q b^{u^c}}}\frac{t_{\beta}}{\tilde{t}_{\beta}}\varepsilon^2\bar\varepsilon, \;\;
(\delta^u_{LR})_{13} \approx \frac{A_{0}v}{m_{0}^2} \frac{\tilde a^{u}_{13}}{ \sqrt{b^Q b^{u^c}}}\frac{t_{\beta}}{\tilde{t}_{\beta}}\varepsilon^2\bar\varepsilon, \;\;
(\delta^u_{LR})_{23} \approx \frac{A_{0}v}{m_{0}^2}  \frac{\tilde a^{u}_{23}}{ \sqrt{b^Q b^{u^c}}}\frac{t_{\beta}}{\tilde{t}_{\beta}}\varepsilon^2,\nn \\
\label{deltaLRpred}
&&(\delta^d_{LR})_{12} \approx \frac{A_{0}v}{m_{0}^2} \frac{\tilde a^{d}_{12}}{ \sqrt{b^Q b^{d^c}}}\frac{1}{\tilde{t}_{\beta}}\overline\varepsilon^3, \quad
(\delta^d_{LR})_{13} \approx \frac{A_{0}v}{m_{0}^2} \frac{\tilde a^{d}_{13}}{ \sqrt{b^Q b^{d^c}}}\frac{1}{\tilde{t}_{\beta}}\overline\varepsilon^3, \quad\,
(\delta^d_{LR})_{23} \approx \frac{A_{0}v}{m_{0}^2}  \frac{\tilde a^{d}_{23}}{ \sqrt{b^Q b^{d^c}}}\frac{1}{\tilde{t}_{\beta}}\overline\varepsilon^2,\nn \\
&&(\delta^e_{LR})_{12} \approx \frac{A_{0}v}{m_{0}^2} \frac{\tilde a^{e}_{12}}{ \sqrt{b^L b^{e^c}}}\frac{1}{\tilde{t}_{\beta}}\overline\varepsilon^3 \; , \quad
(\delta^e_{LR})_{13} \approx \frac{A_{0}v}{m_{0}^2} \frac{\tilde a^{e}_{13}}{ \sqrt{b^L b^{e^c}}}\frac{1}{\tilde{t}_{\beta}}\overline\varepsilon^3 \; , \quad
(\delta^e_{LR})_{23} \approx \frac{A_{0}v}{m_{0}^2}  \frac{\tilde a^{e}_{23}}{ \sqrt{b^L b^{e^c}}}\frac{1}{\tilde{t}_{\beta}}\overline\varepsilon^2, \nn \\
\eea
while for the flavour-diagonal entries one has:
\bea
&&(\delta^u_{LR})_{11} \approx \frac{A_{0}v}{m_{0}^{2}} \frac{\tilde a^{u}_{11}}{ \sqrt{b^Q b^{u^c}}}\frac{t_{\beta}}{\tilde{t}_{\beta}}\varepsilon^2\bar\varepsilon^{2}, \quad
(\delta^u_{LR})_{22} \approx \frac{A_{0}v}{m_{0}^2} \frac{\tilde a^{u}_{22}}{ \sqrt{b^Q b^{u^c}}}\frac{t_{\beta}}{\tilde{t}_{\beta}}\varepsilon^2, \nn\\
&&(\delta^d_{LR})_{11} \approx \frac{A_{0}v}{m_{0}^2} \frac{\tilde a^{d}_{11}}{ \sqrt{b^Q b^{d^c}}}\frac{1}{\tilde{t}_{\beta}}\overline\varepsilon^4, \quad
(\delta^d_{LR})_{22} \approx \frac{A_{0}v}{m_{0}^2} \frac{\tilde a^{d}_{22}}{ \sqrt{b^Q b^{d^c}}}\frac{1}{\tilde{t}_{\beta}}\overline\varepsilon^2, \quad\\
&&(\delta^e_{LR})_{11} \approx \frac{A_{0}v}{m_{0}^2} \frac{\tilde a^{e}_{11}}{ \sqrt{b^L b^{e^c}}}\frac{1}{\tilde{t}_{\beta}}\overline\varepsilon^4 \; , \quad
(\delta^e_{LR})_{22} \approx \frac{A_{0}v}{m_{0}^2} \frac{\tilde a^{e}_{22}}{ \sqrt{b^L b^{e^c}}}\frac{1}{\tilde{t}_{\beta}}\overline\varepsilon^2 \; , \nn
\eea
where $v=174$ GeV and $t_{\beta}\equiv \tan\beta$ and $\tilde{t}_{\beta}\equiv \sqrt{1+\tan^{2}\beta}$. The corresponding $\delta_{RL}$ estimates are obtained from those for $\delta_{LR}$ by replacing $\tilde{a}_{ij}^{f}\to \tilde{a}_{ji}^{f*}$. Notice also that for large $\tan\beta$ there is an extra suppression coming from $\tilde t_{\beta}^{-1}$ in $\delta^{d,e}_{LR}$.
\subsubsection{Effects of running:\label{runningsect}}
One should, however, keep in mind that all these predictions emerge at energies where the family symmetry breaking occurs, typically at the unification scale $M_{G}$. Thus, in order to compare these results with the experimental constraints it is necessary to account for the effects of running down to the electroweak scale. 
\paragraph{Soft masses and $\delta_{LL,RR}$:}\mbox{}\\
Concerning the generic running pattern of the soft mass parameters the key is their approximate diagonality in the SCKM basis. The running of the diagonal elements is governed by the flavour universal gauge interactions while the evolution the off-diagonal entries is driven by the strongly hierarchical Yukawa couplings and/or the soft masses themselves, as can be seen \cite{Chung:2003fi} for instance from the relevant formula for $m_{u^{c}}^{2}$ (omitting the anyway redundant RR and LL subscripts): 
\be\label{murunning}
\frac{\mathrm{d}m_{u^c}^{2}}{\mathrm{d}t}=-\frac{1}{4\pi^{2}}G_{u}\mathbbm{1}+\frac{1}{8\pi^{2}}\left(Y_{u}^{\dagger}Y_{u}m_{u^c}^{2}+m_{u^c}^{2}Y_{u}^{\dagger}Y_{u}+2Y_{u}^{\dagger}m_{Q}^{2}Y_{u}+2m^{2}_{H^{u}}Y_{u}^{\dagger}Y_{u}+2A_{u}^{\dagger}A_{u}\right)+\ldots
\ee
where
$$
G_{u}\equiv\frac{8}{3}g_{3}^{2}m_{3}^{2}+g_{1}^{2}\left\{\frac{8}{15}m_{1}^{2}+\frac{2}{5}\left[m^{2}_{H^{u}}-m^{2}_{H^{d}}+\mathrm{Tr}(m_{Q}^{2}-m^{2}_{\ell}-2m_{u^c}^{2}+m_{d^{c}}^{2}+m_{e^{c}}^{2})\right]\right\}.
$$
Here $m_{1,3}$ and $m_{H^{u,d}}$ stand for the relevant gaugino and Higgs sector mass parameters respectively.

Since the off-diagonal part of $m_{u^c}^{2}$ 'feels' only the second term on the RHS of  (\ref{murunning}), the strong hierarchy of the Yukawas and soft masses therein renders their running strongly suppressed with respect to the effects induced on the diagonal elements by means of the first term. 
Thus, at the leading-log level (i.e. taking the RHS of formula (\ref{murunning}) constant) the 
running  effects of the off-diagonal $\delta$-parameters can be approximated by a multiplicative factor due to the evolution of the average squark or slepton mass (squared) in the denominator of (\ref{deltas}) only:
\be\label{deltaLLRRscaling}
(\delta^{f}_{LL,RR})_{i\neq j}^{\mathrm{exp.}}\sim S^{RG}_{f}\times (\delta^{f}_{LL,RR})_{i\neq j}^{\mathrm{theory}}\;,
\ee
where $(\delta^{f}_{LL,RR})_{i\neq j}^{\mathrm{theory}}$ stands for GUT-scale theory predictions and $S^{RG}_{f}$ is the relevant scaling factor. The actual numbers are quite model dependent, nevertheless, $S^{RG}_{f}$  (which is inversely proportional to $\langle \tilde{m}_{q,l}\rangle^{2}$)
 turns out to be in general significantly smaller than one (because $\langle \tilde{m}_{q,l}\rangle^{2}$ gets bigger towards the low scale since there is a minus sign in the first term on the RHS of (\ref{murunning})). For the typical behaviour $\langle \tilde{m}_{q}(M_{Z})\rangle\sim 5 \langle \tilde{m}_{q}(M_{G})\rangle$ and  $\langle \tilde{m}_{l}(M_{Z})\rangle\sim 2 \langle \tilde{m}_{l}(M_{G})\rangle$ one gets roughly $S^{RG}_{q}\sim 0.05$ and $S^{RG}_{l}\sim 0.2$. 

However, as we have seen in the previous discussion (c.f. section \ref{predictions}), the situation in the model under consideration is slightly simplified by the essential GUT-scale diagonality of the left-handed soft mass matrices. This means that  $(\delta^{f}_{LL})_{i\neq j}^{\mathrm{theory}}= 0$ (to a good precision) and there is no effect along (\ref{deltaLLRRscaling}) in the LL sector. Thus, the leading order off-diagonalities in  $(\delta^{f}_{LL})_{i\neq j}$ come from the second, subleading term in (\ref{murunning}) and correspond to the radiative mechanism in the mSUGRA or CMSSM scenarios.  These effects are known \cite{Bertolini:1990if} to be generally small in the considered class of models and the resulting low-scale $(\delta^{f}_{LL})_{i\neq j}^{\mathrm{exp.}}$ well below the experimental limits. 

To see this explicitly, let us as estimate for instance (in
leading-log approximation) the contribution to the $(\delta^\ell_{LL})_{ij}$
insertions (entering namely the lepton flavour violation amplitudes)
assuming universality at
the high scale, i.e. $(m_{\tilde{l}}^{2})_{LL}\sim \mathbbm{1}$. The off-diagonalities in $(\delta^{\ell}_{LL})_{ij}$ due to the running effects (\ref{murunning}) then read:
\begin{eqnarray}\label{leptonrunning}
(\delta^{\ell,\mathrm{RG}}_{LL})_{i\neq j} \approx - \frac{1}{8 \pi^2} \, \frac{3 m_0^2 + A_{0}^2}{\tilde m^2}\, 
\sum_k(Y_\nu)_{ik} (Y^\dagger_\nu)_{kj}\, \log \left( \frac{M_G}{M_k} \right),
\end{eqnarray}
(here $M_{k}$ correspond to the scales of the three right-handed neutrinos). 
Quantitatively, the relevant off-diagonal terms on the RHS of (\ref{leptonrunning}) obey approximately 
\begin{eqnarray}
&&(Y_\nu Y^\dagger_\nu)_{12} \approx (Y_\nu)_{12} (Y^\dagger_\nu)_{22} \approx (Y_\nu)_{13} (Y^\dagger_\nu)_{32} \approx  {\cal O}(\varepsilon^4\bar\varepsilon^{2} )\; , \nn \\
&&(Y_\nu Y^\dagger_\nu)_{13}\approx (Y_\nu)_{13} (Y^\dagger_\nu)_{33} \approx {\cal O}(\varepsilon^2\bar\varepsilon) \; , \\
&&(Y_\nu Y^\dagger_\nu)_{23}\approx (Y_\nu)_{23} (Y^\dagger_\nu)_{33} \approx {\cal O}(\varepsilon^2\bar\varepsilon) \; \nn.
\end{eqnarray}
For $m_0^2\approx A_0^2 \approx \tilde m^2$ and $M_G=M_{Pl}$ we get $\frac{1}{8 \pi^2}\,\frac{3 m_0^2 + A_{0}^2}{\tilde m^2}\,  \log \left( \frac{M_G}{M_k}\right) \sim {\cal O}(1)$ and one gets: 
\begin{eqnarray}
(\delta^{\ell,\mathrm{RG}}_{LL})_{12} \approx  {\cal O}(\varepsilon^4\bar\varepsilon^{2})\; , \quad
(\delta^{\ell,\mathrm{RG}}_{LL})_{13} \approx  {\cal O}(\varepsilon^2\bar\varepsilon)\; , \quad
(\delta^{\ell,\mathrm{RG}}_{LL})_{23} \approx  {\cal O}(\varepsilon^2\bar\varepsilon).
\end{eqnarray}
The rough estimate shows that the RG induced off-diagonal slepton mass matrix elements are well below the present ``bounds'' on the leptonic $\delta$'s. The same reasoning can be adopted to the other flavour sectors with similar results. Therefore, in what remains we can safely forget about the  LL-sector of the soft mass matrices.
\paragraph{Trilinear couplings and $\delta_{LR}$:}\mbox{}\\
Due to self-renormalization properties of the $A$-terms and the corresponding Yukawas the running effects on the first two generation parameters are in general strongly suppressed and don't lead to any substantial change in their order-of-magnitude estimates given in section \ref{Aandm2inSCKM}. However, the running of the diagonal entries of the soft mass matrices again generates a change in the corresponding $\delta$ parameters as in the case of the $\delta_{LL}$ and $\delta_{RR}$ off-diagonal entries (\ref{deltaLLRRscaling}). Thus, as before, there is a generic extra suppression in theory predictions for $\delta_{LR}$ at low energies.

Concerning the effects on phases relevant for CP violation, (apart from the would-be phase of the $\mu$ term entering the $\delta_{LR/RL}$ parameters discussed in section \ref{spontCPsect}) there are in general two other contributions to the trilinear coupling running one should take into account -- the contribution coming from gauginos and the purely self-renormalization of the A-terms (and Yukawas, c.f. \cite{Chung:2003fi} and references therein):   
\be\label{Aurunning}
\frac{\mathrm{d}A_{u}}{\mathrm{d}t}=\frac{1}{16\pi^{2}}\left(G^{A}_{u}A_{u}+G^{Y}_{u}Y_{u}+\mathrm{terms\,\, cubic\,\, in\,\,}A^{f},Y^{f}\right)+\ldots
\ee
where
\bea
G_{u}^{A} & \equiv & -\left(\frac{16}{3}g_{3}^{2}+3g_{2}^{2}+\frac{13}{15}g_{1}^{2}\right)\mathbbm{1}\;, \nn \\
G_{u}^{Y} & \equiv &  \left(\frac{16}{3}g_{3}^{2}m_{3}+3g_{2}^{2}m_{2}+\frac{13}{15}g_{1}^{2}m_{1}\right)\mathbbm{1}\;.
\eea
While the cubic purely self-renormalization terms can be in most cases safely neglected, this is no longer the case of the gauge- and gaugino-induced terms giving rise to $G_{u}^{A}$ and $G_{u}^{Y}$. In particular, the gaugino loops are able to regenerate radiatively the low-scale trilinear couplings even if we put $A_{0}=0$ at the high scale : $A_{u}({M_{Z}})\sim Y_{u}m_{3}({M_{Z}})\log\frac{M^{G}}{M_{Z}}$. For a non-zero initial $A_{0}$, the running of the trilinears strongly resembles the behaviour of the Yukawa couplings. Thus, the net effect is generally rather mild (apart from the possible non-linearities in the third family) and we shall neglect it. This means that also the magnitudes of the $\delta_{LR/RL}$ factors will evolve according to 
\be\label{deltaLRscaling}
(\delta^{f}_{LR,RL})_{i\neq j}^{\mathrm{exp.}}\sim S^{RG}_{f}\times (\delta^{f}_{LR,RL})_{i\neq j}^{\mathrm{theory}}.
\ee

However, as far as the phases are concerned, the effects in the 12 sector are screened by the small Yukawas, that are, moreover, real and diagonal in the SCKM basis (which is to be used as the high-scale initial condition). That means that any would-be extra phase from running can be generated at higher loops only, and thus can hardly compete with the net phases in the trilinear couplings, that (as we shall see in section \ref{refined2x2analysis}) are suppressed utmost by second generation Yukawa-like trilinear coupling and an extra Cabibbo factor $\lambda$.
Thus, we can safely ignore the effects of running on the phases in the relevant trilinear terms. 
\subsection{Discussion\label{discussion}}
With all this at hand we can now compare the experimental constraints discussed in brief in section \ref{experiment} to the estimates of section \ref{predictions}. We shall take the high scale universal soft scalar mass $m_{0}=100$ GeV (which, indeed, is a rather conservative value translating into $\langle\tilde{m}_{u,d}\rangle_{XY}\sim 500$ and GeV $\langle\tilde{m}_{l}\rangle_{XY}\sim 200$ at the low scale respectively) and similarly $A_{0}=100$ GeV. A would-be different choice of the initial condition for  $m_{0}$ and $A_{0}$ shall be, whenever appropriate, accounted for in the displayed formulae by the relevant powers of the $(500\; \mathrm{GeV}/{\langle\tilde{m}_{u,d}\rangle_{LR}})$, $(200\; \mathrm{GeV}/{\langle\tilde{m}_{l}\rangle_{LR}})$ and $(A_{0}/100\; \mathrm{GeV})$ factors respectively. For all the relevant squark-sector $\delta$-parameters (i.e. $\delta_{LR/RL}$ and $\delta_{RR}$ in the current context) we shall assume that the effects of running correct the high-scale predictions only by an extra factor of $S_{u,d}^{RG}\sim 0.05$ for the squark and about $S_{l}^{RG}\sim 0.2$ for the slepton sector quantities, c.f. formulae (\ref{deltaLLRRscaling}) and (\ref{deltaLRscaling}).  
\subsubsection{Limits on off-diagonal $\delta$'s\label{sectionoffdiagonalities}}
\paragraph{Kaons:}\mbox{}\\
In the squark sector, the most serious constraints on the flavour-violating $\delta_{LR/RL}$ and $\delta_{RR}$ parameters come from the CP and flavour violation in the neutral Kaon system that is rather sensitive to $(\delta^{d}_{RR})_{12}$ and $(\delta^{d}_{LR/RL})_{12}$ entries. Taking into account the effects of running and keeping all the relevant $b$-coefficients at ${\cal O}(1)$, the model under consideration predicts (for $\varepsilon\sim 0.05$, $\overline{\varepsilon}\sim 0.15$) numerically about\footnote{Concerning the CP violation in the neutral kaon system (the $\varepsilon_{K}$ and $\varepsilon_{K}'$ parameters) the 'naive' prediction obtained from the absolute value of $(\delta^{d}_{LR})_{12}$ is slightly above the experimental limits and a more detailed analysis (namely of the phase structure of the relevant terms) is needed. It shall be provided in section \ref{CPinKaons}.} 
\bea
|(\delta^{d}_{LR/RL})_{12}|&\sim & 0.25\times S^{RG}_{d}(A_{0}/100\text{ GeV})(500\; \mathrm{GeV}/{\langle\tilde{m}_{d}\rangle_{LR}})^{2} (10/\tan\beta)  \bar\varepsilon^{3}\sim 4\times 10^{-5}\nn \\ 
|(\delta^{d}_{RR})_{12}|&\sim &1\times S^{RG}_{d} (500\; \mathrm{GeV}/{\langle\tilde{m}_{d}\rangle_{RR}})^{2} \bar\varepsilon^{3} \sim 2\times 10^{-4}
\eea 
at the low scale (for $\tan\beta=10$, $A_{0}=100$ GeV and ${\langle\tilde{m}_{d}\rangle_{LR,RR}}=500$ GeV), which are compatible with the experimental limits in both the LR and RR sectors (even for the most stringent bound imposed on its imaginary part\footnote{Recall the essential diagonality of the LL sector that alleviates the 'combined' bound on $\sqrt{\delta_{LL}\delta_{RR}}$  displayed in Table \ref{Tabledeltad} by roughly an order of magnitude.  
}), c.f. Table \ref{Tabledeltad}. Recall, however, that the numbers in Table  \ref{Tabledeltad} are only approximate and a dedicated analysis is needed to quantify the amount of tension (if any) in the very specific setup under consideration.  
\paragraph{$b\to s\,\gamma$ and $b\to s\,l^{+}l^{-}$:}\mbox{}\\
The situation in $b\to s\gamma$ and $b\to s l^{+}l^{-}$ is even better as we predict approximately:
\bea
|(\delta^{d}_{RR})_{23}|&\sim &1\times S^{RG}_{d} (500\; \mathrm{GeV}/{\langle\tilde{m}_{d}\rangle_{RR}})^{2} \bar\varepsilon^{2} \sim 1\times 10^{-3}
\eea
 at the low scale (for ${\langle\tilde{m}_{d}\rangle_{RR}}=500$ GeV), well below the
current limits. Note that the CP-violating phase in $(\delta_{RR}^{d})_{23}$ arises already at the leading level in \eq{canbcoefficients} and thus there is no extra suppression associated with ${\rm Im}(\delta_{RR}^{d})_{23}$.
Concerning the more stringent limits on $(\delta_{LR,RL}^{d})_{23}$, the extra suppressions associated with relatively large $\tan\beta$ regime drives the low-scale prediction 
$$
|(\delta^{d}_{LR/RL})_{23}|\sim 0.25\times S^{RG}_{d}(A_{0}/100\text{ GeV})(500\; \mathrm{GeV}/{\langle\tilde{m}_{d}\rangle_{LR}})^{2} (10/\tan\beta)  \bar\varepsilon^{2}\sim 2\times 10^{-4}
$$
(for $\tan\beta=10$, $A_{0}=100$ GeV and ${\langle\tilde{m}_{d}\rangle_{LR,RR}}=500$ GeV), c.f.  also (\ref{deltaLRpred}), well below the experimental bounds.
\paragraph{$B_{d}-\bar B_{d}$  mixing:}\mbox{}\\
There is a relatively strong limit on the 13-transitions in down-type squark mass matrix associated with the $B_{d}-\bar B_{d}$ system, c.f. Table \ref{Tabledeltad}. The $SU(3)$ model under consideration yields at the low scale: 
\bea
|(\delta^{d}_{RR})_{13}|&\sim &1\times S^{RG}_{d} (500\; \mathrm{GeV}/{\langle\tilde{m}_{d}\rangle_{RR}})^{2} \bar\varepsilon^{3} \sim 2\times 10^{-4}
\\
|(\delta^{d}_{LR/RL})_{13}|&\sim & 0.25\times S^{RG}_{d}(A_{0}/100\text{ GeV})(500\; \mathrm{GeV}/{\langle\tilde{m}_{d}\rangle_{LR}})^{2} (10/\tan\beta)  \bar\varepsilon^{3}\sim 4\times 10^{-5}\nn
\eea
(again for $\tan\beta=10$, $A_{0}=100$ GeV and ${\langle\tilde{m}_{d}\rangle_{LR,RR}}=500$ GeV), both at least one order of magnitude below the experimental constraints.

\paragraph{Off-diagonalities in the up-type squark sector:}\mbox{}\\
Due to the elusiveness of the (s)top-sector, the only quantities worth commenting on are $(\delta_{LR}^{u})_{12}$ and $(\delta_{RR}^{u})_{12}$. Due to the extra suppression associated with $\varepsilon\sim 0.05$ the predictions of the current flavour model  (again for the low-scale value ${\langle\tilde{m}_{u}\rangle_{LR,RR}}=500$ GeV; $\tan\beta$ effects in the up-sector are negligible for $\tan\beta\sim 10$) at the low scale read:
\bea
|(\delta^{u}_{RR})_{12}|&\sim &1\times S^{RG}_{u} (500\; \mathrm{GeV}/{\langle\tilde{m}_{u}\rangle_{RR}})^{2}\varepsilon^{2}\bar\varepsilon \sim 2\times 10^{-5}
\\
|(\delta^{u}_{LR})_{12}|&\sim & 2.5\times S^{RG}_{u}(A_{0}/100\text{ GeV})(500\; \mathrm{GeV}/{\langle\tilde{m}_{u}\rangle_{LR}})^{2}\varepsilon^{2}\bar\varepsilon\sim 5\times 10^{-5}\nn
\eea
As before, these numbers are several orders of magnitude below the experimental limits given in Table \ref{Tabledeltad}.
\paragraph{$\mu\to e\,\gamma$ and $\mu\to e$ nuclear conversions:}\mbox{}\\
As far as the lepton flavour violation is concerned, we should look namely at\footnote{Apart from $(\delta^{\ell}_{LL})_{12}$ that is generated entirely due to the running effects from the diagonal high-scale $(m_{\tilde l}^{2})_{LL}$, c.f. formula (\ref{leptonrunning}) and discussion around.} and $(\delta^{\ell}_{LR})_{12}$ driving $\mu\to e\gamma$ and nuclear $\mu\to e$ conversions. The current model predicts at the low scale:
\be
|(\delta^{\ell}_{LR})_{12}|\sim  0.25\times S^{RG}_{l}(A_{0}/100\text{ GeV})(200\; \mathrm{GeV}/{\langle\tilde{m}_{l}\rangle_{LR}})^{2} (10/\tan\beta)  \bar\varepsilon^{3}\sim 2\times 10^{-4}
\ee
(for $\tan\beta=10$, $A_{0}=100$ GeV and ${\langle\tilde{m}_{l}\rangle_{LR}}=200$ GeV) which is one order of magnitude above the experimental limit for $\mu\to e\gamma$ given in Table \ref{Tabledeltal}. Recall, however, that the numbers given therein are only approximate and the actual numerical bound for the setting under consideration is subject to many effects. For example, the slepton spectrum as low as $200$ GeV is a very conservative choice and the actual bound gets considerably weakened if this value is lifted. Next, since the $LR/RL$ bounds are essentially $\tan\beta$-independent, one can lower the tension by factor of 5 if large $\tan\beta$ regime is employed\footnote{On top of that, the latest fits of the Yukawa sector of the $SU(3)$ model under consideration point towards smaller values of $\bar\varepsilon$ \cite{Ross:2007az} that can account for an extra suppression of about half order of magnitude.}. Last, $A_{0}$ can easily be smaller than $m_{0}$, thus accounting for further relaxation of the tension.

\vskip 5mm
Last remark concerns the effects of a would-be nonzero phase of $\sigma$ as discussed in section \ref{sectionrealsigma}, that would affect\footnote{Remarkably enough, this is the only sector (apart from CP violation in the CKM matrix) that could be actually sensitive to a would-be phase of $\sigma$; c.f. also section \ref{refined2x2analysis}. } (through the relevant changes in the SCKM rotations) the particular phase structure of $(\delta^{f}_{RR})_{ij}$ is immaterial as the relevant bounds (on imaginary parts) are satisfied because of the suppression in the magnitudes rather than the interplay of the phases.  
\vskip 5mm
\noindent
To conclude, the tri-bimaximal $SU(3)$ flavour model under consideration is (up to a certain tension in $\mu\to e\gamma$) compatible with the SUSY flavour violation limits.
\paragraph{CP violation:}\mbox{}\\
As far as the CP violation is concerned, let us first comment on the origin of the CP-violating phases in the model under consideration and identify the potential issues of the simple approach of the preceding sections.
\subsection{Spontaneous CP violation and $SU(3)$ family symmetry\label{spontCPsect}}
$SU(3)$ family symmetry can provide a solution to the SUSY CP puzzle if
CP is broken only spontaneously
when the family symmetry is broken. In this case all
the relevant parameters in the Lagrangian are real and the CP
phases enter only through VEVs of the flavon fields. This in
particular implies that the phase on the $\mu$ parameter is zero
$\phi_{\mu}=0$ (at a high scale where the flavour
symmetry is exact; however, the radiative effects do not lead to
significant departures from zero even after the family symmetry
breaking, c.f. for instance \cite{Ross:2004qn} and references
therein). Similarly, all the dimensionless order unity
coefficients in the original operator expansions, 
namely $y_i^f$, $a_i^f$, $b_i^f$, $k_i^f$, are all real.
Since in this approach the $A^{f}$-terms and the
corresponding Yukawas originate from a common source also the CP
phases of $A^{f}$ and $Y^{f}$ become naturally strongly correlated and
thus the transition to the Super-CKM basis can lead to suppression of
phases in $\tilde{A}_{f}$.
\paragraph{CP violation in the stop/sbottom sector:}\mbox{}\\
Let us start with the CP violation in the stop/sbottom sector. It is
well known \cite{Pilaftsis:1998dd,Pilaftsis:1999qt,Lee:2003nt} that a
would-be large phase in the stop trilinear couplings can have a big
impact on the MSSM Higgs physics - even in case of a CP conserving
MSSM Higgs sector, the stop loops can induce relatively large CP
effects in the relevant Higgs amplitudes due to only mild suppression
in the third generation Yukawa couplings. However, as one can see from
(\ref{aijSCKM}), the imaginary parts of all the $(\tilde{A}^{f})_{33}$
elements\footnote{Though the subleading contributions to the 33 elements are not explicitly displayed in (\ref{aijSCKM}), it is sufficient to recall that they come from the same operator as the leading order 23 entries and have essentially the same structure.}
are screened with respect to the real leading order $\sim
{\cal O}(1)$ contributions associated with the third generation Yukawas:
\be Arg (\tilde{A}^{f})_{33}\sim
Arg\left(e^{2i\phi_{3}}
\frac{a_{4}{\varepsilon_2^{f2}}}{a_{3}\varepsilon_{3}^{2}}\right)\sim \frac{{\varepsilon_2^{f2}}}{\varepsilon_{3}^{2}} \sin 2\phi_{3} \ee where the small factor comes from the
hierarchy of the leading and subleading contributions to $(\hat
A^f)_{33}$ in the Eq. \ref{Eq:A}.  Numerically, one gets (taking $\varepsilon^{u}=\varepsilon\sim 0.05$, $\varepsilon^{d}=\overline\varepsilon\sim 0.15$ and $\varepsilon_{3}\sim 0.5$)
\be 
Arg (\tilde{A}^{t})\sim 10^{-2}\sin 2\phi_{3}, \qquad
Arg (\tilde{A}^{b})\sim 10^{-1}\sin 2\phi_{3} 
\ee
Moreover, the self-renormalization feature of the trilinear
couplings renders the would-be phases induced by running
negligible. Thus, the current model does not lead to any substantial
deviations from the 'standard' CP-conserving MSSM Higgs physics.

\paragraph{CP violation in the 1-2 sector (EDM's, $\varepsilon_{K}'/\varepsilon_{K}$ of neutral Kaons):}\mbox{}\\
Concerning the leading order predictions for the 1-2 block of the relevant trilinear couplings one can see from (\ref{deltaLRpred}) that (in the SCKM basis) the overall magnitudes of the $(A^{u,d,l})_{11}$ (relevant for the dipole moments) as well as $(A^{d})_{12}$ (driving the CP violation in the Kaon system) overshoot the relevant experimental constraints in Table \ref{Tabledeltad} by several orders of magnitude. 
However, since these entries are real\footnote{For complex $\sigma$ as discussed in section \ref{sectionrealsigma} the situation is less clear and shall be addressed in detail in section \ref{refined2x2analysis}.} at the leading order, c.f. (\ref{aijSCKM}), one can not draw any conclusion unless the relevant phases are revealed by means of a more detailed next-to-leading-order analysis focusing on these issues, as we now discuss.

\subsubsection{$2\times 2$ next to leading order approximation\label{refined2x2analysis}}
In the hierarchical case we can focus on the contributions coming from the first two generations since the admixture due to the potentially large third generation diagonal entry is screened by two powers of the $(U_{L,R}^{f})_{13}$ mixing factors that is enough to make it negligible. 

\paragraph{Technical prerequisites:}\mbox{}\\
Thus, let us inspect in detail the 12 blocks of the Yukawa matrices. Recalling first the leading order result (\ref{Yukawacan}) (restoring for conciseness the original set of expansion parameters):
\be\label{2by2YukawaLO}
Y^{f}_{2\times 2,LO}=\left(   
\begin{matrix} 
 0 &
 y^{f}_{12} \varepsilon^f_{1}\varepsilon_{2}^{f}  \\
 y^{f}_{21} \varepsilon^f_{1}\varepsilon_{2}^{f} & 
 y^{f}_{22} \varepsilon_{2}^{f2} 
   \end{matrix}
\right)+\ldots
\ee
(which is entirely real) one should look at next-to leading order contributions to reveal the dominant phases. Inspecting the higher order terms coming (in the defining basis) from the second and third term in formula (\ref{Eq:Y}) one can see that only the 22 entry of $Y_{2\times 2}$ receives a phase (at next-to-leading order) and the matrix form of the NLO correction to (\ref{2by2YukawaLO}) is (before canonical normalization)
\be
\Delta \hat Y^{f}_{2\times 2, NLO}=
\begin{pmatrix} 
 0 &
0  \\
 0& 
\varepsilon_{1}^{f}\varepsilon_{2}^{f} (y_{1}^{f}+y_{2}^{f})e^{i\phi_{1}}  
   \end{pmatrix}
\ee
However, canonical normalization contributes only by a pure rescaling factor $R^{f}$ (at leading order, c.f. (\ref{yijcan})) because, as we have seen in section \ref{cansection}, the unitary piece of $P_{f,f^{c}}$ is not powerful enough to affect the 12 block. Thus, defining 
\be\label{DeltayNLO}
\Delta y_{22}^{f}\equiv R^{f} (y_{1}^{f}+y_{2}^{f})
\ee  
along the lines of (\ref{yijcan}) one can write the (canonically normalized) NLO correction to (\ref{2by2YukawaLO}) as follows: 
\be\label{2by2YukawaNLO}
\Delta Y^{f}_{2\times 2, NLO}=
\begin{pmatrix} 
 0 &
0  \\
 0& 
\Delta y_{22}^{f} \varepsilon_{1}^{f}\varepsilon_{2}^{f}e^{i\phi_{1}}
   \end{pmatrix}
\ee
Collecting  (\ref{2by2YukawaLO}) and  (\ref{2by2YukawaNLO}) one arrives to the desired formula for the relevant part of the canonically normalized 2$\times$2 Yukawa matrix
\be\label{2by2Yukawa}
Y^{f}_{2\times 2}=\left(   
\begin{matrix} 
      0 & \alpha^{f}_{1} \\
      \alpha_{2}^{f} & \beta^{f} e^{i\phi^{f}} \\
   \end{matrix}
\right)+\ldots
\ee
where the following shorthand notation is used: 
\be\label{defalphabeta}
\alpha^{f}_{1}\equiv \varepsilon_{1}^{f}\varepsilon_{2}^{f} y_{12}^{f},\qquad 
\alpha^{f}_{2}\equiv \varepsilon_{1}^{f}\varepsilon_{2}^{f} y_{21}^{f},\qquad 
\beta^{f} e^{i\phi^{f}}=y_{22}^{f}\varepsilon_{2}^{f2}+\Delta y_{22}^{f} \varepsilon_{1}^{f}\varepsilon_{2}^{f}e^{i\phi_{1}}
\ee
Since $\alpha_{1}^{f}$, $\alpha_{2}^{f}$ and $\beta^{f}$ are all real it is easy to derive the generic shape of the relevant (2$\times$ 2 parts of the) SCKM rotation matrices $U_{L}$ and $U_{R}$ (dropping for a while the flavour index):
\be\label{2x2SCKMrotations}
U_{L}=e^{i\omega}\left(   
\begin{matrix} 
      \cos\theta_{2} & e^{i\rho}\sin\theta_{2} \\
      -\sin\theta_{2} e^{i\phi} &  e^{i(\phi+\rho)}\cos\theta_{2} \\
   \end{matrix}
\right)
,\qquad 
U_{R}=e^{i\omega}\left(   
\begin{matrix} 
      \cos\theta_{1} e^{i\phi}&  e^{i(\phi+\rho)}\sin\theta_{1} \\
      -\sin\theta_{1}  & e^{i\rho} \cos\theta_{1} \\
   \end{matrix}
\right)
\ee
where 
\be\label{thetaangles}
\theta_{1}=\theta+\Delta\theta,\quad \theta_{2}=\theta-\Delta\theta,\;\;\;\mathrm{provided}\;\;\; 
\tan 2\theta = \frac{\alpha_{2}+\alpha_{1}}{\beta},\quad \tan 2\Delta\theta = \frac{{\alpha_{2}-\alpha_{1}}}{\beta}.
\ee 
The requirement of diagonality does not impose any further constraints on the remaining phases angles $\omega$ and $\rho$. However, an extra input is provided by the phase-fixing convention we impose on the CKM matrix $V_{CKM}=(U^{u}_{L})^{\dagger}U_{L}^{d}$ (i.e. the standard form with 11 and 12 entries real). Distinguishing the up and down-sector angles and phases by the relevant superscripts one obtains\footnote{Concerning the $\rho^{u}$ phase, one can get a constraint similar to (\ref{Vckm12}) by considering the 21 element of $V_{CKM}$, but in what follows we shall not need $\rho^{u}$.}:
\bea
e^{i(\omega^{d}-\omega^{u})}\left(\cos\theta_{2}^{u}\cos\theta_{2}^{d}+e^{i(\phi^{d}-\phi^{u})}\sin\theta_{2}^{u}\sin\theta_{2}^{d}\right)\in\mathbbm{R}\label{Vckm11}\\
e^{i(\omega^{d}-\omega^{u}+\rho^{d})}\left(\cos\theta_{2}^{u}\sin\theta_{2}^{d}-e^{i(\phi^{d}-\phi^{u})}\sin\theta_{2}^{u}\cos\theta_{2}^{d}\right)\in\mathbbm{R}\label{Vckm12}
\eea
This admits fixing the free phases $\rho_{d}$ and $\omega^{d}-\omega^{u}$: 
it is clear from (\ref{Vckm11}) that $\omega^{d}-\omega^{u}$ is tiny (the phase in (\ref{Vckm11}) enters through a suppressed second term only) while $\rho^{d}$ receives a decent correction from (\ref{Vckm12}). Numerically, one can estimate:
\be\label{rhoinsin}
\sin \rho^{d}\sim -\mathrm{Arg}\left(\cos\theta_{2}^{u}\sin\theta_{2}^{d}-e^{i(\phi^{d}-\phi^{u})}\sin\theta_{2}^{u}\cos\theta_{2}^{d}\right)\sim \sin(\phi^{d}-\phi^{u})\frac{\sin\theta_{2}^{u}}{\sin\theta_{2}^{d}}
\ee
which is very small because of the screening of both $\phi^{u,d}$ and also $\theta_{2}^{u}<\theta_{2}^{d}$. From (\ref{defalphabeta}) we find: 
\be\label{sinrho}
\sin(\phi^{d}-\phi^{u})=\left(\frac{\Delta y_{22}^{d}}{y_{22}^{d}}\frac{\varepsilon_{1}^{d}}{\varepsilon_{2}^{d}}-\frac{\Delta y_{22}^{u}}{y_{22}^{u}}\frac{\varepsilon_{1}^{u}}{\varepsilon_{2}^{u}}\right)\sin\phi_{1}\sim \left(\frac{\Delta y_{22}^{d}}{y_{22}^{d}}-\frac{\Delta y_{22}^{u}}{y_{22}^{u}}\right)\bar\varepsilon\sin\phi_{1}.
\ee
while the ratio of the mixing angles in the quark sector yields ${\sin\theta_{2}^{u}}/{\sin\theta_{2}^{d}}\approx {C_{d}}/{C_{u}}=\tfrac{1}{2}$.
\paragraph{Trilinear couplings in the SCKM basis:}\mbox{}\\
With all this at hand we can approach the SCKM rotations of the trilinear couplings. Repeating  the arguments that brought us to (\ref{2by2YukawaNLO}) one can again parametrize the canonically normalized NLO A-terms (in the $2\times 2$ case) as follows:
\be
A^{f}_{2\times 2}=A_{0}\left(   
\begin{matrix} 
      0 & \gamma^{f}_{1} \\
      \gamma^{f}_{2} & \delta^{f} e^{i\psi^{f}} \\
   \end{matrix}
\right)
\ee
where as in (\ref{defalphabeta}):
\be\label{Aparameters}
\gamma^{f}_{1}\equiv \varepsilon_{1}^{f}\varepsilon_{2}^{f} a_{12}^{f},\qquad 
\gamma^{f}_{2}\equiv \varepsilon_{1}^{f}\varepsilon_{2}^{f} a_{21}^{f},\qquad 
\delta^{f} e^{i\psi^{f}}=a_{22}^{f}\varepsilon_{2}^{f2}+\Delta a_{22}^{f} \varepsilon_{1}^{f}\varepsilon_{2}^{f}e^{i\phi_{1}}
\ee
again with 
\be\label{DeltaaNLO}
\Delta a_{22}^{f}\equiv R^{f} (a_{1}^{f}+a_{2}^{f}).
\ee  
Recall that  $\gamma_{i}^{f}$ and $\delta^{f}$ are again real parameters.
The quantity of major importance is then $\tilde A=U_{L}^{\dagger}A U_{R}$ (with the SCKM rotations taken from (\ref{2x2SCKMrotations})) which obeys:
\be
\tilde A \approx A_{0}\label{AtermSCKM2x2}
\left(   
\begin{matrix} 
      \cos\theta_{2} & e^{i\rho}\sin\theta_{2} \\
      -\sin\theta_{2} e^{i\phi} &  e^{i(\phi+\rho)}\cos\theta_{2} \\
   \end{matrix}
\right)^{\dagger}
\left(   
\begin{matrix} 
      0 & \gamma_{1} \\
      \gamma_{2} & \delta e^{i\psi} \\
   \end{matrix}
\right)
\left(   
\begin{matrix} 
      \cos\theta_{1} e^{i\phi}&  e^{i(\phi+\rho)}\sin\theta_{1} \\
      -\sin\theta_{1}  & e^{i\rho} \cos\theta_{1} \\
   \end{matrix}
\right)
\ee
\paragraph{Extracting the physical parameters:}\mbox{}\\
In what follows we shall focus on the imaginary parts of the 11 and 12 entries of $\tilde A$ driving the SUSY contributions to EDMs and CPV in rare 2nd generation decays respectively.
For the sake of that, it is convenient to write
\bea\label{calA11expansion}
{\rm Im}(\tilde A)_{11}= {\rm Im}\left[U^{L\dagger}_{11}A_{11}U^{R}_{11}+U^{L\dagger}_{12}A_{21}U^{R}_{11}+ U^{L\dagger}_{11}A_{12}U^{R}_{21}+U^{L\dagger}_{12}A_{22}U^{R}_{21}\right]+\ldots \\
{\rm Im}(\tilde A)_{12}= {\rm Im}\left[U^{L\dagger}_{11}A_{11}U^{R}_{12}+U^{L\dagger}_{12}A_{21}U^{R}_{12}+ U^{L\dagger}_{11}A_{12}U^{R}_{22}+U^{L\dagger}_{12}A_{22}U^{R}_{22}\right]+\ldots \label{calA12expansion}
\eea
Taking into account the hierarchical nature of the SCKM rotations, $A_{11}=0$ is very welcome as the potentially dangerous first terms in (\ref{calA11expansion}) and (\ref{calA12expansion})  drop out.

\subsubsection{Electric dipole moments}
Concerning the other contributions to ${\rm Im}(\tilde A)_{11}$, the second and third terms in (\ref{calA11expansion}) are completely $\rho$ and $\omega$ blind and actually drop out too:
\be
{\rm Im}\left[U^{L\dagger}_{12}A_{21}U^{R}_{11}+ U^{L\dagger}_{11}A_{12}U^{R}_{21}\right]={\rm Im}\left[-\sin\theta_{2}e^{-i\phi}\gamma_{2}\cos\theta_{1}e^{i\phi}-\cos\theta_{2}\gamma_{1}\sin\theta_{1}\right]=0
\ee
All that remains is the last term in  (\ref{calA11expansion}) that yields 
\be\label{sinphiminuspsi}
{\rm Im}(\tilde{A})_{11}\approx {\rm Im}\left[U^{L\dagger}_{12}A_{22}U^{R}_{21}\right]
=
A_{0}\delta \sin\theta_{1}\sin\theta_{2} \sin(\psi-\phi)
\ee
where $\delta$ is the magnitude of the 22 entry of $A$, c.f. (\ref{Aparameters}),  and $\theta_{i}$ are the relevant SCKM mixing angles given by formula (\ref{thetaangles})
The $\sin(\psi-\phi)$ factor\footnote{Notice that (as expected) it is actually the phase difference of the 22 entry of the Yukawa and trilinear couplings that drives the EDM's here. Thus, an overall phase associated with $\sigma$ in order to resolve the $\delta_{CKM}$ issue (c.f. section \ref{sectionrealsigma}) does not alter the current predictions.} can be readily estimated from (\ref{defalphabeta}) and (\ref{Aparameters}): 
\be\label{sinEDMs}
\sin(\psi-\phi)\approx\left(\frac{\Delta a_{22}}{a_{22}}-\frac{\Delta y_{22}}{y_{22}}\right)\frac{\varepsilon_{1}}{\varepsilon_{2}}\sin\phi_{1}= \left(\frac{\Delta a_{22}}{a_{22}}-\frac{\Delta y_{22}}{y_{22}}\right)\bar\varepsilon\sin\phi_{1}.
\ee
Restoring the flavour index and using $\delta^{f}=a_{22}^{f}\varepsilon_{2}^{f2}$ one can write at leading order:
\be
{\rm Im}(\tilde{A}^{f})_{11}\approx A_{0}\sin\theta_{1}^{f}\sin\theta_{2}^{f}\varepsilon_{1}^{f}\varepsilon_{2}^{f}\left(\Delta a^{f}_{22}-\frac{a_{22}^{f}}{y_{22}^{f}}\Delta y_{22}^{f}\right)\sin\phi_{1}
\ee
and recast (using identification (\ref{phenoepsilons})) everything in terms of the phenomenological parameters as follows:
\bea\label{EDMsmainresult}
{\rm Im}(\tilde{A}^{u})_{11} & \approx & A_{0}\left(\Delta a^{u}_{22}-\frac{a_{22}^{u}}{y_{22}^{u}}\Delta y_{22}^{u}\right)\frac{1}{4}\bar\varepsilon^{3}\varepsilon^{2}\sin\phi_{1}+\ldots \nn \\
{\rm Im}(\tilde{A}^{d})_{11} & \approx & A_{0}\left(\Delta a^{d}_{22}-\frac{a_{22}^{d}}{y_{22}^{d}}\Delta y_{22}^{d}\right)\overline\varepsilon^{5}\sin\phi_{1}+\ldots \\
{\rm Im}(\tilde{A}^{e})_{11} & \approx & A_{0}\left(\Delta a^{e}_{22}-\frac{a_{22}^{e}}{y_{22}^{e}}\Delta y_{22}^{e}\right)\frac{1}{9}\overline\varepsilon^{5}\sin\phi_{1}+\ldots\nn 
\eea
Here we made use of the fact that the Cabibbo-like angles are given approximately by\footnote{The relative factors of $2$ and $3$ between the down-quark and up-quark and charged lepton mixings come from $y_{22}^{u}:y_{22}^{d}:y_{22}^{e}=2:1:3$, i.e. the Georgi-Jarlskog type of hierarchy in the relevant Yukawa matrices, c.f. formulae (\ref{LHrotations}) and (\ref{ySigma4}).} $\sin\theta_{1,2}^{u,d,e}\sim \bar\varepsilon/2$, $\overline\varepsilon$, $\overline\varepsilon/3$ respectively and $\varepsilon^{u}_{1}\sim \varepsilon\bar\varepsilon$, $\varepsilon^{u}_{2}\sim \varepsilon$, $\varepsilon^{d,e}_{1}\sim \overline\varepsilon^{2}$, $\varepsilon_{2}^{d,e}\sim \overline \varepsilon$.
  
It is obvious that if the $A$-terms and the Yukawa couplings are linearly dependent (in the matrix sense) these quantities vanish as expected. 
Note also that the fact $\alpha_{i}$ and $\gamma_{i}$ are real in a chosen basis does not play any role and the argument can be generalized to any other basis, because the effect is driven by $\sin(\phi-\psi)$ which is sensitive only to the relative phases between the trilinear and Yukawa couplings.
\paragraph{Simple numerical estimate:}\mbox{}\\
If CP is broken (spontaneously) only by the flavon VEVs the imaginary part of the relevant mass-insertion parameter  $|\mathrm{Im}(\delta_{LR}^{f})_{11}|$ is given from (\ref{deltadefinition}) by
\be
|\mathrm{Im}(\delta_{LR}^{u,d,l})_{11}|\sim \frac{v_{u,d}}{\langle\tilde{m}_{u,d}\rangle^{2}_{LR}}|\mathrm{Im}(\tilde A_{u,d,e})_{11}|+\ldots
\ee
using the relevant formulae for the relevant SCKM trilinear couplings (\ref{EDMsmainresult}) one gets approximately (at the GUT scale):
\bea\label{properpowers}
|\mathrm{Im}(\delta_{LR}^{u})_{11}| & \sim &  \frac{A_{0}v}{m_{0}^{2}}\frac{\tan\beta}{\sqrt{1+\tan^{2}\beta}}\left|\Delta a_{22}^{u}-\frac{a_{22}^{u}}{y_{22}^{u}}\Delta y_{22}^{u}\right| \frac{1}{4}\bar\varepsilon^{3}\varepsilon^{2}\sin\phi_{1} \\
|\mathrm{Im}(\delta_{LR}^{d})_{11}| & \sim & \frac{A_{0}v}{m_{0}^{2}}\frac{1}{\sqrt{1+\tan^{2}\beta}}\left|\Delta a_{22}^{d}-\frac{a_{22}^{d}}{y_{22}^{d}}\Delta y_{22}^{d}\right|\overline\varepsilon^{5}\sin\phi_{1}\nn\\
|\mathrm{Im}(\delta_{LR}^{\ell})_{11}| & \sim &  \frac{A_{0}v}{m_{0}^{2}}\frac{1}{\sqrt{1+\tan^{2}\beta}}\left|\Delta a_{22}^{e}-\frac{a_{22}^{e}}{y_{22}^{e}}\Delta y_{22}^{e}\right|\frac{1}{9}\overline\varepsilon^{5}\sin\phi_{1}\nn
\eea
which (taking into account the running effects, c.f. section \ref{runningsect}, and choosing $\tan\beta=10$, $A_{0}=100$ GeV, $m_{0}=100$ GeV yielding at the low scale ${\langle\tilde{m}_{u,d}\rangle_{LR}}\sim 500$ GeV and ${\langle\tilde{m}_{l}\rangle_{LR}}\sim 200$) leads to the following approximate predictions at the low ($M_{Z}$) scale:
\bea
\!\!\!\!\!\!\!\!|\mathrm{Im}(\delta_{LR}^{u})_{11}| \!& \! \sim \! & 3\!\times\! 10^{-7}\frac{A_{0}}{100\text{ GeV}}\!\!\left(\frac{500\,\mathrm{GeV}}{\langle\tilde{m}_{u}\rangle_{LR}}\right)^{2}\!\!\left(\frac{\overline{\varepsilon}}{0.15}\right)^{3}\!\!\!\left(\frac{\varepsilon}{0.05}\right)^{2}\left|\Delta a_{22}^{u}-\frac{a_{22}^{u}}{y_{22}^{u}}\Delta y_{22}^{u}\right| \sin\phi_{1}\nn \\
\!\!\!\!\!\!\!\!|\mathrm{Im}(\delta_{LR}^{d})_{11}| \!& \! \sim \! &1 \!\times\!  10^{-6}\frac{A_{0}}{100\text{ GeV}}\!\!\left(\frac{500\,\mathrm{GeV}}{\langle\tilde{m}_{d}\rangle_{LR}}\right)^{2}\!\!\left(\frac{\overline{\varepsilon}}{0.15}\right)^{5}\!\!\frac{10}{\tan\beta}\left|\Delta a_{22}^{d}-\frac{a_{22}^{d}}{y_{22}^{d}}\Delta y_{22}^{d}\right|\sin\phi_{1} \label{numbers}\\
\!\!\!\!\!\!\!\!|\mathrm{Im}(\delta_{LR}^{\ell})_{11}| \!& \! \sim \! & 4\!\times\!  10^{-7}\frac{A_{0}}{100\text{ GeV}}\!\!\left(\frac{200\,\mathrm{GeV}}{\langle\tilde{m}_{e}\rangle_{LR}}\right)^{2}\!\!\left(\frac{\overline{\varepsilon}}{0.15}\right)^{5}\!\!\frac{10}{\tan\beta}\left|\Delta a_{22}^{e}-\frac{a_{22}^{e}}{y_{22}^{e}}\Delta y_{22}^{e}\right|\sin\phi_{1}\nn
\eea
Thus, one can satisfy the experimental limits on $d_{n}$, $d_{Hg}$  and $d_{e}$ (c.f. Tables \ref{Tabledeltad}, \ref{Tabledeltau} and \ref{Tabledeltal}) even without extra fine-tuning (for the Mercury EDM this can be achieved for somewhat larger $\tan\beta\sim 50$ and/or $\overline{\varepsilon}$ below $0.15$ which is, however, compatible with the Yukawa fits, c.f. \cite{Ross:2007az}).
\paragraph{EDMs in the CMSSM approach:}\mbox{}\\
Let us compare our predictions for the EDMs with the traditional CMSSM to the SUSY CP problem sketched at the beginning of the section.
Taking into account the suppressions entering the first generation Yukawa eigenvalues $|\tilde{y}_{11}^{d}|\sim \overline\varepsilon^{4}$ and $|\tilde{y}_{11}^{u}|\sim \varepsilon^{2}\bar\varepsilon^{2}$ one can recast the generic (GUT-scale) mSUGRA (or CMSSM) relations (\ref{mSUGRAEDMs}) in the form very similar to the one of (\ref{properpowers}):
\bea\label{mSUGRAEDMs2}
|\mathrm{Im}(\delta_{11}^{u})_{LR}| & \sim &\frac{ \mathrm{Im}(A_{0})v}{m_{0}^{2}}\frac{\tan\beta}{\sqrt{1+\tan^{2}\beta}}\varepsilon^{2}\bar\varepsilon^{2}  \\
|\mathrm{Im}(\delta_{11}^{d,l})_{LR}| & \sim & \frac{ \mathrm{Im}(A_{0})v}{m_{0}^{2}}\frac{1}{\sqrt{1+\tan^{2}\beta}}\overline\varepsilon^{4}\nn
\eea
Comparing (\ref{mSUGRAEDMs2}) to (\ref{properpowers})
one can easily see that the effective approach gains at least one power of extra Cabibbo suppression with respect to the mSUGRA (or CMSSM) ansatz at worst, without any reference to the details of the particular SUSY breaking mechanism\footnote{Here, as before, we implicitly assume that the effective soft SUSY-breaking sector does conform the $SU(3)$ family symmetry.}. 
\paragraph{Understanding the results:}\mbox{}\\
The decent total suppression of $\mathrm{Im}(\delta_{LR}^{f})_{11}$ in the $SU(3)$ model under consideration comes from several sources: {\bf 1.} The leading contribution containing a phase comes from the 22 term of $A$ that requires two Cabibbo-like mixings $\sim{\cal O}\left(\overline\varepsilon^{2}\right)$ in the down quark sector and typically even more $\sim{\cal O}\left(\tfrac{1}{4}\bar\varepsilon^{2}\right)$ in the up quark sector to work its way to 11 entry via SCKM rotations.
{\bf 2.} The magnitude of the 22 entry of $A$ is suppressed by ${\cal O}(\overline\varepsilon^{2})$ and ${\cal O}(\varepsilon^{2})$ as for the corresponding Yukawa sector 22 element. 
{\bf 3.} The phase of the 22 term is also suppressed because it comes from the $\langle\phi_{123}\rangle_{2}\langle\phi_{23}\rangle_{2}$ term that is further $\sim{\cal O}\left(\overline\varepsilon\right)$  suppressed with respect to any would-be common phase in the leading part coming from $\langle\phi_{23}\rangle_{2}\langle\phi_{23}\rangle_{2}$. Thus, both $\sin\phi$ and $\sin\psi$ are naturally in $\sim{\cal O}\left(\overline\varepsilon\right)$ region. All this together, this accounts for the fifth power of the small parameters in the formulae (\ref{properpowers}).
\paragraph{EDMs in non-tri-bimaximal $SU(3)$ models:}\mbox{}\\
In non-tribimaximal $SU(3)$ family models there is an ${\cal O}(1)$ phase in either the 12 or 21 entry of $A$, and in addition, the phase of the 22 entry of $A$ relative to that of the Yukawa matrix element is not suppressed. This results in the EDMs predicted in these models being typically of order $\overline\varepsilon^{4}$ as in the CMSSM. The reason why the tri-bimaximal $SU(3)$ models have an additional suppression as discussed above is due to the smaller number of leading order operators entering the 12 sector of $A$. Indeed, the tri-bimaximal model involves fewer physical phases as compared to the more complicated non-tri-bimaximal setups where the 12 block contains contributions from larger number of higher order operators with more phases.

\subsubsection{CP violation in neutral Kaon system - $\varepsilon_{K}' / \varepsilon_{K}\label{CPinKaons}$}
While the rephasing needed to bring $V_{CKM}$ into the standard form leaves $(\tilde A_{u,d})_{11}$ (driving the EDMs) intact, it does indeed affect the $(\tilde A_{d})_{12}$ entry (with possible effects on $\varepsilon_{K}'/\varepsilon_{K}$). 

Taking into account the smallness of $\omega^{u}-\omega^{d}$ and the relevant formulae for $\rho^{d}$ (\ref{rhoinsin}) and (\ref{sinrho}), the net effect generated onto $(\tilde A_{d})_{12}$ is given from (\ref{AtermSCKM2x2}):
\bea
(\tilde A^{d})_{12}&\sim& A^{d}_{12}e^{i\rho^{d}}\cos\theta_{1}^{d}\cos\theta_{2}^{d}- A^{d}_{21}e^{i\rho^{d}}\sin\theta_{1}^{d}\sin\theta_{2}^{d}-A^{d}_{22}e^{i(\rho^{d}-\phi^{d})}\cos\theta_{1}^{d}\sin\theta_{2}^{d}+\ldots\nn\\
& = &A_{0}\left[\gamma_{1}^{d}e^{i\rho^{d}}\cos\theta_{1}^{d}\cos\theta_{2}^{d}-\delta^{d}e^{i(\rho^{d}+\psi^{d}-\phi^{d})}\cos\theta_{1}^{d}\sin\theta_{2}^{d}+{\cal O}(\sin^{2}\theta_{1,2}^{d})\right]+\ldots \label{A12sckm}
\eea
which in terms of $\varepsilon$, $\overline{\varepsilon}$ gives roughly (note that one can not {\it a priori} neglect either of these two terms, because the Cabibbo suppression $\propto \sin\theta_{2}^{d}$ in the second term is compensated by the hierarchy between $\delta$ and $\gamma_{1}$, c.f. (\ref{Aparameters}) so they can both generate comparable contributions)
\be
\mathrm{Re}(\tilde A^{d})_{12}\sim A_{0}\overline{\varepsilon}^{3},\qquad 
\mathrm{Im}(\tilde A^{d})_{12}\lesssim A_{0}\,\overline{\varepsilon}^{4}\sin\phi_{1}
\ee
where we have used\footnote{Notice the apparent rephasing invariance of our results, in particular  (\ref{sinphiminuspsi}), (\ref{sinrho}) and (\ref{A12sckm})  - indeed, all physics depends only on differences of the relevant phase angles! This also justifies the simplification of working with real VEV of the Georgi-Jarlskog field ($\sigma$) only, c.f. section \ref{sectionrealsigma}. } (for small $\rho^{d}$ and $\psi^{d}-\phi^{d}$)
\be
\sin(\rho^{d}+\psi^{d}-\phi^{d})\approx \sin\rho^{d}+\sin(\psi^{d}-\phi^{d})+\ldots
\ee
together with formulae (\ref{sinEDMs}), (\ref{rhoinsin}) and (\ref{sinrho}). Substituting into (\ref{deltas}) one obtains (at the GUT-scale):
\bea\label{epsilonprimegen}
|\mathrm{Re}(\delta_{LR}^{d})_{12}|  & \approx & \frac{A_{0}v}{m_{0}^{2}}\frac{\overline{\varepsilon}^{3}}{\sqrt{1+\tan^{2}\beta}}\,,\\
|\mathrm{Im}(\delta_{LR}^{d})_{12}|  &\lesssim &\frac{A_{0}v}{m_{0}^{2}}\frac{\overline{\varepsilon}^{4}}{\sqrt{1+\tan^{2}\beta}}\sin\phi_{1}\,.\nn
\eea
Taking into account the running effects discussed in section \ref{runningsect} this leads to the low-scale prediction  (choosing $\tan\beta=10$, $A_{0}=100$ GeV, $m_{0}=100$ GeV yielding at the low scale ${\langle\tilde{m}_{d}\rangle_{LR}}\sim 500$ GeV):
\bea\label{epsilonprimenumbers}
|\mathrm{Re}(\delta_{LR}^{d})_{12}| & \sim & 4\times 10^{-5}\frac{A_{0}}{100\text{ GeV}}\!\!\left(\frac{500\,\mathrm{GeV}}{\langle\tilde{m}_{d}\rangle_{LR}}\right)^{2}\!\!\left(\frac{\overline{\varepsilon}}{0.15}\right)^{3}\!\!\frac{10}{\tan\beta}\\
|\mathrm{Im}(\delta_{LR}^{d})_{12}| \!& \! \sim \! & 6 \times 10^{-6}\frac{A_{0}}{100\text{ GeV}}\!\!\left(\frac{500\,\mathrm{GeV}}{\langle\tilde{m}_{d}\rangle_{LR}}\right)^{2}\!\!\left(\frac{\overline{\varepsilon}}{0.15}\right)^{4}\!\!\frac{10}{\tan\beta}\sin\phi_{1}\nn
\eea
We can see that the detailed analysis alleviated the apparent tension in $|\mathrm{Im}(\delta_{LR}^{d})_{12}|$ (identified previously in section \ref{discussion}) by almost an order of magnitude. The numbers (\ref{epsilonprimenumbers}) are indeed well within the current experimental limits, c.f. Table \ref{Tabledeltad}.

\section{Conclusions}
The Flavour Problem of the Standard Model has become
even more intriguing following the discovery of neutrino mass
and mixing. The solution to the Flavour Problem may well
call for a spontaneously broken Family Symmetry, with non-Abelian
Family Symmetries emerging as being particularly suitable for
accounting for large lepton mixing angles via the see-saw
mechanism with sequential dominance and vacuum alignment
of flavon VEVs. In particular tri-bimaximal neutrino
mixing could originate from constrained sequential dominance.
Such non-Abelian family symmetries, when combined appropriately with SUSY,
also control the structure of the soft mass matrices, 
leading to suppressed SUSY induced flavour changing neutral
currents. 
For example $SU(3)$ family symmetry predicts universal
soft scalar mass squared matrices in the symmetry limit.
If CP is spontaneously broken by flavon VEVs then such a scenario
also leads to suppressed SUSY induced CP violation, since CP
is preserved in the symmetry limit. In the real world where
$SU(3)$ family symmetry is spontaneously broken by flavon VEVs,
non-universal soft masses and CP violating effects may be determined in terms of powers of the symmetry breaking flavon VEVs, leading to 
suppressed and calculable effects.

We have analysed this possibility in some detail here,
focussing on the case of $SU(3)$ family symmetry with tri-bimaximal
neutrino mixing from constrained sequential dominance.
Using a bottom-up approach with the only assumption
that the SUSY breaking sector follows the constraints imposed by the (yet unbroken) family symmetry, we expanded
the Yukawa and soft trilinear and scalar mass squared matrices and
kinetic terms in powers of the flavons used to
spontaneously break the $SU(3)$ family symmetry, and the canonically
normalized versions of these matrices were constructed.
The soft mass matrices were then expressed in the Super-CKM basis,
and the leading order mass insertion parameters were calculated,
and are shown to satisfy the experimental constraints from 
flavour changing neutral current processes. 
At each stage in the calculation
we keep track of the dimensionless coefficients,
enabling a full detailed analysis that was not previously
possible for the more complicated previous
version of the $SU(3)$ model without tri-bimaximal mixing.

Assuming that the flavon VEVs break CP spontaneously, the next-to-leading
order effects responsible for CP violation were then estimated,
and the predictions for electric dipole moments were shown to
have an additional Cabibbo suppression compared to both the 
non-tribimaximal $SU(3)$ models and that predicted from the CMSSM,
and may be further suppressed if the high energy trilinear soft 
parameter is assumed to be relatively small. However, in the 
absence of such additional suppression, we expect that $\mu\to e \gamma$ and 
EDMs should be observed soon, if the approach presented here is correct.
We also predict that, unlike in the CMSSM,
$\varepsilon_{K}'/\varepsilon_{K}$ in the neutral Kaon system may be dominated by the SUSY operator $O_8$.
We also discussed the additional constraints from unification,
which can lead to further predictions for flavour changing in our
scheme.

It is interesting to compare the $SU(3)$ approach here to the CMSSM.
In the CMSSM there is no understanding of the origin of flavour,
and instead an {\it ad hoc} assumption is made that the
soft mass matrices take a universal form at some high energy
scale such as the GUT or Planck scale. However the universal
soft masses and the $\mu$ parameter are complex in general,
leading to CP violation and large contributions to EDMs.
In the $SU(3)$ approach, such universality is achieved only
in the exact symmetry limit where the Yukawa couplings are zero.
In the real world the Yukawa couplings originate from flavon VEVs
which break the family symmetry, and these flavon VEVs simultaneously
induce non-universalities in the soft mass matrices and also
spontaneously break the CP symmetry. 

The result is that, at the high
energy scale where flavour emerges from the flavon VEVs,
the soft mass matrices will already contain small off-diagonal 
entries as well as deviations from universality at the diagonal positions, suppressed by parameters similar to those that govern the
small Yukawa couplings. In other words the high scale soft masses
are not predicted to be universal, but the non-universal soft masses
are suppressed by Yukawa couplings, leading to highly suppressed
non-universal soft masses associated with the first and second
families, but large non-universal soft masses associated with the
third family. The CP violation is similarly suppressed,
with the highly suppressed first family CP violation leading to
EDMs being an order of magnitude more suppressed than in the 
CMSSM, yet relatively large CP violation in the $b$-sector, 
much larger than in the CMSSM.

In our bottom-up approach, the Yukawa and soft mass matrices 
all emerge from independent uncorrelated operator expansions
expressed in terms of powers of flavon fields. A similar
expansion for the kinetic form also leads to non-canonical
kinetic terms, and we have rotated these matrices to the canonical
basis of kinetic terms before interpreting the results.
In the canonical basis, the non-universality of the soft trilinear and
scalar mass squared matrices is suppressed by ratios of powers of flavon fields
to the typical mass scale in the messenger sector responsible for the non-renormalizable
operators.  

In this paper we have assumed that the messenger sector
responsible for the soft operators is the same as that responsible for
the Yukawa couplings and kinetic terms, leading to expressions for the
soft mass matrices in terms of the same expansion parameters
$\varepsilon , \bar{\varepsilon}$ which describe the Yukawa matrices.
This is an important assumption since it leads to quite predictive
soft mass matrices in the canonical basis, expressed as powers
of these expansion parameters (with generic ${\cal O}(1)$ coefficients which depend on 
the degree to which the kinetic operators and Yukawa and soft operators
are misaligned) and non-universality controlled by the expansion
parameters $\varepsilon , \bar{\varepsilon}$. With these assumptions, the smallness of FCNCs and CP violation then 
originates entirely from the underlying $SU(3)$ family symmetry breaking pattern. In particular, it leads to an expectation that most of the flavour violation should come from the RR and LR/RL sectors.

However, in specific 
SUSY breaking schemes there may be additional suppressions of
non-universality, corresponding to the case where the 
SUSY breaking effects factorize, clearly seen in the
structure of the coefficients in the canonical basis,
as in mSUGRA for example. In the Super-CKM basis, this possible
additional suppression effect is seen most manifestly.
Although we have not considered such an additional suppression 
here, it may be necessary to invoke it
when dealing with other family symmetries such as $SO(3)$
which only acts on the left-handed sector, leading to 
unsuppressed FCNCs from the right-handed sector unless
such additional suppression is present. However, with $SU(3)$
family symmetry, no such additional suppression is required, but could be desirable for $\mu\to e \gamma$ and EDMs.

In summary, $SU(3)$ family symmetry is capable of providing
a good solution to the flavour problem, not only
in the Standard Model, but also in its SUSY extensions,
leading to predictions for FCNCs and CP violation compatible with the current
limits, some of them ($\mu\to e \gamma$ and EDM's in particular) capable of being observed soon. In particular there
is no need for a very heavy first and second family of squarks and sleptons
in order to obtain EDMs compatible with the current bounds, and it is quite possible to have these
sparticles being relatively light, as required in order to account for the
anomalous magnetic moment of the muon. On the other hand, the slight tension in $\mu\to e \gamma$ is alleviated in the large $\tan\beta$ regime, quite along the lines of the full third family Yukawa unification.
Moreover, a non-universal
third family of squarks and sleptons is a generic prediction
of this approach \cite{Ramage:2003pf}, and it might be the case that the third
family sparticles could be heavier than the first two families.
However CP violation in the third family of squarks and sleptons
is expected to be small, with third family 
CP violating phases being only of order a few per cent,
which implies that electroweak baryogenesis is probably not
viable in this approach, and CP violation in the Higgs sector
will not be observed in any SUSY model of this kind.
The results presented here apply to a wide class of SUSY models,
and are not restricted to the MSSM, or to any particular 
type of SUSY breaking, only relying on the $SU(3)$
symmetry. If the constraints of unification
such as $SO(10)$ are imposed then this leads to even more 
tightly constrained
predictions relating the squark and slepton masses and flavour
violation, as we have discussed.

\section*{Acknowledgements}
The authors are grateful to Stefano Bertolini for clarifications about the current status of $\varepsilon_{K}'/\varepsilon_{K}$ and to Graham G. Ross and Oscar Vives for invaluable discussions. We acknowledge partial support from the following grants:
PPARC Rolling Grant PPA/G/S/2003/00096;
EU Network MRTN-CT-2004-503369;
NATO grant PST.CLG.980066;
EU ILIAS RII3-CT-2004-506222.

\appendix
\section*{Appendices}
\addcontentsline{paper.toc}{section}{Appendices}
\section{Higher order $\phi_{3}$, $\bar\phi_{3}$ insertions in the soft masses and K\"ahler potential \label{Appendixhigherorder}}
The full set of $SU(3)$-allowed contractions that could enter the expansion of the soft SUSY-breaking mass parameters and the K\"ahler potential up to fourth order in number of flavon insertions (with utmost two occurences of the ``small'' VEVs of the $\phi_{123}$, $\phi_{23}$ type) reads: 
\bea\label{basicexpansion}
({C_{0}})_{j}^{i} & : & \underline{\delta^{i}_{j}} \\ 
& & \nn \\
({C_{2}})_{j}^{i}& : & \underline{(\phi_{3})_{j}(\phi_{3}^{*})^{i}}, \underline{(\phi_{23})_{j}(\phi_{23}^{*})^{i}},\underline{(\phi_{123})_{j}(\phi_{123}^{*})^{i}}, \delta_{j}^{i}(\phi_{3}.\phi_{3}^{*}), \delta_{j}^{i}(\phi_{23}.\phi_{23}^{*}), \delta_{j}^{i}(\phi_{123}.\phi_{123}^{*}),  \nn \\
& &  (\bar\phi_{3}^{*})_{j}(\bar\phi_{3})^{i}, (\bar\phi_{23}^{*})_{j}(\bar\phi_{23})^{i},(\bar\phi_{123}^{*})_{j}(\bar\phi_{123})^{i}, \delta_{j}^{i}(\bar\phi_{3}^{*}.\bar\phi_{3}), \delta_{j}^{i}(\bar\phi_{23}^{*}.\bar\phi_{23}), \delta_{j}^{i}(\bar\phi_{123}^{*}.\bar\phi_{123}) \nn \\
& & \nn 
\\
({C_{4}})_{j}^{i} & : & \delta_{j}^{i}(\phi_{3}.\phi_{3}^{*})(\phi_{3}.\phi_{3}^{*}), \delta_{j}^{i}(\phi_{23}.\phi_{23}^{*})(\phi_{3}.\phi_{3}^{*}), \delta_{j}^{i}(\phi_{123}.\phi_{123}^{*})(\phi_{3}.\phi_{3}^{*}), \delta_{j}^{i}(\phi_{23}.\phi_{3}^{*})(\phi_{3}.\phi_{23}^{*}),  \nn \\
&  & \delta_{j}^{i} (\phi_{123}.\phi_{3}^{*})(\phi_{3}.\phi_{123}^{*}), (\phi_{3})_{j}(\phi_{3}^{*})^{i}(\phi_{3}.\phi_{3}^{*}), (\phi_{3})_{j}(\phi_{3}^{*})^{i}(\phi_{23}.\phi_{23}^{*}), (\phi_{3})_{j}(\phi_{3}^{*})^{i}(\phi_{123}.\phi_{123}^{*}),
\nn \\
&  & (\phi_{23})_{j}(\phi_{23}^{*})^{i}(\phi_{3}.\phi_{3}^{*}), (\phi_{123})_{j}(\phi_{123}^{*})^{i}(\phi_{3}.\phi_{3}^{*}), \underline{(\phi_{23})_{j}(\phi_{3}^{*})^{i}(\phi_{3}.\phi_{23}^{*})}, \underline{(\phi_{3})_{j}(\phi_{23}^{*})^{i}(\phi_{23}.\phi_{3}^{*})}, \nn \\
&  & \underline{(\phi_{123})_{j}(\phi_{3}^{*})^{i}(\phi_{3}.\phi_{123}^{*})}, \underline{(\phi_{3})_{j}(\phi_{123}^{*})^{i}(\phi_{123}.\phi_{3}^{*})}, \nn \\
&  &\underline{\varepsilon_{jkl} \varepsilon^{imn}(\phi_{23})_{k}(\phi_{3})_{l}(\phi_{23}^{*})^{m}(\phi_{3}^{*})^{n}}, \underline{\varepsilon_{jkl} \varepsilon^{imn}(\phi_{123})_{k}(\phi_{3})_{l}(\phi_{123}^{*})^{m}(\phi_{3}^{*})^{n}}, \nn \\
& & \delta_{j}^{i}(\bar\phi_{3}.\bar\phi_{3}^{*})(\bar\phi_{3}.\bar\phi_{3}^{*}), \delta_{j}^{i}(\bar\phi_{23}.\bar\phi_{23}^{*})(\bar\phi_{3}.\bar\phi_{3}^{*}), \delta_{j}^{i}(\bar\phi_{123}.\bar\phi_{123}^{*})(\bar\phi_{3}.\bar\phi_{3}^{*}), \delta_{j}^{i}(\bar\phi_{23}.\bar\phi_{3}^{*})(\bar\phi_{3}.\bar\phi_{23}^{*}),  \nn \\
&  & \delta_{j}^{i} (\bar\phi_{123}.\bar\phi_{3}^{*})(\bar\phi_{3}.\bar\phi_{123}^{*}), (\bar\phi_{3}^{*})_{j}(\bar\phi_{3})^{i}(\bar\phi_{3}.\bar\phi_{3}^{*}), (\bar\phi_{3}^{*})_{j}(\bar\phi_{3})^{i}(\bar\phi_{23}.\bar\phi_{23}^{*}), (\bar\phi_{3}^{*})_{j}(\bar\phi_{3})^{i}(\bar\phi_{123}.\bar\phi_{123}^{*}),
\nn \\
&  & (\bar\phi_{23}^{*})_{j}(\bar\phi_{23})^{i}(\bar\phi_{3}.\bar\phi_{3}^{*}), (\bar\phi_{123}^{*})_{j}(\bar\phi_{123})^{i}(\bar\phi_{3}.\bar\phi_{3}^{*}), (\bar\phi_{23}^{*})_{j}(\bar\phi_{3})^{i}(\bar\phi_{3}^{*}.\bar\phi_{23}), (\bar\phi_{3}^{*})_{j}(\bar\phi_{23})^{i}(\bar\phi_{23}^{*}.\bar\phi_{3}), \nn \\
&  & (\bar\phi_{123}^{*})_{j}(\bar\phi_{3})^{i}(\bar\phi_{3}^{*}.\bar\phi_{123}), (\bar\phi_{3}^{*})_{j}(\bar\phi_{123})^{i}(\bar\phi_{123}^{*}.\bar\phi_{3}), \nn \\
&  &\varepsilon_{jkl} \varepsilon^{imn}(\bar\phi_{23}^{*})_{k}(\bar\phi_{3}^{*})_{l}(\bar\phi_{23})^{m}(\bar\phi_{3})^{n}, \varepsilon_{jkl} \varepsilon^{imn}(\bar\phi_{123}^{*})_{k}(\bar\phi_{3}^{*})_{l}(\bar\phi_{123})^{m}(\bar\phi_{3})^{n}, \nn \\
%
& & \delta_{j}^{i}(\phi_{3}.\bar\phi_{3})(\phi_{3}^{*}.\bar\phi_{3}^{*}), 
(\phi_{3})_{j}(\bar\phi_{3})^{i}(\phi_{3}^{*}.\bar\phi_{3}^{*}),  (\phi_{3}^{*})_{j}(\bar\phi_{3}^{*})^{i}(\phi_{3}.\bar\phi_{3}), \nn \\
& & (\phi_{23})_{j}(\bar\phi_{3})^{i}(\bar\phi_{3}^{*}.\phi_{23}^{*}),  (\bar\phi_{23}^{*})_{j}(\phi_{3}^{*})^{i}(\phi_{3}.\bar\phi_{23}), (\phi_{3})_{j}(\bar\phi_{23})^{i}(\bar\phi_{23}^{*}.\phi_{3}^{*}),  (\bar\phi_{3}^{*})_{j}(\phi_{23}^{*})^{i}(\phi_{23}.\bar\phi_{3}), \nn \\
& & (\phi_{123})_{j}(\bar\phi_{3})^{i}(\bar\phi_{3}^{*}.\phi_{123}^{*}),  (\bar\phi_{123}^{*})_{j}(\phi_{3}^{*})^{i}(\phi_{3}.\bar\phi_{123}), \nn \\
& & (\phi_{3})_{j}(\bar\phi_{123})^{i}(\bar\phi_{123}^{*}.\phi_{3}^{*}),  (\bar\phi_{3}^{*})_{j}(\phi_{123}^{*})^{i}(\phi_{123}.\bar\phi_{3}), \nn \\
&  &\varepsilon_{jkl} \varepsilon^{imn}(\phi_{23})_{k}(\bar\phi_{3}^{*})_{l}(\phi_{23}^{*})^{m}(\bar\phi_{3})^{n},  \varepsilon_{jkl} \varepsilon^{imn}(\bar\phi_{23}^{*})_{k}(\phi_{3})_{l}(\bar\phi_{23})^{m}(\phi_{3}^{*})^{n},  \nn \\
&  &\varepsilon_{jkl} \varepsilon^{imn}(\phi_{123})_{k}(\bar\phi_{3}^{*})_{l}(\phi_{123}^{*})^{m}(\bar\phi_{3})^{n},  \varepsilon_{jkl} \varepsilon^{imn}(\bar\phi_{123}^{*})_{k}(\phi_{3})_{l}(\bar\phi_{123})^{m}(\phi_{3}^{*})^{n} \nn \}
\eea
However, there is no need to take all these terms into account as it is easy to see that only the effects of the underlined terms are nontrivial in the sense that they can not be mimicked by changing the Wilson coefficient of some of the other underlined terms.

Note also that if the family symmetry was purely continuous the $D$-flattness aligns {\it all} flavour charges of $\phi$'s and $\bar\phi$'s to be conjugated\footnote{Unlike the discrete charges that (being irrelevant for D-flatness) need not be aligned.}. In such a case the list above can be extended to account also for the $\phi\to \bar\phi^{\dagger}$ symmetry. This, however, does not affect the shape of the ``irreducible'' set of operators.

\bibliographystyle{JHEP.bst}
\providecommand{\href}[2]{#2}\begingroup\raggedright\endgroup

\end{document}